\documentclass[aps,amsmath,preprint,epsf,superscriptaddress,nofootinbib]{revtex4-1}
\pdfoutput=1
\usepackage{graphicx,epsfig,amssymb,amsmath,color,hyperref,setspace,physics}

\begin{document}
\draft \tighten \preprint{IPMU19-0046}
\title{Fast-Rolling Relaxion}
\author{Masahiro Ibe}
\affiliation{Institute for Cosmic Ray Research (ICRR),
The University of Tokyo, Kashiwa, Chiba 277-8582, Japan}
\affiliation{Kavli IPMU (WPI), UTIAS
The University of Tokyo, Kashiwa, Chiba 277-8583, Japan}
\author{Yutaro Shoji}
\affiliation{Kobayashi-Maskawa Institute for the Origin
of Particles and the Universe, Nagoya University, Nagoya 464-8602, Japan}
\author{Motoo Suzuki}
\affiliation{Institute for Cosmic Ray Research (ICRR),
The University of Tokyo, Kashiwa, Chiba 277-8582, Japan}
\affiliation{Kavli IPMU (WPI), UTIAS
The University of Tokyo, Kashiwa, Chiba 277-8583, Japan}

\newcommand{\comment}[1]{$\bigstar$ {\bf #1}$\bigstar$}

\begin{abstract}
We discuss new mechanisms to stop the relaxion field during inflation.
They can be realized in a generic model, including the original model
but in a quite different parameter region.  We consider a {\it
fast-rolling} relaxion field, which can go over the bumps created by
QCD-like dynamics.  Then, in one of the mechanisms, we stop it with a
parametric resonance of the Higgs field. The mechanisms are free from a
super-Planckian field excursion or a gigantic number of $e$-folds of
inflation.  The relaxion has a mass around the weak scale and mixes with
the Higgs boson, which enhances the testability of our mechanisms.
\end{abstract}
\maketitle

\section{Introduction}
The succession of null results of new physics search at the Large Hadron
Collider (LHC) might imply that the energy scale of new physics is
higher than the TeV scale.  If this is the case, the Higgs mass
parameter seems to be severely fine-tuned since it is highly sensitive
to the energy scale of new physics.  The gap between the two scales is
worth considering since a mechanism might reside behind it.

Among various possibilities, it is interesting that the electroweak (EW)
scale is determined dynamically with a new field, called the
relaxion~\cite{Graham:2015cka}.  It scans the Higgs mass squared during
inflation, rolling down a linear potential.  After the relaxion passes
the {\it critical point}, where the Higgs mass squared becomes zero, the
Higgs field develops a vacuum expectation value (VEV).  It triggers a
back-reaction and the relaxion stops around the critical point. Since
the Higgs mass squared is determined independently of its initial value,
we can naturally explain the gap between the two scales. The
back-reaction can be implemented by using an axion-like coupling of the
relaxion to strong dynamics, which generates bumps that depend on the
value of the Higgs field.  Although the idea is in an early stage of
development, there are already a number of papers discussing its
phenomenology or its variants
\cite{Gupta:2019ueh,Banerjee:2019epw,Wang:2018ddr,Hook:2016mqo,
Choi:2016luu,Fonseca:2018xzp,Tangarife:2017vnd,Choi:2016kke,Tangarife:2017rgl,
Matsedonskyi:2017rkq,Abel:2018fqg,Banerjee:2018xmn,Geller:2018xvz,Fonseca:2018kqf,
Frugiuele:2018coc,Davidi:2018sii,Gupta:2018wif,Fonseca:2017crh,Jeong:2017gdy,
Nelson:2017cfv,Batell:2017kho,You:2017kah,Lalak:2016mbv,McAllister:2016vzi,
Flacke:2016szy,Kobayashi:2016bue,Evans:2016htp,Fowlie:2016jlx,Ibanez:2015fcv,
DiChiara:2015euo,Marzola:2015dia,Gupta:2015uea,Jaeckel:2015txa,Antipin:2015jia,
Patil:2015oxa,Hardy:2015laa,Kobakhidze:2015jya}.

In the original model, the relaxion is assumed to slow-roll so that it
stops immediately after the bumps match the slope of the linear
potential. Since the potential barrier is very low at that point, the
first selected vacuum quickly decays or hops into the lower vacuum
through quantum tunneling or the Hubble fluctuations.  The relaxion
continues to decay, or hop for the first bumps, to lower vacua until it
finds a vacuum whose lifetime is much longer than the age of the
universe.  This process, however, requires a gigantic number of
$e$-folds since the Hubble volume during inflation is very small.  To be
consistent with the age of the universe, the bubble nucleation rate in
the final vacuum, $\gamma$, should satisfy
\begin{equation}
 \gamma\ll H_0^4\ ,
\end{equation}
with $H_0$ being the Hubble constant of the current universe. For such a
vacuum to decay during the inflation era, we need a gigantic number of
$e$-folds, $\mathcal N_e$, as
\begin{equation}
 \mathcal N_e>\frac{H^4}{\gamma}\gg\left(\frac{H}{H_0}\right)^4\ ,
\end{equation}
with $H$ being the Hubble constant during inflation.  It easily exceeds
$10^{150}$ even with a low scale inflation, which causes problems in the
inflation sector.  If one considers slow-roll inflation during the
relaxation, for example, the number of $e$-folds is expected to be lower
than $10^{24}$ to avoid fine-tuning problems
\cite{Iso:2015wsf,Dine:2011ws,German:2001tz,Choi:2016luu,Beauchesne:2017ukw}.
Alternatively, one may assume eternal inflation during the relaxation.
It, however, induces multiverse, which obscures the virtue of the
relaxion mechanism.\footnote{ It has been also argued that eternal
inflation is generically incompatible with the (refined) de Sitter
swampland
conjecture~\cite{Matsui:2018bsy,Dimopoulos:2018upl,Kinney:2018kew}.  The
hilltop eternal inflation is marginally consistent with the refined de
Sitter swampland conjecture, but the Hubble constant naturally lies
around the Planck scale in such a case.}

A simple solution to the above problem is to violate one of the
slow-roll conditions and allow the relaxion to fly over the bumps with
its kinetic energy.\footnote{ Another solution is make $\epsilon=0$
after the relaxation, which is discussed in the context of solving the
strong CP problem in the QCD relaxion model \cite{Graham:2015cka}.  } In
compensation, we need another mechanism to stop the relaxion at a
desired position. Several alternative mechanisms have already been
proposed. For example,
\cite{Hook:2016mqo,Fonseca:2018xzp,Tangarife:2017vnd,Choi:2016kke,
Tangarife:2017rgl,Matsedonskyi:2017rkq} use particle production and
\cite{Wang:2018ddr} uses potential instability to stop the
relaxion. Such mechanisms can also avoid a super-Planckian excursion of
the relaxion, which is often troublesome when one UV-completes the
model.  In addition, it also reduces the number of $e$-folds during the
slow-rolling phase.

In this paper, we discuss two stopping mechanisms, which can be
realized without extending the original model. One is to use Higgs
coherent oscillation caused by a parametric resonance, and the other is
rather close to the original mechanism.  In this paper, we focus on the
former one and provide a detailed analysis. For the latter, we only
sketch the idea and give an example in the Appendix
\ref{apx_the_other_way}.  In both mechanisms, the relaxion mass lies
around the weak scale and mixes with the Higgs boson, which enhances the
testability of these mechanisms. In addition, classical rolling always
dominates over quantum fluctuations during the relaxation. Thus, the
Higgs VEV is determined almost identically over different Hubble
patches, which cures the {\it oddity} raised in the original model
\cite{Graham:2015cka}.

This paper is organized as follows. In Section \ref{sec_model}, we
briefly review the original relaxion model with QCD-like dynamics, which
we use throughout this paper.  An overview of our first mechanism is
given in Section \ref{sec_mechanism}. Since we use a parametric
resonance to stop the relaxion, we explain it in Section \ref{sec_edge}.
Then, we discuss resonant particle production in Section
\ref{sec_resonant_pp}.  After a short review of a vacuum decay rate in
Section \ref{sec_vacuum}, we summarize theoretical and experimental
constraints in Section \ref{sec_constraints}. Then, we show an example
parameter region in Section \ref{sec_param_sp}. In Section
\ref{sec_cosmo}, we discuss how an appropriate potential of the relaxion
can be obtained, identifying it with a pseudo Nambu-Goldstone boson
(NGB).  We also discuss the origin of the fast-rolling in the latter
part of the section.  The final section is devoted to our conclusions.

\section{Model}\label{sec_model}
We review the original non-QCD relaxion model \cite{Graham:2015cka} in
this section.

The Lagrangian is defined as
\begin{equation}
 \mathcal{L} = \sqrt{-g}\left[\frac{1}{2}(\partial X)^2
 +|D\Phi|^2
 - V(\Phi,X)\right]+\mathcal{L}_{\rm SM}\ ,
\end{equation}
where
\begin{equation}
\label{eq_potential}
 V(\Phi,X) = (M^2-\epsilon X)|\Phi|^2
 -r\epsilon M^2 X
 +\Lambda^4(|\Phi|^2)\cos\left(\frac{X}{f}\right)
 +\frac{\lambda}{4}|\Phi|^4\ ,
\end{equation}
and $\mathcal L_{\rm SM}$ is the SM Lagrangian without the Higgs
potential.  Here, we denote the relaxion field and the Higgs doublet as
$X$ and $\Phi$, respectively.  The Higgs quartic coupling, $\lambda$, is
almost the same as that in the SM, while the Higgs mass, $M^2$, is
assumed to lie around a new physics scale. To relax the hierarchy
between the two scales, we introduce a coupling between the Higgs boson
and the relaxion, where the small coupling constant, $\epsilon$, is
technically natural (see Section \ref{sec_cosmo}).  Since $\epsilon$
breaks the shift symmetry of the relaxion, we expect a tadpole term of
the relaxion with $|r|\gtrsim1/16\pi^2$. We assume $M^2,~\epsilon$ and
$r$ are positive.  We have another source of the shift symmetry breaking
due to non-QCD strong dynamics, which we assume to have the form of
\begin{equation}
\label{eq_dynamics0}
 \Lambda^4(|\Phi|^2) = \frac{\Lambda_0^4}{2}
 +\Lambda^2_h|\Phi|^2\ .
\end{equation}

The simplest example of the new strong sector is given in
\cite{Graham:2015cka}, where we introduce new light fermions having the
same SM charges as a right handed neutrino, $N$ and $N^c$, and new heavy
fermions having the same Standard Model (SM) charges as a lepton doublet, $L$ and
$L^c$.\footnote{
In this paper, all the fermions without a dagger
represent left-handed Weyl fermions.}
They are charged under a new $SU(3)$ group, whose field strength is
denoted as $G^a_{\mu\nu}$. The relevant part of the Lagrangian is given
by
\begin{equation}
\label{eq_dynamics}
 \mathcal L_{\rm UV} = \mathcal L_{\rm kin.}
 -\frac{1}{32\pi^2}\frac{X}{f}G^a_{\mu\nu}{\tilde G}^{a\mu\nu}
 +m_LLL^c
 +m_NNN^c
 +(y\Phi LN^c
 +\tilde y\Phi^\dagger L^cN + h.c.)\ ,
\end{equation}
where $f$ is a decay constant of the relaxion, $y$ and $\tilde y$ are
Yukawa couplings, and $\mathcal L_{\rm kin}$ includes the kinetic terms.
After integrating out $L$,  the mass of $N$ is given by
\begin{equation}
m^{(\rm eff)}_N \simeq m_N
 -\frac{y\tilde y}{8\pi^2}m_L\ln\frac{M^2}{m_L^2}
 -\frac{y\tilde y}{m_L}|\Phi|^2\ ,
 \label{eq_massN}
\end{equation}
at the one-loop level.  If $m^{(\rm eff)}_N$ is smaller than the
dynamical scale of the new strong dynamics, $\Lambda_c$, $NN^c$
condensates and generates
\begin{align}
 \Lambda_0^4&\simeq 
 2\left(m_N-\frac{y\tilde y}{8\pi^2}m_L\ln\frac{M^2}{m_L^2}\right)\Lambda_c^3\ ,\\
 \Lambda_h^2&\simeq
 -\frac{y\tilde y}{m_L}\Lambda_c^3\ .
 \label{eq_lambda_h}
\end{align}
In this paper, we assume that $\Lambda_c$ is smaller than $m_L$.
\section{Mechanism}\label{sec_mechanism}
In this section, we give an overview of our first mechanism. The main
difference from the original model is that the relaxion does not
slow-roll due to
\begin{equation}
H\lesssim\left|\frac{\ddot X}{\dot X}\right|\ ,
\end{equation}
where $H$ is the Hubble constant during inflation and the dot indicates
the time derivative.  In this case, the original stopping mechanism does
not work since the kinetic energy of the relaxion is not negligible.
Instead, we use the following mechanism.
\begin{enumerate}
 \item The initial Higgs mass squared is assumed to be positive at the
       onset of relaxation. The relaxion rolls down the potential with
       its terminal velocity until the Higgs mass squared becomes
       smaller than $\Lambda_h^2$.
 \item When the Higgs mass squared decreases down to a certain positive
       value, the Higgs field starts to oscillate homogeneously due to a
       parametric resonance, which is explained in the next section.
       The amplitude of the oscillation grows {\it gradually} as the Higgs
       mass becomes smaller.
 \item The Higgs oscillation is then fed back to the relaxion roll in
       the following ways; (i) the height of the bumps of the relaxion
       potential oscillates, (ii) the relaxion slows down due to the
       additional Hubble friction acting on the Higgs field. Since they
       hinder the roll, the relaxion eventually hits a bump and bounces
       back.
 \item Just after the bounce of the relaxion, the Higgs field finds its
       mass is negative due to the negative contribution from the second
       term in Eq.\,(\ref{eq_dynamics0}), and develops a VEV. The sudden
       development of the Higgs VEV plays a role of an anchor, which
       secures the relaxion in a potential well between the bumps.
 \item Once the anchor bites, the Higgs field can not return to the
       symmetric point due to the Hubble friction, and the oscillation
       around the VEV dumps quickly. It finalizes the relaxation.
\end{enumerate}

It should be noticed that the Higgs mass squared is typically {\it
positive} when the relaxion stops unlike in the original mechanism.
However, it does not matter since the Higgs mass squared has been
reduced to a value smaller than $\Lambda_h^2$, and the negative Higgs
mass squared is provided by $\Lambda_h^2\cos(X/f)$.

\section{Edge solution}\label{sec_edge}
In this section, we show the behavior of the Higgs field around the
critical point, which triggers our stopping mechanism.

Before going into discussion, let us summarize the equations of motion
for the homogeneous modes of the relaxion and the Higgs field, which are
denoted by ${\bar X}(t)$ and ${\bar h}(t)$, respectively. They are given
by
\begin{align}
 \ddot {\bar X}+3H\dot {\bar X}&=
 \epsilon\left(rM^2+\frac{{\bar h}^2}{2}\right)
 +\frac{\Lambda^4_0+\Lambda_h^2{\bar h}^2}{2f}\sin\frac{{\bar X}}{f}\ ,\label{eq_full_eom_X}\\
\ddot {\bar h}+3H\dot {\bar h}&=
 -(M^2-\epsilon {\bar X}){\bar h}
 -\frac{\lambda}{4} {\bar h}^3
 -\Lambda_h^2{\bar h}\cos\frac{{\bar X}}{f}\ ,
 \label{eq_full_eom_h}
\end{align}
where we choose a basis of the Higgs field as
\begin{equation}
 \Phi=\frac{1}{\sqrt{2}}
\begin{pmatrix}
 \eta^1+i\eta^2\\
 h+i\eta^3
\end{pmatrix}\ .
\end{equation}
Here, $\eta^a$'s are taken so that their expectation values are always
zero without loss of generality.

One might think that nothing happens during the roll of the relaxion
since the simplest solution to the equations of motion is
\begin{align}
 \bar X(t,{\bf x})&\simeq\frac{r\epsilon M^2}{3H}(t-t_0)\ ,\label{eq_simple_X}\\
 \bar h(t,{\bf x})&\simeq0\ ,\label{eq_simple_h}
\end{align}
with $t_0$ being a constant.  However, it is not always a stable
solution and there appears another branch of stable solutions where
the oscillation of $\bar h$ grows gradually.
\subsection{Simplified System}
Let us first see that the solution of ~\eqref{eq_simple_X} and
\eqref{eq_simple_h} is not stable using a simple differential
equation. Plugging Eq.~\eqref{eq_simple_X} into
Eq.~\eqref{eq_full_eom_h}, one obtains a differential equation that
looks like
\begin{align}
 \ddot y+\left(m^2(t)+\Lambda^2\cos\omega t\right)y
 +\frac{\lambda}{4}y^3&=0\ ,
 \label{eq_mathieulike}
\end{align}
with $y \sim \bar{h}$ and 
\begin{eqnarray}
 m^2(t)&=&m_0^2-\delta m^2 t\ .
\end{eqnarray}
Here, $\omega,~m_0^2$ and $\delta m^2$ are the constants determined by
the model parameters and we have ignored the Hubble friction acting on
the Higgs field for simplicity.

\begin{figure}[t]
 \begin{center}
  \begin{minipage}{0.48\linewidth}
   \includegraphics[width=\linewidth]{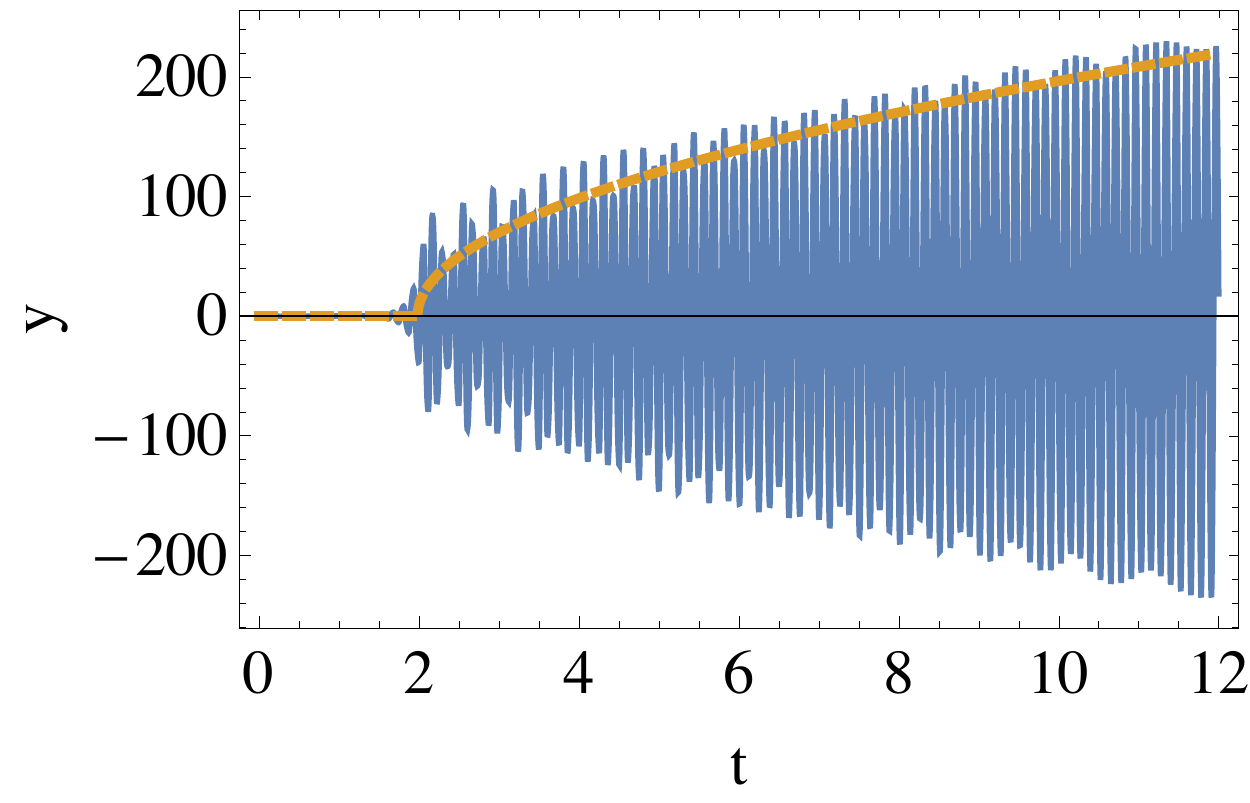}
  \end{minipage}
  \begin{minipage}{0.48\linewidth}
   \includegraphics[width=0.85\linewidth]{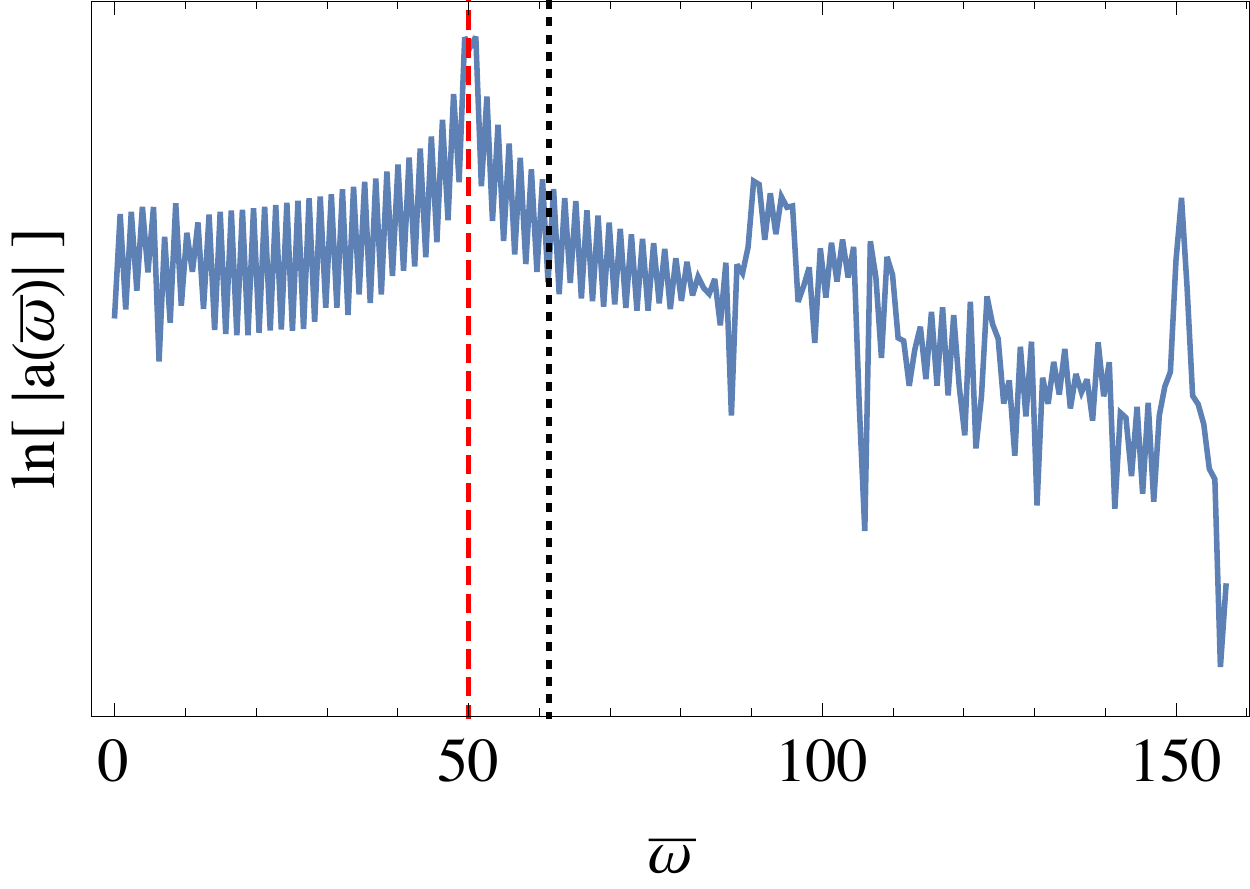}
  \end{minipage}

  \caption{A solution to Eq.~\eqref{eq_mathieulike} with the parameters
  given in \eqref{eq_coh_params}. {\it Left:} The solid blue line shows
  the numerical solution and the dashed orange line shows the amplitude
  obtained with Eq.~\eqref{eq_resoCondHiggs}. The solution represents
  the evolution of the Higgs field in a simplified treatment. {\it
  Right:} The coefficients of the Fourier cosine series,
  $a(\bar\omega)$, with $\bar\omega$ being a frequency of $y(t)$. It is
  calculated for $8<t<12$. The dashed red line indicates
  $\bar\omega=\omega/2$ and the dotted black line indicates
  $\bar\omega=\sqrt{\overline m^2}$.}  \label{fig_edge}
 \end{center}
\end{figure}

A numerical solution to Eq.~\eqref{eq_mathieulike} is shown in the left
panel of Fig.~\ref{fig_edge}. The solid blue line corresponds to the
solution with
\begin{equation}
 \lambda=0.52,~m_0^2=4700,~\delta m^2=470,~\Lambda=60,~\omega=100\ ,\label{eq_coh_params}
\end{equation}
where the unit of scale is arbitrary. The initial conditions of $y$ are
irrelevant if it is small enough but non-zero.

We can see that $y$ starts to oscillate around $t\sim2$ and the
amplitude grows gradually afterwards, which shows that the solution
described by \eqref{eq_simple_X} and \eqref{eq_simple_h} is unstable
with the above parameter set.  We call the solutions that behave like
this the {\it edge solutions} and explain how they are stabilized below.

If we ignore the non-linear term and the time dependence of $m^2$ in
Eq.\,(\ref{eq_mathieulike}), the equation becomes the so-called Mathieu
equation, which exhibits instability for\footnote{Here, we consider the
first resonance band of the Mathieu equation. We will comment on the
effects of the higher resonance bands later.}
\begin{equation}
 \frac{\omega^2-2\Lambda^2}{4}\lesssim m^2\lesssim\frac{\omega^2+2\Lambda^2}{4}\
  .
\end{equation}
Since $m^2$ decreases monotonically due to the motion of the relaxion,
it enters the above resonance band and $y$ starts to grow exponentially.

The exponential growth, however, stops immediately after the non-linear
term in Eq.\,(\ref{eq_mathieulike}) becomes important.  To understand
the effect of the non-linear term, let us analyze the differential
equation given by
\begin{align}
 \ddot {\tilde y}(t)+m^2{\tilde y}(t)+\frac{\lambda}{4}{\tilde y}^3(t)&=0\ ,
\end{align}
assuming $m^2$ is constant.
A relevant oscillating solution is given by
\begin{align}
 {\tilde y}(t)&=\mathcal A\,{\rm sn}
 \left(
  \sqrt{m^2+\frac{\lambda}{8}\mathcal A^2} t,-\frac{\mathcal A^2\lambda}{8m^2+\mathcal A^2\lambda}
 \right) \nonumber\\
 &\simeq \mathcal A\sin(\overline m(\mathcal A)t)\ ,
\end{align}
where ${\rm sn}(u,k)$ is the Jacobi elliptic sine function and
\begin{equation}
 \overline m^2(\mathcal A)=
 \frac{\pi^2(8m^2+\mathcal A^2\lambda)}{32\left[K\left(-\frac{\mathcal A^2\lambda}{8m^2+\mathcal A^2\lambda}\right)\right]^2}
 \simeq m^2+\frac{3}{16}\lambda\mathcal A^2\ .
\end{equation}
Here, $K(k)$ is the complete elliptic integral of the first kind, which
is defined as
\begin{equation}
 K(k^2)=\int_0^{2/\pi}\frac{1}{\sqrt{1-k^2\sin^2\theta}}d\theta\ .
\end{equation}

It motivates us to approximate the original differential equation as
\begin{equation}
 \ddot Y+\left(\overline m^2(\mathcal A)+\Lambda^2\cos\omega  t\right)Y=0\ ,\label{eq_simple_assumption}
\end{equation}
with $\mathcal A$ being the amplitude of the oscillation.  Let us
describe the behavior of the solution as follows.  When $\overline m^2$
enters the resonance band, $\mathcal A$ starts to grow
exponentially. Since it makes $\overline m^2$ larger, the growth stops
immediately.  However, since $m^2$ decreases due to the motion of the
relaxion, it re-enters the resonance band and the amplitude grows until
$\overline m^2$ gets out of the resonance band. These occur repeatedly
and the amplitude is kept around the {\it edge} of the resonance band.

Let us discuss it more quantitatively and justify the above
description. Assuming $\overline m^2$ is a constant, one can construct a
solution to Eq.~\eqref{eq_simple_assumption} that behaves like
\begin{equation}
 Y(t)=e^{\frac{i}{2}\nu\omega t}u(t)\ ,
\end{equation}
where $u(t)$ is an $\mathcal O(1)$ function having a periodicity
$\omega$. As discussed in Appendix \ref{apx_mathieu}, $\nu$ can be
approximated as
\begin{equation}
 \nu\simeq 1\pm \frac{i}{2}\sqrt{\frac{4\Lambda^4}{\omega^4}
 \left(1-\frac{3\Lambda^4}{8\omega^4}\right)
 -\left(\frac{4\overline m^2}{\omega^2}-1
 +\frac{3\Lambda^4}{2\omega^4}\right)^2}\ ,
\end{equation}
around the first resonance band.  When $\nu$ has an imaginary part,
the amplitude grows exponentially. Assuming $\overline m^2$ is kept
around the edge of the resonance band, {\it i.e.} ${\Im(\nu)\simeq 0}$ , we
can predict the amplitude as
\begin{equation}
  m^2+\frac{3}{16}\lambda\mathcal A^2\simeq\frac{\omega^2}{4}
 \left(1-\frac{3}{2}\frac{\Lambda^4}{\omega^4}\right)
 +\frac{\Lambda^2}{2}\sqrt{1-\frac{3}{8}\frac{\Lambda^4}{\omega^4}}\ .\label{eq_resoCondHiggs}
\end{equation}

We show the amplitude, ${\mathcal A}$, determined by
Eq.~\eqref{eq_resoCondHiggs} with the dashed orange line in the left
panel of Fig.~\ref{fig_edge}. Notice that the time dependence of
$\mathcal A$ comes from $m^2(t)$.  As we can see from the figure, the
predicted amplitude agrees well with the numerical solution, supporting
our intuitive description.

Another justification comes from the frequency of the amplified mode.
Since the real part of $\nu$ is one, we expect that the frequency of
$y(t)$ peaks around $\omega/2$. In the right panel of
Fig.~\ref{fig_edge}, we plot the coefficients of the Fourier cosine
series for $8<t<12$. It clearly shows that the frequency of the
amplified mode is $\omega/2$ as indicated with the red dashed line. For
comparison, we show $\sqrt{\overline m^2}$ determined by
Eq.~\eqref{eq_resoCondHiggs} with the black dotted line. Notice that
$m^2$ is irrelevant since $-30^2\lesssim m^2\lesssim 30^2$ in this
interval.

\subsection{Relaxion-Higgs System}
The solution obtained in the previous subsection can not be directly
used in our analysis since it does not include the Hubble friction and
feedback to the relaxion dynamics.

We search for an edge solution, which has the form of
\begin{align}
 {\bar X}(t)&\simeq\frac{M^2-m_{\Phi}^2}{\epsilon}+f\omega_X t\ ,\label{eq_bar_X_sol}\\
 {\bar h}(t)&\simeq\mathcal A_h\cos\left(\frac{\omega_X}{2}t-\alpha\right)\ ,\label{eq_bar_h_sol}
\end{align}
where $\alpha,~\omega_X,~m_\Phi^2$ and $\mathcal A_h$ are treated as
constants for $|t|\ll |m_\Phi^2/(\epsilon f\omega_X)|$. The time
dependence of these constants will be taken into account by shifting
$m_\Phi^2$ so that it cancels the last term of Eq.~\eqref{eq_bar_X_sol}.
In Appendix \ref{apx_edge}, we determine $\alpha$ as
\begin{equation}
 \sin2\alpha\simeq-\frac{3H\omega_X}{\Lambda_h^2}\ ,~\cos2\alpha<0\ ,\label{eq_sin2a}
\end{equation}
and obtain relations among $\omega_X,~m_\Phi^2$ and $\mathcal A_h$ as
\begin{align}
 m_\Phi^2+\frac{3\lambda}{16}\mathcal A_h^2&\simeq\frac{\omega_X^2}{4}
 +\frac{\Lambda_h^2}{2}-\frac{\Lambda_h^2(32\Lambda_0^4+17\Lambda_h^2\mathcal A_h^2)}{128f^2\omega_X^2}\nonumber\\
 &\hspace{3ex}-\frac{(8\Lambda_h^2-3\lambda\mathcal A_h^2)(8\Lambda_h^2-\lambda\mathcal A_h^2)}{512\omega_X^2}
 -\frac{9\omega_X^2}{4\Lambda_h^2}H^2\ ,\label{eq_approx_Ah}\\
\omega_X&\simeq\frac{r\epsilon M^2}{3Hf}-\frac{\mathcal A_h^2\omega_X}{8f^2}
 -\frac{\lambda\mathcal A_h^4}{128f^2\omega_X}
 +\frac{3\mathcal A_h^2\omega_X^3}{16f^2\Lambda_h^4}H^2\ , \label{eq_approx_omega}
\end{align}
ignoring $\mathcal O(1/\omega_X^3),~\mathcal O(H^2/\omega_X)$ and
$\mathcal O(H^3)$ corrections.  It should be noted that the Higgs
oscillation does not grow due to the Hubble friction in the case of
$3H\omega_X/\Lambda_h^2 \gtrsim 1$.

\begin{figure}[t]
 \begin{center}
  \begin{minipage}{0.48\linewidth}
   \includegraphics[width=\linewidth]{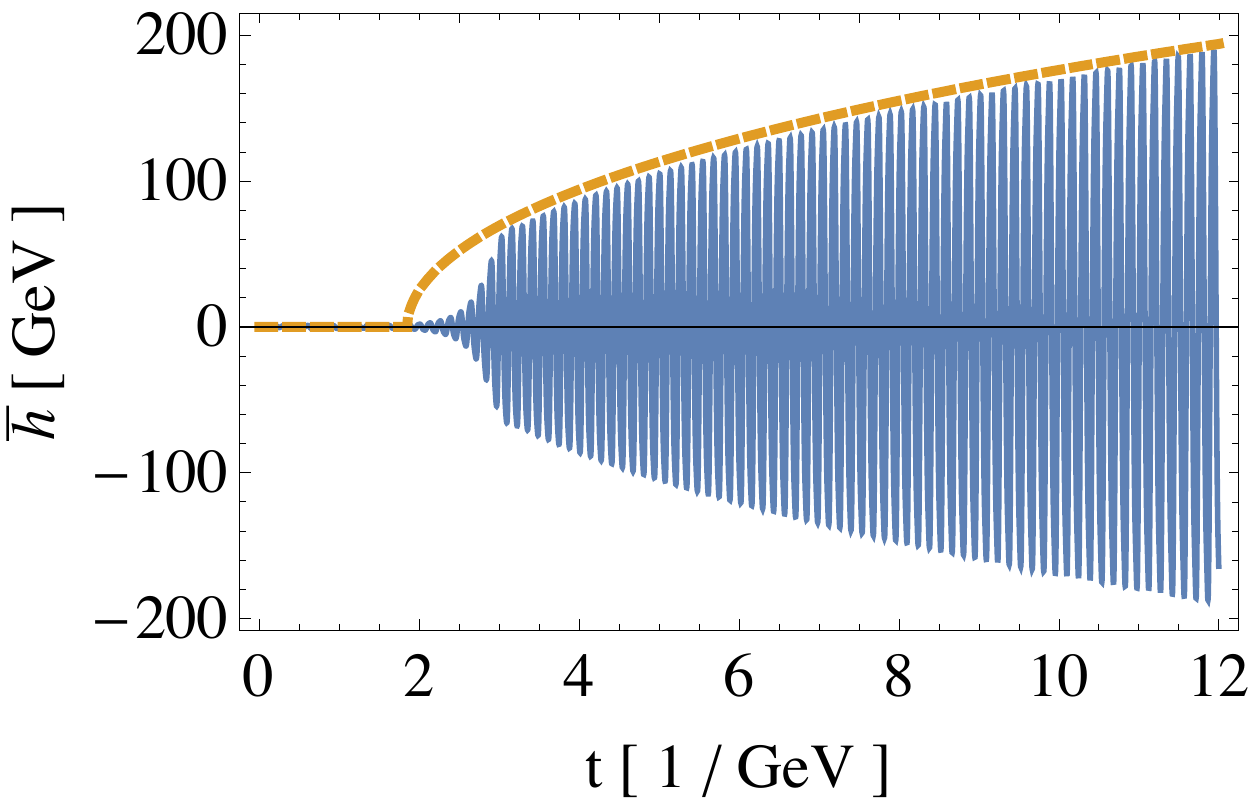}
  \end{minipage}
  \begin{minipage}{0.48\linewidth}
   \includegraphics[width=\linewidth]{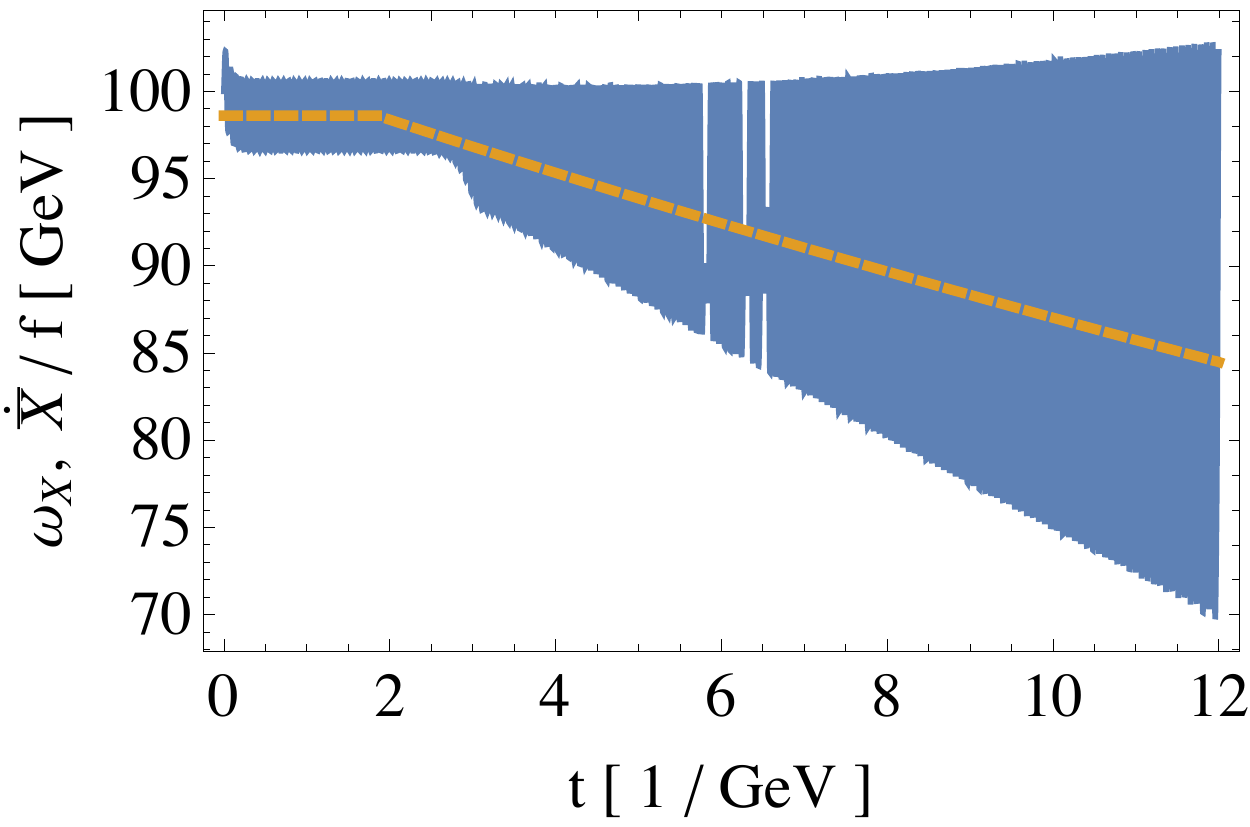}
  \end{minipage}

  \caption{A comparison between a numerical solution to
Eqs.~\eqref{eq_full_eom_X} and \eqref{eq_full_eom_h} and the analytic
result given in Eqs.~\eqref{eq_approx_Ah} and
\eqref{eq_approx_omega}. {\it Left:} The blue line shows ${\bar h}$ and
the orange line shows $\mathcal A_h$. {\it Right:} The blue line shows
$\dot {\bar X}/f$ and the orange line shows $\omega_X$.}

 \label{fig_full}
 \end{center}
\end{figure}

In Fig.~\ref{fig_full}, we compare a numerical solution to
Eqs.~\eqref{eq_full_eom_X} and \eqref{eq_full_eom_h} with the analytic
results given in Eqs.~\eqref{eq_approx_Ah} and
\eqref{eq_approx_omega}. The left panel shows $\bar h$ (blue) and
$\mathcal A_h$ (orange), and the right panel shows $\dot{\bar X}/f$
(blue) and $\omega_X$ (orange). We take
\begin{align}
 &\lambda=0.52\ ,~H=5~{\rm GeV}\ ,~\Lambda_0=\Lambda_h=60~{\rm GeV}\ ,~\epsilon=0.026~{\rm GeV}\ ,~r=0.01\ ,\nonumber\\
 &f=180~{\rm GeV}\ ,~M=32~{\rm TeV}\ .
\end{align}
At $t=0$, we set $M^2-\epsilon \bar X\simeq4700\,{\rm GeV}^2$ and take
$\dot {\bar X}$ around the terminal velocity.  To mimic quantum
fluctuations of ${\bar h}$, we take\footnote{The behavior of the
solution is almost independent of this value.} ${\bar h}\simeq H/(2\pi)$ at
$t=0$ and switch off the Hubble friction when $|\bar h|$ is smaller than
$H/(2\pi)$.  As we can see from the left panel, the amplitude is well
approximated by Eqs.~\eqref{eq_approx_Ah} and \eqref{eq_approx_omega}
once the amplitude reaches the dashed line. From the right panel, we see
that the predicted $\omega_X$ roughly reproduces the time average of
$\dot{\overline X}/f$.

\begin{figure}[t]
 \begin{center}
  \begin{minipage}{0.48\linewidth}
   \includegraphics[width=\linewidth]{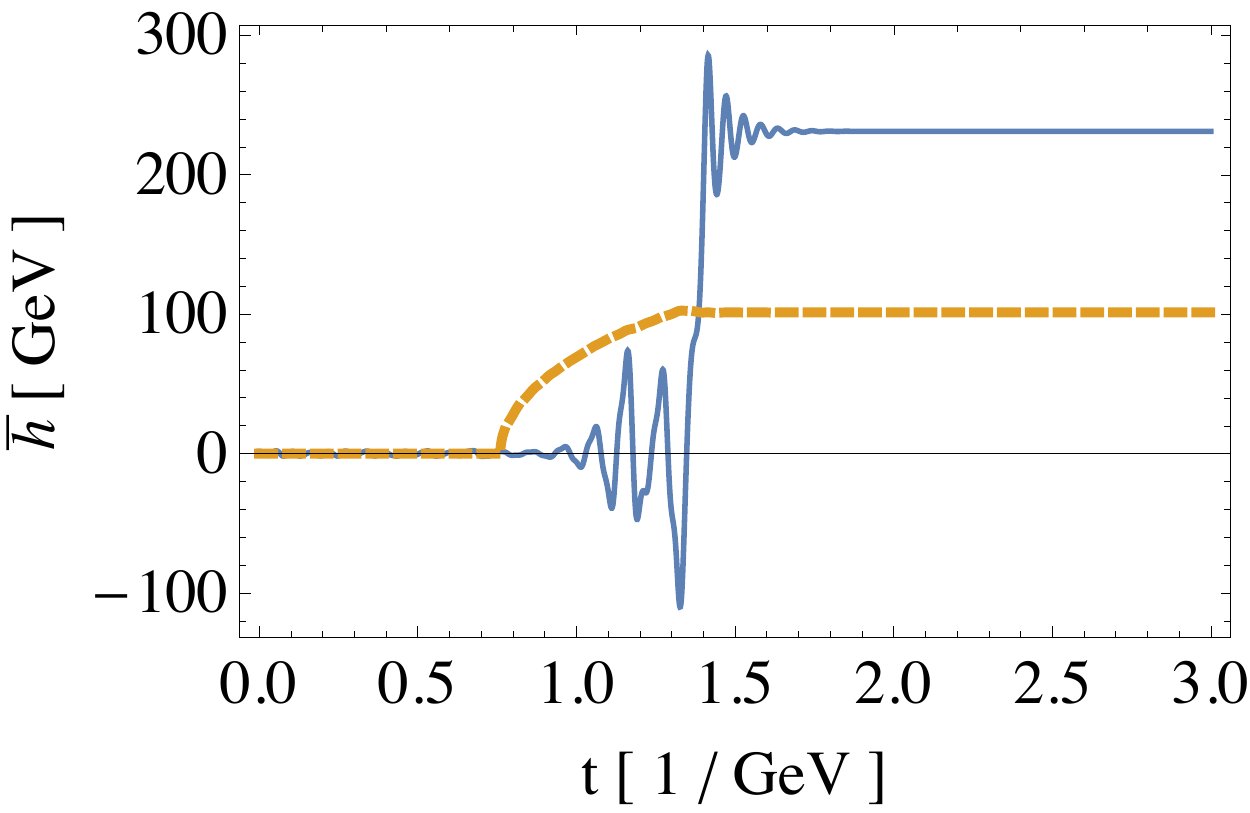}
  \end{minipage}
  \begin{minipage}{0.48\linewidth}
   \includegraphics[width=\linewidth]{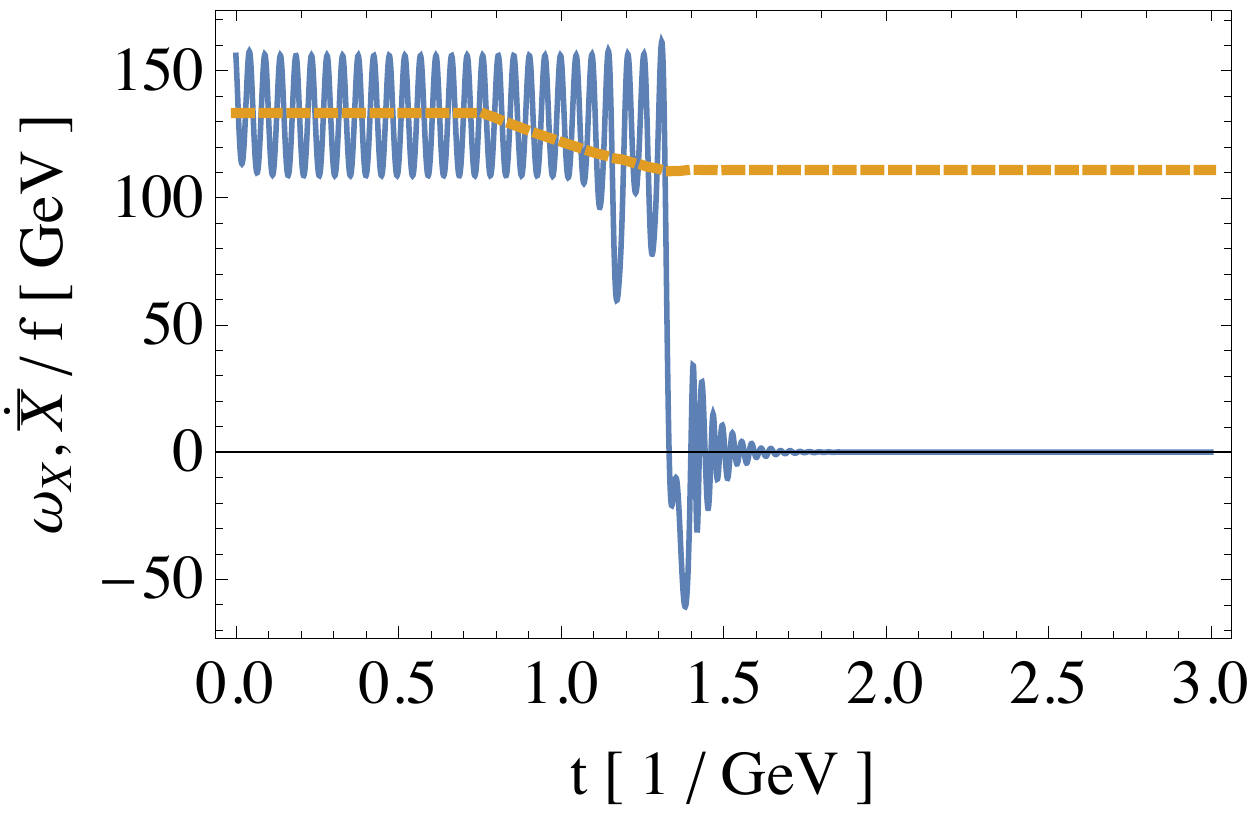}
  \end{minipage}
  \caption{The same figures as in Fig.~\ref{fig_full} with the parameter set given in Eq.~\eqref{eq_ex_param}.}
 \label{fig_ex_1}
 \end{center}
\end{figure}

In Fig.~\ref{fig_ex_1}, we show another example where the relaxion hits
a bump and gets trapped in a potential well.  Here, we take the
following parameters.
\begin{align}
 &\lambda=0.52,~H=10~{\rm GeV},~\Lambda_0=80~{\rm GeV},~\Lambda_h=100~{\rm GeV},~\epsilon=0.8~{\rm GeV},~r=0.001,\nonumber\\
 &f=80~{\rm GeV},~M=20~{\rm TeV}.\label{eq_ex_param}
\end{align}
We set $M^2-\epsilon \bar X\simeq13800~{\rm GeV}^2$ at $t=0$.  As the
figure shows, the relaxion hits a bump at around $t \sim
1.3$\,GeV$^{-1}$ and the Higgs field quickly develops a VEV just like
an anchor, securing the relaxion. After the relaxation, we have
$M^2-\epsilon \bar X\simeq3000~{\rm GeV}^2$ and the negative Higgs mass
squared is provided by $\Lambda_h^2\cos(\bar X/f)$.

Finally, let us comment on other resonance bands. So far, we have used the
first resonance band for the edge solution, but similar solutions with
the second and higher resonance bands should also exist. One could
eliminate them by taking a large enough Hubble constant since they are
much weaker than the first one. However, it is not necessary because the
Higgs field does not develop its VEV after the relaxion hits a
bump. Since the anchor does not bite, the relaxion continues to roll.

\section{Suppression of resonant particle production}\label{sec_resonant_pp}
In the previous section, we have analyzed the homogeneous modes of the
relaxion and the Higgs field. However, in field theory, we have to take
into account the dynamics of inhomogeneous modes as well. In particular,
resonant particle production can affect our mechanism in the following
ways.
\begin{itemize}
 \item Thermal bath induces thermal potential.
 \item Thermal bath behaves like friction for the Higgs field and
       the relaxion.
 \item Large fluctuations of the relaxion may average out the bumps.
 \item Increase of the temperature may cause phase transition and erase
       the bumps.
\end{itemize}
Although they do not always spoil our mechanism, we seek a parameter space
where the particle production is suppressed to keep our analysis as simple as possible.

Let us start with classical field theory. We separate the fields into
the homogeneous modes and fluctuations around them as
\begin{align}
 X(x,t)&={\bar X}(t)+\varphi(x)\ ,\\
 h(x,t)&={\bar h}(t)+\chi(x)\ .
\end{align}
In the Fourier space, the fluctuations are expanded as
\begin{align}
 \varphi(x)&=\int\frac{d^3k}{(2\pi)^3}\varphi_{\bf k}(t)e^{i{\bf k\cdot x}}\ ,\\
 \chi(x)&=\int\frac{d^3k}{(2\pi)^3}\chi_{\bf k}(t)e^{i{\bf k\cdot x}}\ .
\end{align}
For small fluctuations, they satisfy
\begin{align}
 \ddot \varphi_{\bf k}+3H\dot\varphi_{\bf k}
 +\left(m_\varphi^2+\frac{|{\bf k}|^2}{a^2(t)}\right)\varphi_{\bf k}
 +\delta m_\varphi^2\varphi_{\bf k}+\delta m_{\rm mix}^2\chi_{\bf k}&=0\ ,\label{eq_eom_phi_k}\\
 \ddot \chi_{\bf k}+3H\dot\chi_{\bf k}+\left(m_\chi^2+\frac{|{\bf k}|^2}{a^2(t)}\right)\chi_{\bf k}
 +\delta m_\chi^2\chi_{\bf k}+\delta m_{\rm mix}^2\varphi_{\bf k}&=0\ ,\label{eq_eom_chi_k}
\end{align}
where\footnote{The sign of $\delta m_{\rm mix}^2$ depends on that of
$\alpha$, but it is not important.}
\begin{align}
 m_\varphi^2&\simeq \frac{\Lambda_h^2\mathcal A_h^2}{8f^2}\ ,\\
 \delta m_\varphi^2&\simeq -\frac{2\Lambda_0^4+\Lambda_h^2\mathcal A_h^2}{4f^2}\cos\omega_Xt
 +\frac{\Lambda_h^2\mathcal A_h^2}{8f^2}\cos2\omega_Xt\ ,\\
 m_\chi^2&\simeq m_\Phi^2+\frac{3}{8}\lambda\mathcal A_h^2\ ,\\
 \delta m_\chi^2&\simeq \left(\Lambda_h^2-\frac{3\lambda}{8}\mathcal A_h^2\right)\cos\omega_Xt\ ,\\
 \delta m_{\rm mix}^2&\simeq  -\frac{\Lambda_h^2\mathcal A_h}{2f}
 \left(\cos\frac{\omega_X}{2}t-\cos\frac{3\omega_X}{2}t\right)\ .
\end{align}
with $a(t)= e^{Ht}$ being the scale factor. Here, we substituted the
edge solution given by Eqs.~\eqref{eq_bar_X_sol} and
\eqref{eq_bar_h_sol}. We have ignored inhomogeneous terms since they can
be erased by using special solutions. We assume that $m_\Phi^2$ and
$\omega_X$ are almost constant in the timescale of $1/H$.

We estimate the effect of resonant particle production in quantum field
theory by comparing $\varphi_{\bf k}$ and $\chi_{\bf k}$ with ``free''
solutions, $\varphi^{\rm free}_{\bf k}$ and $\chi^{\rm free}_{\bf k}$,
which satisfy the same differential equations but without $\delta
m^2_{\varphi},~\delta m^2_\chi$ or $\delta m^2_{\rm mix}$. Notice that
$\chi^{\rm free}_{\bf k}$ and $\varphi^{\rm free}_{\bf k}$ should give
the zero point fluctuations with appropriate normalization.
Since $|{\bf k}|^2/a^2$ dominates over all the mass terms when
$t\to-\infty$, we set
\begin{equation}
 \lim_{t\to-\infty}\frac{\chi_{\bf k}}{\chi_{\bf k}^{\rm free}}=
 \lim_{t\to-\infty}\frac{\varphi_{\bf k}}{\varphi_{\bf k}^{\rm free}}=1\ .
\end{equation}

To suppress the resonant particle production, we require\footnote{We do
not consider the modes with $|{\bf k}|<3H/2$ since they are
super-horizon modes.}
\begin{equation}
 N_{\chi}\left(\frac{|\bf k|^2}{a^2}\right)\lesssim\mathcal O(1)\ ,
 ~N_{\varphi}\left(\frac{|\bf k|^2}{a^2}\right)\lesssim\mathcal O(1)\ ,
\end{equation}
for all $|{\bf k}|>3H/2$. Here,
\begin{equation}
 N_{\chi}\left(\frac{|\bf k|^2}{a^2}\right) \equiv
 \frac{1}{2}\left|\frac{\chi_{\bf k}}{\chi^{\rm free}_{\bf k}}\right|^2-\frac{1}{2}\ ,
 ~N_{\varphi}\left(\frac{|\bf k|^2}{a^2}\right)\equiv
 \frac{1}{2}\left|\frac{\varphi_{\bf k}}{\varphi^{\rm free}_{\bf k}}\right|^2-\frac{1}{2}\ ,
\end{equation}
which we call the {\it occupation numbers}.

In the WKB approximation, $\chi_{\bf k}^{\rm free}$ and $\varphi_{\bf
k}^{\rm free}$ are calculated as
\begin{align}
 \chi_{\bf k}^{\rm free}&\simeq
 \frac{C_\chi}{\left(m_\chi^2+\frac{|{\bf k}|^2}{a^2(t)}-\frac{9}{4}H^2\right)^{1/4}}
 e^{i\int^t\sqrt{m_\chi^2+\frac{|{\bf k}|^2}{a^2(t')}-\frac{9}{4}H^2}dt'-\frac{3}{2}Ht}\ ,\\
 \varphi_{\bf k}^{\rm free}&\simeq
 \frac{C_\varphi}{\left(m_\varphi^2+\frac{|{\bf k}|^2}{a^2(t)}-\frac{9}{4}H^2\right)^{1/4}}
 e^{i\int^t\sqrt{m_\varphi^2+\frac{|{\bf k}|^2}{a^2(t')}-\frac{9}{4}H^2}dt'-\frac{3}{2}Ht}\ ,
\end{align}
where $C_\chi$ and $C_\varphi$ are constants.

On the other hand, $\chi_{\bf k}$ and $\varphi_{\bf k}$ are well
approximated by
\begin{align}
 \chi_{\bf k}&\simeq\frac{C_\chi}{\sqrt{\mu_\chi(t)}}F_\chi(t)e^{i\int^t\mu_\chi(t')dt'-\frac{3}{2}Ht}\ ,\\
 \varphi_{\bf k}&\simeq\frac{C_\varphi}{\sqrt{\mu_\varphi(t)}}F_\varphi(t)e^{i\int^t\mu_\varphi(t')dt'-\frac{3}{2}Ht}\ ,
\end{align}
where $\mu_\chi$ and $\mu_\varphi$ are the characteristic exponents,
which will be given explicitly later. Here, $F_\chi$ and $F_\varphi$
are $\mathcal O(1)$ functions satisfying
\begin{equation}
 \lim_{t\to-\infty}F_\chi(t)=\lim_{t\to-\infty}F_\varphi(t)=1\ .
\end{equation}
One can easily check these are in good agreement with numerical solutions.

At the first approximation, the occupation numbers are given by
\begin{align}
 N_{\chi}\left(\frac{|\bf k|^2}{a^2}\right)&\simeq\frac{1}{2}
 \exp\left[\int^{\infty}_{|{\bf k}|^2/a^2(t)}\frac{|\Im(\mu_\chi)|}{H}\frac{dK}{K}\right]-\frac{1}{2}\ ,\label{eq_n_chi}\\
 N_{\varphi}\left(\frac{|\bf k|^2}{a^2}\right)&\simeq\frac{1}{2}
 \exp\left[\int^{\infty}_{|{\bf k}|^2/a^2(t)}\frac{|\Im(\mu_\varphi)|}{H}\frac{dK}{K}\right]-\frac{1}{2}\ \label{eq_n_varphi},
\end{align}
where
\begin{equation}
 K(t')=\frac{|{\bf k}|^2}{a^2(t')}\ .
\end{equation}
Since the smallest $|{\bf k}|/a$ gives the strongest constraints, we check
\begin{equation}
 N_\chi^{\max}\equiv N_{\chi}\left(\frac{9H^2}{4a^2}\right)\lesssim\mathcal O(1)\ ,~ N_\varphi^{\max}\equiv N_{\varphi}\left(\frac{9H^2}{4a^2}\right)\lesssim\mathcal O(1)\ .
\end{equation}

\subsection{Before the Higgs oscillation}
Before the Higgs oscillation, Eqs.~\eqref{eq_eom_phi_k} and
\eqref{eq_eom_chi_k} are independent of each other and have the form of
the Mathieu equation. In Appendix \ref{apx_mathieu}, we obtain the
characteristic exponents for the first and the second resonance bands, which
are the strongest and the second strongest resonance bands, respectively.

\begin{figure}[t]
 \begin{center}
  \begin{minipage}{0.48\linewidth}
   \includegraphics[width=\linewidth]{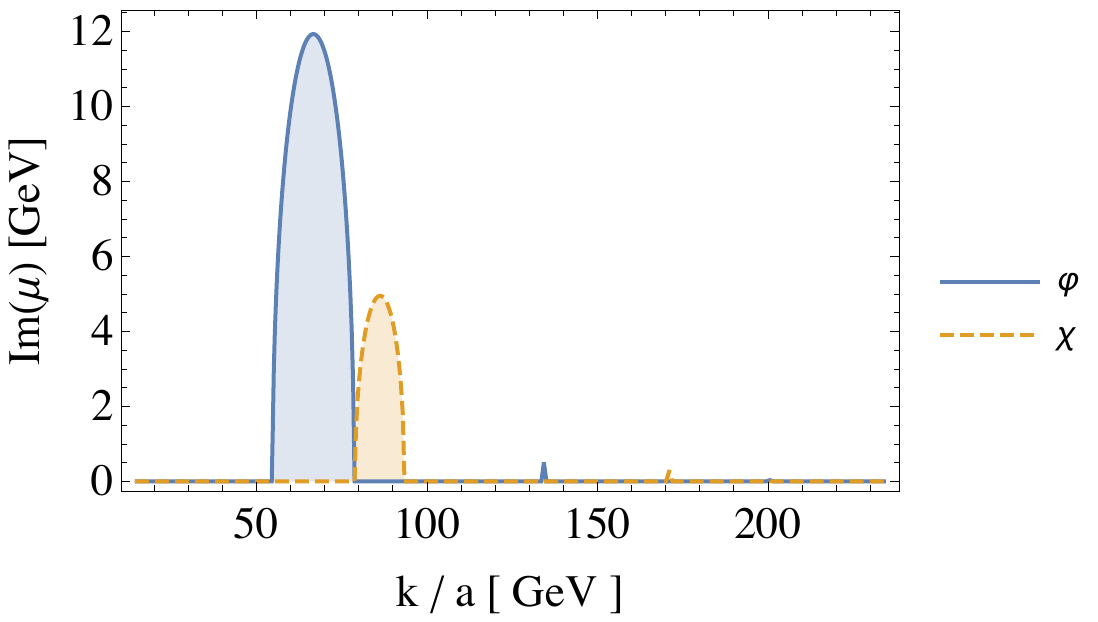}
  \end{minipage}
  \begin{minipage}{0.48\linewidth}
   \includegraphics[width=0.8\linewidth]{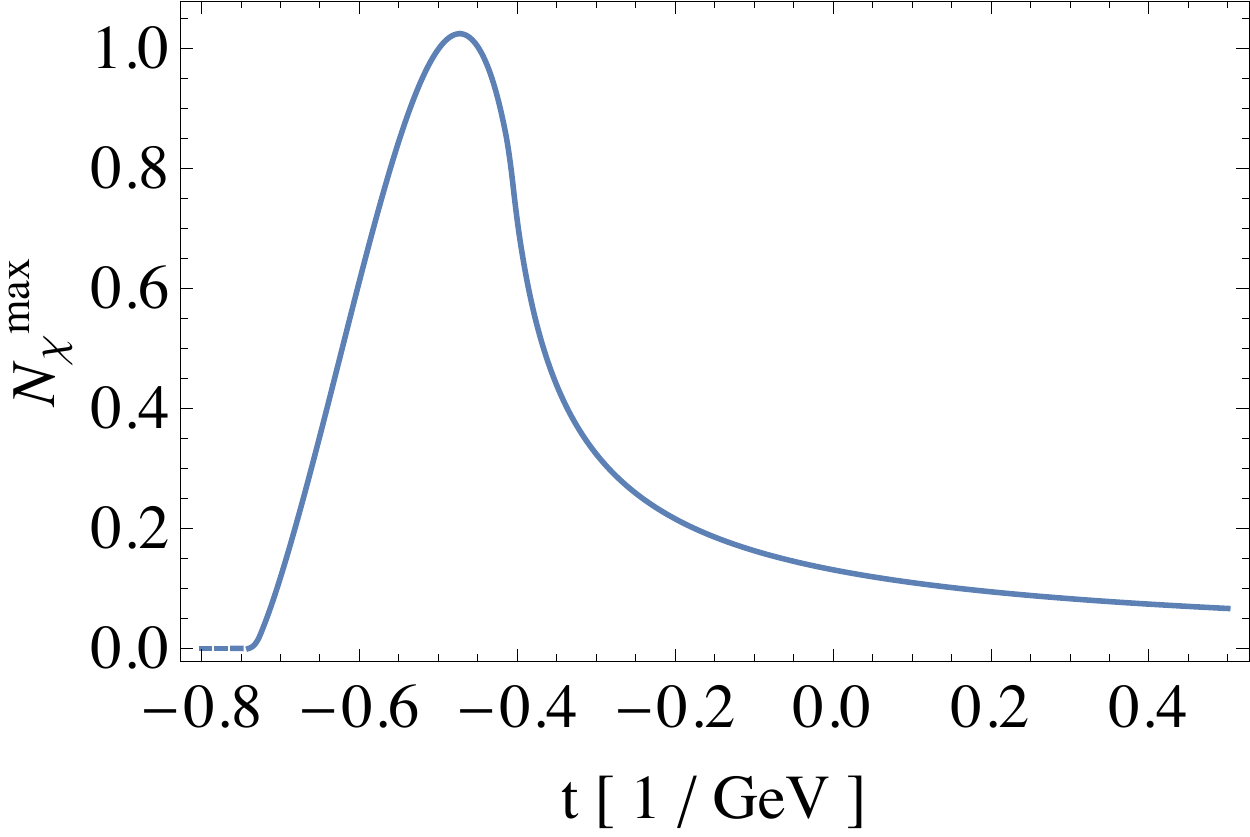}
  \end{minipage}

  \caption{{\it Left:} The characteristic exponents evaluated at
  $t=0.3$. {\it Right:} the maximum occupation number of $\chi$ for each
  $t$.}  \label{fig_reso_ex_1}
 \end{center}
\end{figure}

In the left panel of Fig.~\ref{fig_reso_ex_1}, we plot the imaginary
part of the characteristic exponents evaluated at $t=0.3$, when the
Higgs oscillation has not started yet. We evaluate them by directly
solving the recurrence formula for $\Delta(0)$ in Appendix
\ref{apx_mathieu}. The largest peaks of $\mu_{\varphi}$ corresponds to
the first resonance band and that of $\mu_\chi$ corresponds to the
second resonance band. We can see that there are tiny peaks in the right
of the largest peaks. They correspond to the second resonance band of
$\mu_\varphi$ and the third resonance band of $\mu_\chi$. Notice that
the first resonance band for $N_{\chi}$ is what we use for the Higgs
coherent oscillation, and hence it can not contribute to the particle
production.

For the maximum occupation number of $\varphi$, we obtain
$N_\varphi^{\max}\simeq0.5$.  As for that of $\chi$, we show its time
dependence in the right panel of Fig.~\ref{fig_reso_ex_1}. For
$t\lesssim-0.7$, the resonance condition can not be satisfied due to a
large Higgs mass. It has a peak structure and we get
$N_\chi^{\max}\lesssim1.0$. Since they are small enough, we can safely
ignore the effect of particle production with this parameter set.

In the subsequent sections, we use the approximations given in Appendix
\ref{apx_mathieu}. They are given by
\begin{equation}
 |\Im(\mu_\varphi(t))|\simeq\frac{\omega_X}{4}\Re\left[\sqrt{\frac{\Lambda_0^8}{f^4\omega_X^4}\left(1-\frac{3}{32}\frac{\Lambda_0^8}{f^4\omega_X^4}\right)-\left(\frac{4}{\omega_X^2}\left(\frac{|{\bf k}|^2}{a^2(t)}-\frac{9}{4}H^2\right)-1+\frac{3}{8}\frac{\Lambda_0^8}{f^4\omega_X^4}\right)^2}\right]\ ,
\end{equation}
for the first resonance band of $N_{\varphi}$, and 
\begin{equation}
 |\Im(\mu_\chi(t))|\simeq\frac{\omega_X}{8}\Re\left[\sqrt{\frac{\Lambda_h^8}{\omega_X^8}-\left(\frac{4}{\omega_X^2}\left(m_\Phi^2-\frac{9}{4}H^2+\frac{|{\bf k}|^2}{a^2(t)}\right)-4-\frac{2}{3}\frac{\Lambda_h^4}{\omega_X^4}\right)^2}\right]\ ,
\end{equation}
for the second resonance band of $N_{\chi}$. Notice that the integration
in Eqs.~\eqref{eq_n_chi} and \eqref{eq_n_varphi} can be executed
analytically.

\subsection{After the Higgs oscillation}
After the Higgs oscillation begins, $\delta m_{\rm mix}^2$ is turned on
and $\varphi_{\bf k}$ and $\chi_{\bf k}$ are influenced by each
other. In addition, the Higgs oscillation can cause resonant production
of the W boson and the Z boson.

By solving Eqs.~\eqref{eq_eom_phi_k} and \eqref{eq_eom_chi_k} directly,
one finds that these differential equations have broad and strong
instabilities.  To avoid such instabilities, we restrict the number of
the Higgs oscillations during the amplification to be $\mathcal
O(1)$. Then, $\mathcal A_h,~m_\Phi^2$ and $\omega_X$ are different for
each oscillation and we can avoid the exponential amplification of mode
functions.

Another concern is that a single oscillation could produce too many
particles, which might disturb the Higgs oscillation.  Let us consider a
generic particle, $\zeta$, which has a mass of
\begin{equation}
 m_\zeta^2=g^2 \mathcal A_h^2\cos^2\left(\frac{\omega_X}{2}t\right)\ ,
\end{equation}
where $g$ is a coupling constant. In the following, we estimate how much
$\zeta$ is produced for each oscillation. The adiabaticity condition
is violated for momenta satisfying \cite{Allahverdi:2010xz}
\begin{equation}
 \frac{|\bf k|^2}{a^2}\lesssim\frac{1}{2\pi}g\mathcal A_h\omega_X\ ,
\end{equation}
whose occupation number becomes around one for a single oscillation. The
energy density of the created $\zeta$ particles is evaluated as
\begin{equation}
 \varepsilon_\zeta\simeq\frac{1}{8\pi^2}\left(\frac{g\mathcal A_h\omega_X}{2\pi}\right)^2\ .
\end{equation}
Since it is much less than that of the Higgs oscillation,
\begin{equation}
 \varepsilon_h\simeq\frac{\omega_X^2\mathcal A_h^2}{4}\ ,
\end{equation}
we can safely neglect the effect of particle production.

As we can see from Fig.~\ref{fig_ex_1}, the $\mathcal O(1)$ number of
oscillations is indeed realized with the parameter set given in
\eqref{eq_ex_param}.

\section{Vacuum decay rate}\label{sec_vacuum}
In this section, we review a vacuum decay rate, which will be used to
evaluate the stability of a vacuum after the relaxation.  As pointed out
in the introduction, the lifetime of the first selected vacuum should be
much longer than the age of the universe, otherwise one needs an
unacceptably large number of $e$-folds to find a sufficiently long-lived
vacuum.

We approximate the potential around the EW vacuum by a one-dimensional
potential along $X$, which is given by
\begin{equation}
 V\simeq-r\epsilon M^2X-\frac{\Lambda_0^4+\Lambda_h^2v^2}{2}\cos\frac{X}{f}\ ,
\end{equation}
where $v\simeq246~{\rm GeV}$ is the Higgs VEV. 

The origin of $X$ is redefined so that the correct EW vacuum corresponds
to $X\simeq0$, which we label as $X_F$. In the following, we evaluate
the tunneling rate into the deeper vacuum around $X\simeq 2\pi f$, which
we label as $X_T$. The decay rate is given by the bubble nucleation rate
\cite{Coleman:1977py,Callan:1977pt}, which is expressed as
\begin{equation}
 \gamma=Ae^{-B}\ .
\end{equation}
Here, 
\begin{equation}
 B=S_E[X_B]-S_E[X_F]\ ,
\end{equation}
and $X_B$ is the bounce solution. The prefactor $A$ is assumed to be
$(100~{\rm GeV})^4$ in our analysis\footnote{Since all the parameters
are around the EW scale, we expect $A$ is not so far from the EW
scale.}. For the evaluation of $B$, we use the thin wall approximation
\cite{Coleman:1977py}, which is given by
\begin{equation}
 B_{\rm thin}\equiv\frac{27\pi^2}{2}\frac{\sigma^4}{(V[X_F]-V[X_T])^3}\ ,
\end{equation}
where
\begin{equation}
 \sigma=\int_{X_*}^{X_F}dX\sqrt{2(V[X]-V[X_F])}\ .
\end{equation}
Here, $X_*$ is a constant determined by
\begin{equation}
 V[X_*]=V[X_F],~X_*\in[X_F,X_T]\ .
\end{equation}

Since it gives a lower bound on $B$, we get an upper bound on the
bubble nucleation rate as
\begin{equation}
 \gamma_{\rm ub}\equiv (100~{\rm GeV})^4e^{-B_{\rm thin}}\ .
\end{equation}
The actual constraint of vacuum stability will be discussed in the
following section together with other constraints.

\section{Viable Parameter Space for the Relaxion Mechanism}\label{sec_constraints}
In this section, we summarize constraints on the parameters. We divide
them into four categories; (i) those for the successful relaxation, (ii)
those from consistency of our analysis, (iii) those from the explicit
model of the strong sector, and (iv) those from experiments.  The
constraints of (i) are essential and independent of what are behind the
relaxion potential, while those of (iii) depend on the detail of the
strong sector. Some of the constraints of (ii) may be removed, but it is
beyond the scope of this paper.
\subsection{Successful Relaxation}
\subsubsection{Roll down}
We first discuss the conditions for the relaxion to roll down the
potential.

The terminal velocity of the relaxion should be large enough so that the
relaxion can go over the bumps, which gives
\begin{equation}
 \frac{r\epsilon M^2}{3H}\gtrsim\Lambda_0^2,~{\rm(Go~over~bumps)}.\label{eq_go_over_bumps}
\end{equation}

To violate the slow-roll condition, we require
\begin{equation}
 \frac{3H\Lambda_0^4}{2fr\epsilon M^2}\gtrsim H,~{\rm(Fast~roll)}.\label{eq_fast_roll}
\end{equation}

The classical rolling should dominate over quantum fluctuations, which gives
\begin{equation}
 \frac{H}{2\pi}\lesssim\frac{r\epsilon M^2}{3H^2},~{\rm(Classical~roll)}.
\end{equation}

\subsubsection{Stopping mechanism}
Next, we discuss the conditions to stop the relaxion at the desired
position.

The existence condition of the edge solution can be read off from
Eqs.~\eqref{eq_sin2a} and \eqref{eq_approx_omega} with $\mathcal A_h=0$. It is given by
\begin{equation}
 \frac{r\epsilon M^2}{\Lambda_h^2f}\lesssim1\ ,~{\rm(Edge~solution)}\ .\label{eq_edge_solution}
\end{equation}

The Higgs oscillation should start before the relaxion scans the correct Higgs
mass. It gives
\begin{equation}
  m_\Phi^2(t_{\rm start})>m_\Phi^2(t_{\rm end})\ ,~{\rm(Do~not~pass~through)}\ ,\label{eq_do_not_pass}
\end{equation}
where
\begin{align}
 m_\Phi^2(t_{\rm start})&\simeq\frac{r^2\epsilon^2M^4}{36f^2H^2}+\frac{\Lambda_h^2}{2}-\frac{r^2\epsilon^2M^4}{4f^2\Lambda_h^2}-\frac{9H^2(2\Lambda_0^4\Lambda_h^2+f^2\Lambda_h^4)}{8r^2\epsilon^2M^4}\ ,\label{eq_mass_start}\\
 m_\Phi^2(t_{\rm end})&\simeq-\frac{\lambda}{4}v^2+\Lambda_h^2\sqrt{1-\left(\frac{2fr\epsilon M^2}{\Lambda_0^4+\Lambda_h^2v^2}\right)^2}\ .
\end{align} 

When the relaxion hits a bump, the Higgs boson should acquire a large
enough VEV to anchor the relaxion.  We require the amplitude of the
Higgs oscillation be smaller than the EW vacuum, {\it i.e.}
\begin{equation}
 \mathcal A_h (t_{\rm end})<v\ ,~{\rm(Anchor)}\ .
\end{equation}

\subsubsection{EW vacua}
Lastly, we discuss the conditions related to the EW vacuum.

There must be a stationary point of the relaxion potential around the EW
vacuum, which gives
\begin{equation}
 \frac{2fr\epsilon M^2}{\Lambda_0^4+\Lambda_h^2v^2}<1\ ,~{\rm(Stationary~point)}\ .
\end{equation}

To avoid fine-tuning, there should be a sufficient number of vacua that
realize the Higgs mass around the EW scale, which gives
\begin{equation}
 2\pi \epsilon f\lesssim v^2\  ,~{\rm(Many~vacua)}\ .
\end{equation}

To make the lifetime of the EW vacuum longer than the age of the
Universe, we require
\begin{equation}
 \gamma_{\rm ub}\lesssim H_0^4\ ,~{\rm(Vacuum~stability)}\ ,
\end{equation}
with $H_0$ being the current value of the Hubble constant. It is
equivalent to $B_{\rm thin}\gtrsim400$.

\subsection{Consistency of Analysis}
Since we assume that the inflaton potential dominates the total energy
density during the relaxation, we require
\begin{equation}
 H\gtrsim\sqrt{\frac{rM^4}{3M_{\rm Pl}^2}}\ ,
\end{equation}
with $M_{\rm Pl}$ being the reduced Planck mass.

For the approximation used in Eq.~\eqref{eq_mass_start} to be reliable,
we need
\begin{equation}
 \frac{r^2\epsilon^2M^4}{36f^2H^2}\gtrsim\left|\frac{\Lambda_h^2}{2}
 -\frac{r^2\epsilon^2M^4}{4f^2\Lambda_h^2}
 -\frac{9H^2(2\Lambda_0^4\Lambda_h^2+f^2\Lambda_h^4)}{8r^2\epsilon^2M^4}\right|.\label{eq_const_q}
\end{equation}

To avoid the resonant particle production before the Higgs oscillation
starts, we require
\begin{equation}
 N_\chi^{\max}\lesssim\mathcal O(1)\ ,~N_\varphi^{\max}\lesssim\mathcal O(1)\ .
\end{equation}

To avoid the resonant particle production after the Higgs oscillation
starts, we require
\begin{equation}
  N_{\rm osc}\equiv
 \frac{1}{2}\frac{m_\Phi^2(t_{\rm start})-m_\Phi^2(t_{\rm end})}{2\pi \epsilon f}\lesssim\mathcal O(10) \ .
\end{equation}
Notice that not
all of them are responsible for the particle production since the
amplitude is small for the first several oscillations.

\subsection{Model of Strong Sector}
The parameters coming from the new strong sector, $\Lambda_0$ and $\Lambda_h$,
should be obtained naturally from an explicit model. In our analysis, we
adopt the simplest model described in Section \ref{sec_model}.

Since we need fermions that condensate, we require
\begin{equation}
m^{(\rm eff)}_N\lesssim\Lambda_c\ .\label{eq_str_mn}
\end{equation}

To restrict effects of the new doublet fermions on the strong dynamics, we require
\begin{equation}
 \Lambda_c\lesssim m_L\ .
\end{equation}

The Hubble expansion should not disturb the condensation, which gives
\begin{equation}
 H\lesssim\Lambda_c\ .
\end{equation}

To keep the new Yukawa couplings perturbative, we require
\begin{equation}
 \max(|y|,|\tilde y|)\lesssim 4\pi\ .\label{eq_str_y}
\end{equation}

Since there are tree and quantum contributions in $\Lambda_0^4$, there
can be an implicit cancellation. To avoid a fine-tuning, we define
fine-tuning parameter $\xi$ as
\begin{equation}
 \xi=\frac{\frac{y\tilde y}{8\pi^2}m_L\ln\frac{M^2}{m_L^2}-m_N}{\frac{y\tilde y}{8\pi^2}m_L\ln\frac{M^2}{m_L^2}}\ ,
\end{equation}
and require
\begin{equation}
 |\xi|\gtrsim0.1\ .
\end{equation}

Notice that, once $m_L$ and $\Lambda_c$ are fixed, the other parameters
are determined as
\begin{align}
 y\tilde y&=-\frac{\Lambda_h^2m_L}{\Lambda_c^3}\ ,\\
 m^{(\rm eff)}_N&=\frac{\Lambda_0^4+\Lambda_h^2v^2}{2\Lambda_c^3}\ , \\
 \xi&=\frac{4\pi^2\Lambda_0^4}{m_L^2\Lambda_h^2\ln\frac{M^2}{m_L^2}}\ .\label{eq_finetuning}
\end{align}

\subsection{Experiments}
\subsubsection{Strong sector}
We first discuss experimental constraints on the new strong sector.

From Eq.~\eqref{eq_finetuning}, we expect that $m_L$ is not much
larger than $\Lambda_0$ to avoid a fine-tuning. The constraints on the
new doublet fermions have been discussed in \cite{Beauchesne:2017ukw} in
the context of the original relaxion mechanism.  They evaluate the
contributions to the six EW precision valuables, $(S,T,U,V,W,X)$, and
show the 95\% confidence level (CL) excluded region on the
$(m_L,y=\tilde y)$ plane. For example, for $m_L\gtrsim200~{\rm GeV}$, we
can take $|y\tilde y|\lesssim0.1$, and for $m_L\gtrsim500~{\rm GeV}$, we
can take $|y\tilde y|\lesssim0.4$. They also discuss the constraints
from collider searches, which can be avoided if $m_L\gtrsim200~{\rm
GeV}$.

Next, we discuss the constraints on the mesons in the new strong
sector. In our mechanism, all the new mesons naturally have masses
around or larger than the Higgs mass. This is because the Higgs mass is
determined mainly by $\Lambda_h$, and $\Lambda_c$ is typically larger
than $\Lambda_h$, as can be seen from Eq.~\eqref{eq_lambda_h}. Since the
anomalous axial symmetry is the only broken symmetry, all the states in
the strong sector are expected to have masses around or larger than
$\Lambda_c$.

If new mesons have mass around the EW scale, some of them can mix with
the Higgs boson and decrease the signal strengths of the Higgs boson.
For simplicity, we avoid such mixing by taking
\begin{equation}
 \Lambda_c\gtrsim 400~{\rm GeV}\ .
\end{equation}
Then, they effectively decouple from Higgs phenomenology.

Lastly, we comment on the dynamics that gives the relaxion decay
constant, $f$. Such a small decay constant is obtained, for
example, from a scalar clockwork mechanism discussed in
Appendix~\ref{sec:clockwork2}. In such a case, we expect new particles
around $f$. Alternatively, one could use fermion condensations for the
building block of the clockwork \cite{Coy:2017yex}. Then, we
expect a rich spectrum around $4\pi f\sim1~{\rm TeV}$ and thus we can
safely ignore their effects. Since it is highly model dependent, we do
not go into details in this paper.
\subsubsection{Relaxion and Higgs boson}
Let us move on to the collider constraints on the relaxion and the
Higgs boson.

Since we assume a rather small $f$, the relaxion easily mixes with the
Higgs boson. We define the mixing angles as
\begin{equation}
 s_X\equiv-\sin\frac{\langle X\rangle}{f}=\frac{2fr\epsilon M^2}{\Lambda_0^4+\Lambda_h^2v^2}\ ,
 ~c_X\equiv-\cos\frac{\langle X\rangle}{f}=\sqrt{1-s_X^2}\ ,
\end{equation}
where $\langle X\rangle$ is the expectation value of the relaxion at the EW vacuum.%
\footnote{In the absence of $\epsilon$, the relaxion settles at $\sin \langle{X}\rangle/f = 0$ and $\cos\langle{X}\rangle/f = -1$.}
 Using these, the mass matrix of the scalars are expressed as
\begin{equation}
 \begin{pmatrix}
  \frac{\lambda v^2}{2}&\frac{\Lambda_h^2v}{f}s_X\\
  \frac{\Lambda_h^2v}{f}s_X&\frac{\Lambda_0^4+\Lambda_h^2v^2}{2f^2}c_X
 \end{pmatrix}\ ,
\end{equation}
in the basis of $(\chi,\varphi)$.%
\footnote{The coupling between the relaxion and the Higgs boson, $\epsilon X |\Phi|^2$, does not affect
the collider phenomenology since $\epsilon$ is small. }

When $\theta$ is small, the mixing angle of $\chi$ and $\varphi$ can be
calculated as
\begin{equation}
 \theta\simeq\frac{-2fv\Lambda_h^2s_X}{(\Lambda_0^4+v^2\Lambda_h^2)c_X-\lambda f^2v^2}\ .
\end{equation}

Here, the mixing angle is defined by
\begin{equation}
 \begin{pmatrix}
  h_{125}\\
  S
 \end{pmatrix}=
\begin{pmatrix}
 c_\theta&-s_\theta\\
 s_\theta&c_\theta
\end{pmatrix}
\begin{pmatrix}
 \chi\\
 \varphi
\end{pmatrix}\ ,
\end{equation}
where $h_{125}$ is the 125GeV Higgs boson and $S$ is the relaxion-like
boson.  The masses of $h_{125}$ and $S$ are approximately given by
\begin{align}
 m_{125}^2&=\frac{\lambda v^2}{2}+2\theta\frac{\Lambda_h^2v}{f}s_X\ ,\\
 m_{S}^2&=\frac{\Lambda_0^4+\Lambda_h^2v^2}{2f^2}c_X-2\theta\frac{\Lambda_h^2v}{f}s_X\ ,
\end{align}
respectively. Since the Higgs mass has been measured, we tune $\lambda$ so
that it reproduces
\begin{equation}
 m_{125}=125~\rm GeV\ .
\end{equation}
Here, we ignore the uncertainty since it has very little effect on our
results.

Since $S$ has couplings to the SM particles through the mixing, it can
be produced at colliders. We assume that $S$ decays only into the SM
particles. The partial decay widths of the relaxion are given by
\begin{align}
 \Gamma(S\to 2h_{125})&\simeq\frac{1}{32\pi}\frac{|g_{Shh}|^2}{m_S}\sqrt{1-\frac{4m_{125}^2}{m_S^2}}\ ,\\
 \Gamma(S\to {\rm SM})&=s_\theta^2\times\Gamma(h_{125}\to {\rm SM})|_{m_{125}\to m_S}\ ,
\end{align}
where 
\begin{equation}
 g_{Shh}=\frac{3\lambda v}{2}c_\theta^2s_\theta-\frac{2v\Lambda_h^2}{f^2}(2c_\theta^2s_\theta-s_\theta^3)c_X
 -\frac{\Lambda_h^2}{f}(2c_\theta s_\theta^2-c_\theta^3)s_X
 -\frac{\Lambda_0^4+\Lambda_h^2v^2}{2f^3}c_\theta s_\theta^2s_X\ ,
\end{equation}
and $\Gamma(h_{125}\to {\rm SM})|_{m_{125}\to m_S}$ is the Higgs partial
decay width with the Higgs mass being replaced by $m_S$.%
\footnote{We neglect possible decay of the relaxion into two photons through the strong sector. }

The experiments at LEP, Tevatron, and LHC have extensively searched for
neutral Higgs bosons. In our analysis, we use {\tt
HiggsBounds}\cite{Bechtle:2013wla,Bechtle:2013gu,Bechtle:2011sb,Bechtle:2008jh,Bechtle:2015pma}
to obtain the 95\% CL excluded region.

Finally, we discuss the Higgs phenomenology. 
To avoid the constraints on the Higgs signal strengths, we require
\cite{Beauchesne:2017ukw,Bechtle:2014ewa}
\begin{equation}
 c_\theta^2\gtrsim0.8\ ,
\end{equation}
as a reference\footnote{By using the P-value of {\tt HiggsSignals}
\cite{Bechtle:2013xfa,Bechtle:2014ewa}, the number is about 0.7 at 95\%
CL. The ATLAS and the CMS collaborations provide global fits of the
signal strengths, which give about 0.95 at 95\%CL since the central
values are rather high \cite{ATLAS:2018doi,CMS:2018lkl}.}.  In
particular, the branching ratio of the Higgs boson into two relaxions
easily dominates over the others due to a large coupling,
\begin{equation}
 g_{hSS}\simeq\frac{v\Lambda_h^2}{f^2}c_X\ .
\end{equation}
Thus, we forbid it kinematically by requiring
\begin{equation}
 \frac{m_{125}}{2}<m_S\ .
\end{equation}

Notice that all the collider constraints can be avoided if the relaxion
mass and the Higgs mass are degenerated. However, we do not consider
such a region to avoid additional fine-tuning.

\section{Parameter region}\label{sec_param_sp}
In this section, we present an example parameter region. To reduce the
number of parameters, we fix the following parameters in this section;
\begin{equation}
 H=10~{\rm GeV}\ ,~\epsilon=0.8~{\rm GeV}\ ,~r=0.002\ ,~M=20~{\rm TeV}\ .\label{eq_fixed_params}
\end{equation}
In addition, we restrict $\Lambda_0=\Lambda_h$.  If not explicitly
specified, $\lambda$ is chosen so that it reproduces the observed Higgs
mass.
%

\begin{figure}[t]
 \begin{center}
   \includegraphics[width=0.7\linewidth]{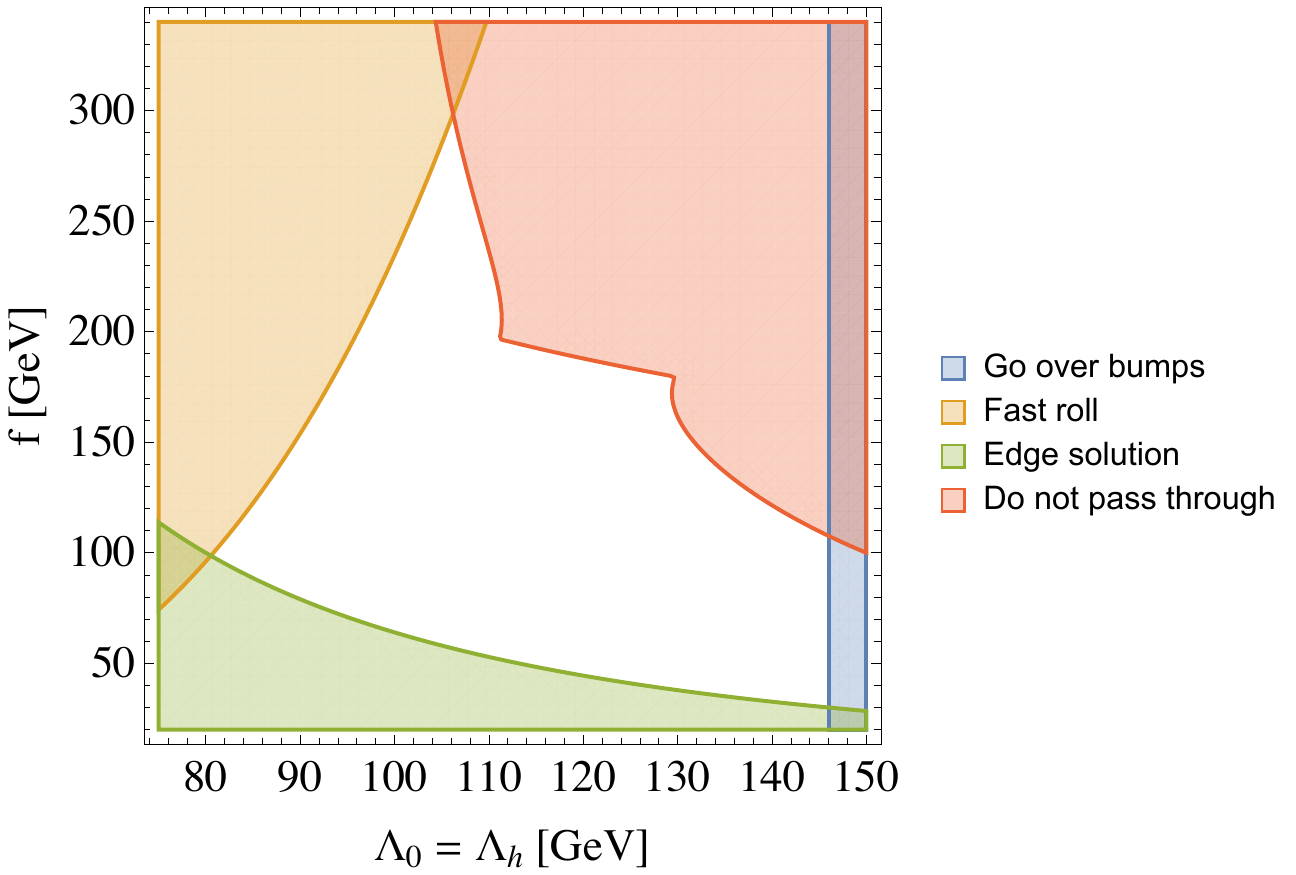} \caption{The
  conditions for the successful relaxation. We plot only the strongest
  ones. The blue shaded region around the right edge is excluded by ``Go
  over bumps'' condition. The orange shaded region around the upper left
  corner is excluded by ``Fast roll'' condition. The green shaded region
  around the bottom left corner is excluded by ``Edge solution''
  condition. The red shaded region around the upper right corner is
  excluded by ``Do not pass through'' condition.}  \label{fig_const_SR}
 \end{center}
\end{figure}

The conditions for the successful relaxation are shown in
Fig.~\ref{fig_const_SR}. We show only those of
Eqs.~\eqref{eq_go_over_bumps}, \eqref{eq_fast_roll},
\eqref{eq_edge_solution}, and \eqref{eq_do_not_pass}, which give the
strongest bounds with the parameter set given in
\eqref{eq_fixed_params}. The shaded regions are excluded by the reasons
described in the legend. As we can see, the allowed region is not small,
so that we do not need to fine-tune the parameters.

\begin{figure}[t]
 \begin{center}
  \begin{minipage}{0.48\linewidth}
   \includegraphics[width=\linewidth]{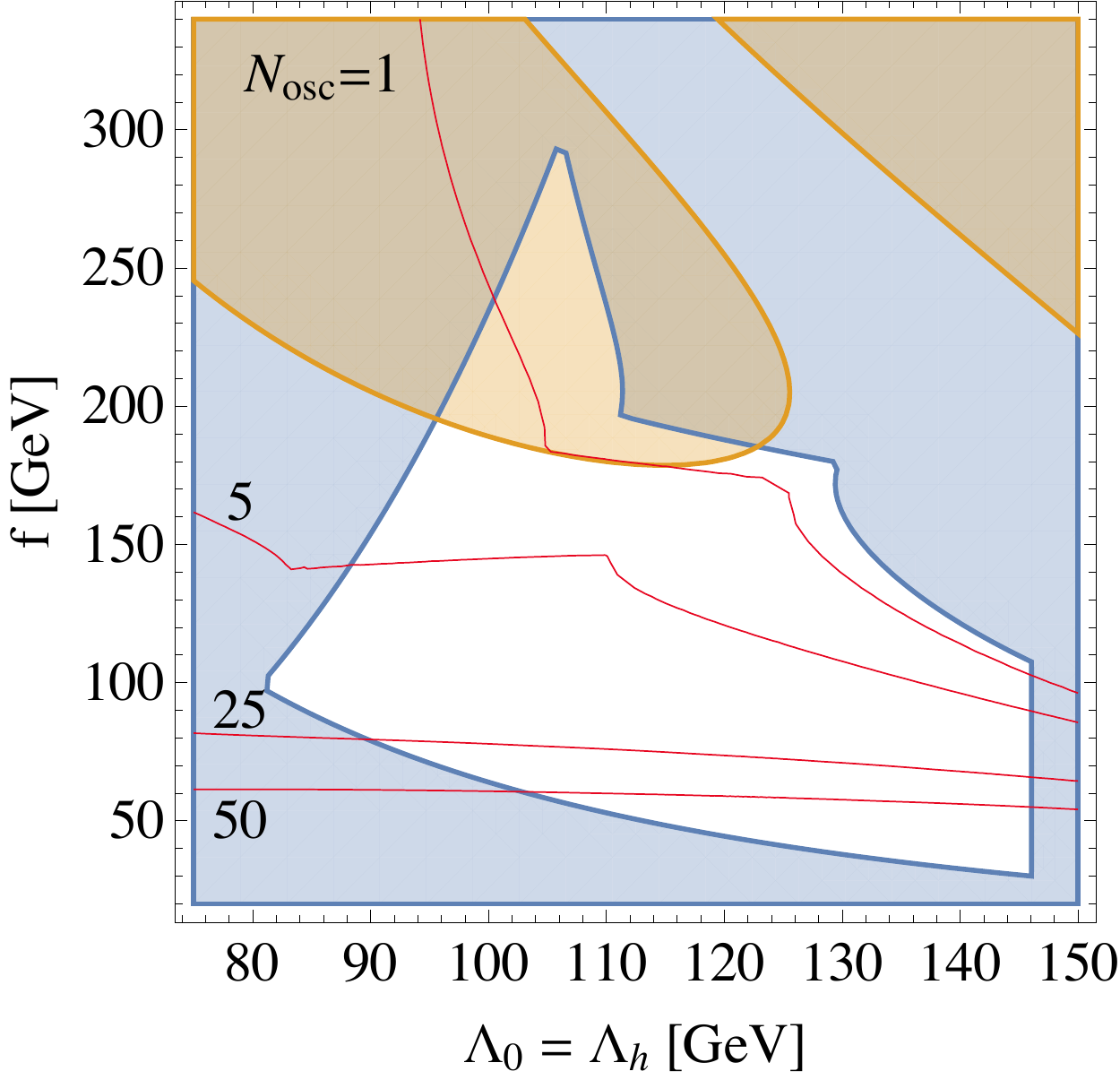}
  \end{minipage}
  \begin{minipage}{0.48\linewidth}
   \includegraphics[width=\linewidth]{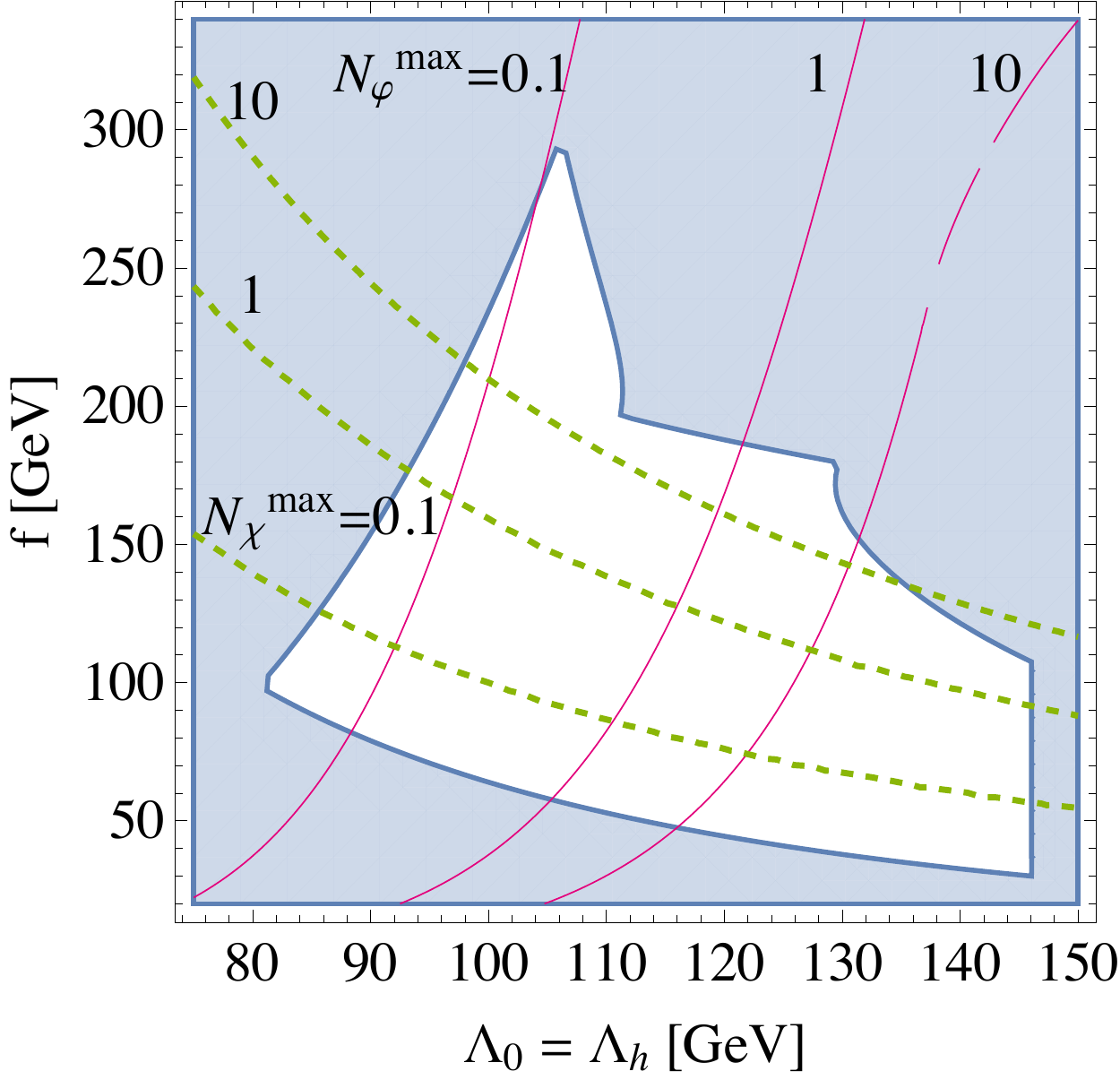}
  \end{minipage}

  \caption{Consistency check of our analysis. The blue shaded region is
  a summary of Fig.~\ref{fig_const_SR}. {\it Left:} The orange shaded
  region violates Eq.~\eqref{eq_const_q}. The red lines show the estimate
  of the number of Higgs oscillation, $N_{\rm osc}=1,~5,~25,~50$. {\it Right:} The
  maximum occupation number produced before the Higgs oscillation. The red
  solid lines correspond to $N_\varphi^{\max}=0.1,~1,~10$ and the green dashed ones
  correspond to $N_\chi^{\max}=0.1,~1,~10$.}  \label{fig_const_C}
 \end{center}
\end{figure}

Let us move on to the constraints from the consistency of our
analysis. In Fig.~\ref{fig_const_C}, we show where our analysis becomes
unreliable. The blue region is the same as in Fig.~\ref{fig_const_SR}
but all the constraints are combined. In the left panel, the orange
shaded region violates Eq.~\eqref{eq_const_q}, where we can not use the
analytic relations for the edge solution. The red lines in the same
panel show the number of Higgs oscillations. To avoid the resonant
particle production, this should not be so large. Since this is a very
rough approximation, we assume
\begin{equation}
 N_{\rm osc}\lesssim25\ ,\label{eq_safe_osc}
\end{equation}
is safe.

The right panel of Fig.~\ref{fig_const_C} shows the maximum occupation
numbers of $\varphi$ and $\chi$ before the Higgs oscillation
starts. These numbers should be small enough so that the assumption of
homogeneous fields is a good approximation. In this analysis, we assume
\begin{align}
 N_\varphi^{\max}\lesssim10\ ,~N_\chi^{\max}\lesssim10\ ,\label{eq_safe_pp}
\end{align}
are safe.

\begin{figure}[t]
 \begin{center}
  \begin{minipage}{0.48\linewidth}
   \includegraphics[width=0.8\linewidth]{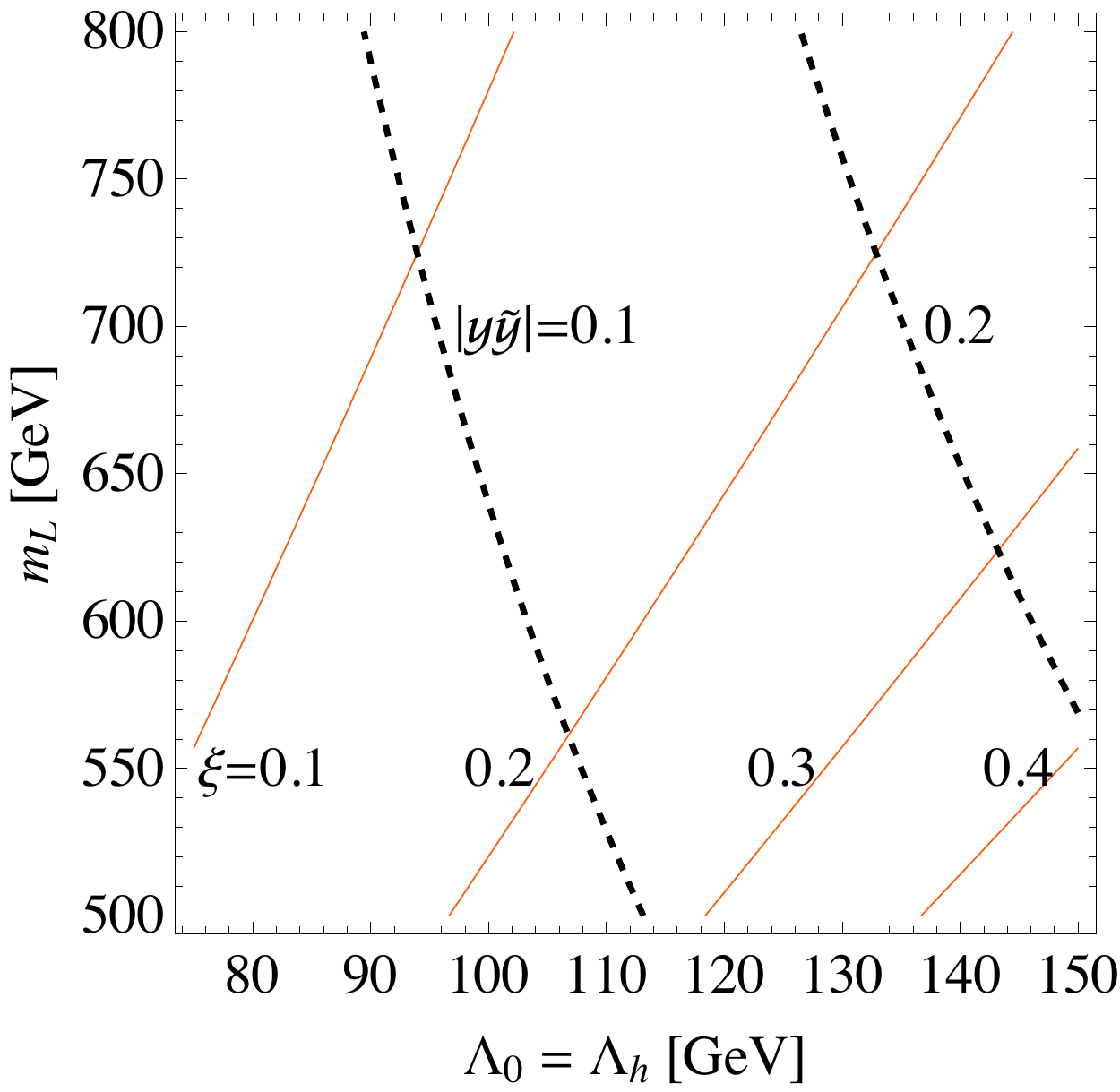}
  \end{minipage}
  \begin{minipage}{0.48\linewidth}
   \includegraphics[width=\linewidth]{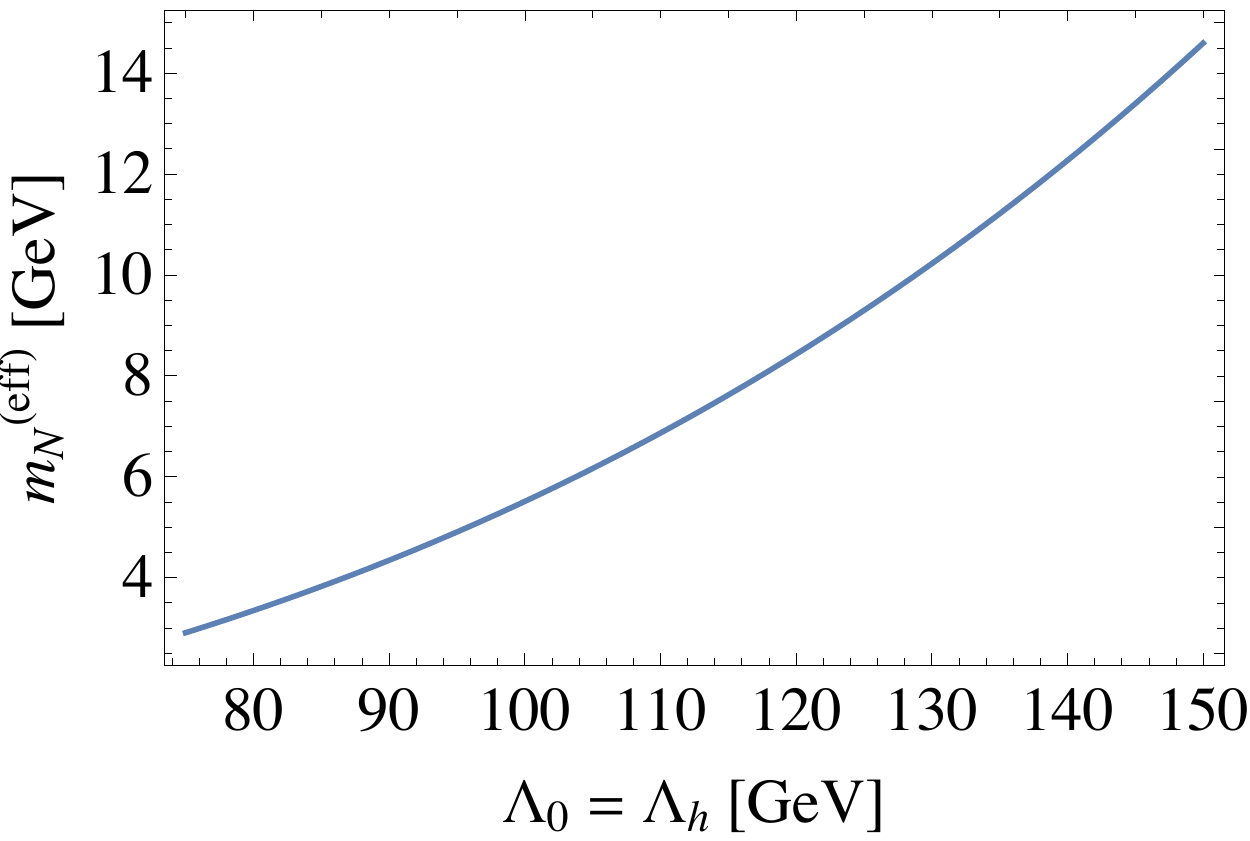}
  \end{minipage}

  \caption{The parameters in the model of strong sector. {\it Left:} The
  black dashed lines indicate $|y\tilde y|=0.1,~0.2$ and the red
  solid lines indicate $\xi=0.1,~0.2,~0.3,~0.4$. {\it Right:} The
  lightest fermion mass in the strong sector.}  \label{fig_const_Str}
 \end{center}
\end{figure}

Next, we discuss the constraints on the strong sector.
To illustrate them, let us take
\begin{equation}
 \Lambda_c=400~{\rm GeV}\ ,~\lambda=0.52\ .
\end{equation}
In Fig.~\ref{fig_const_Str}, we plot $|y\tilde y|$, $\xi$, and
$m_N^{(\rm eff)}$. Since the first two depend on $m_L$, we show also the $m_L$
dependencies in the left panel. As we can see, the fine-tuning
parameter, $\xi$, is $\mathcal O(0.1)$, and thus the parameter is not so
tuned. In addition, $|y\tilde y|$ and $M_N$ are small enough, so that
the conditions of Eqs.~\eqref{eq_str_mn} and \eqref{eq_str_y} are
satisfied. Since the doublet fermions are heavier than $500~{\rm GeV}$,
we can avoid the constraints from the EW precision and the resonance
searches.

\begin{figure}[t]
 \begin{center}
   \includegraphics[width=0.6\linewidth]{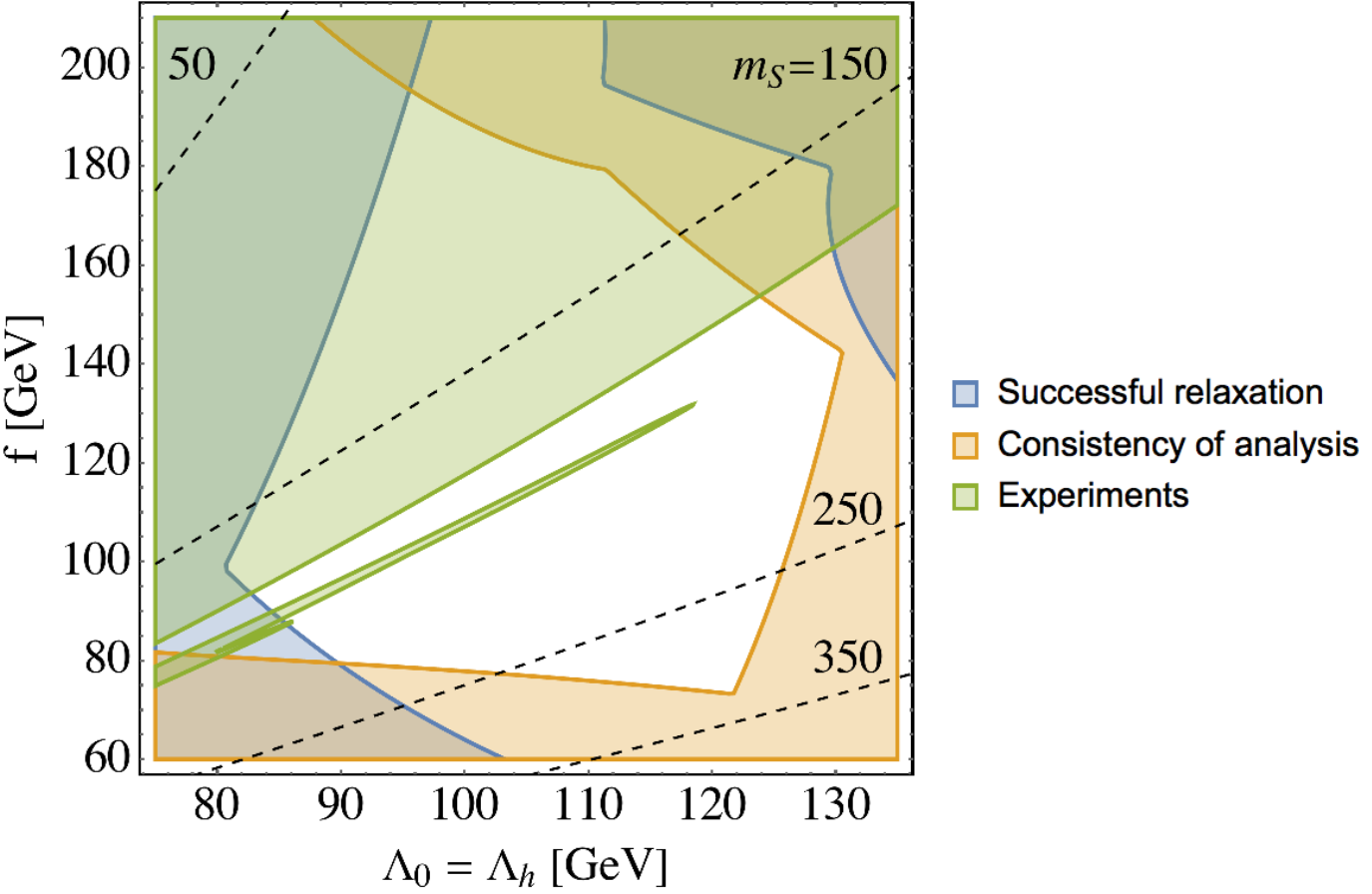} \caption{The
   experimental constraints. The blue and the orange shaded regions are
   summaries of Figs.~\ref{fig_const_SR} and \ref{fig_const_C},
   respectively. The green region is excluded either by the collider
   searches at the 95\% CL, or by the Higgs signal strength
   measurements. The black dashed lines indicates
   $m_S=50,~150,~250,~350$.}  \label{fig_const_Ex}
 \end{center}
\end{figure}

Finally, we show the collider constraints on the relaxion and the Higgs
signal strengths. In Fig.~\ref{fig_const_Ex}, we shade the excluded
region with green, which comes mainly from the searches for the
relaxion. The blue region is the same as in Fig.~\ref{fig_const_Str} and
the orange region violates the consistency conditions, namely,
Eqs.~\eqref{eq_const_q}, \eqref{eq_safe_osc} and \eqref{eq_safe_pp}. We
plot $m_S$ with black dashed lines.  In the allowed region, the main
decay modes of the relaxion are typically WW and ZZ. Around the bottom
right corner of the allowed region, we have a region with $m_S>2m_{\rm
125}$, where the $S\to hh$ channel opens. It tends to dominate the decay
width. The narrow green band comes from the low-mass end of the LHC
constraints.

\begin{figure}[t]
 \begin{center}
  \begin{minipage}{0.48\linewidth}
   \includegraphics[width=0.8\linewidth]{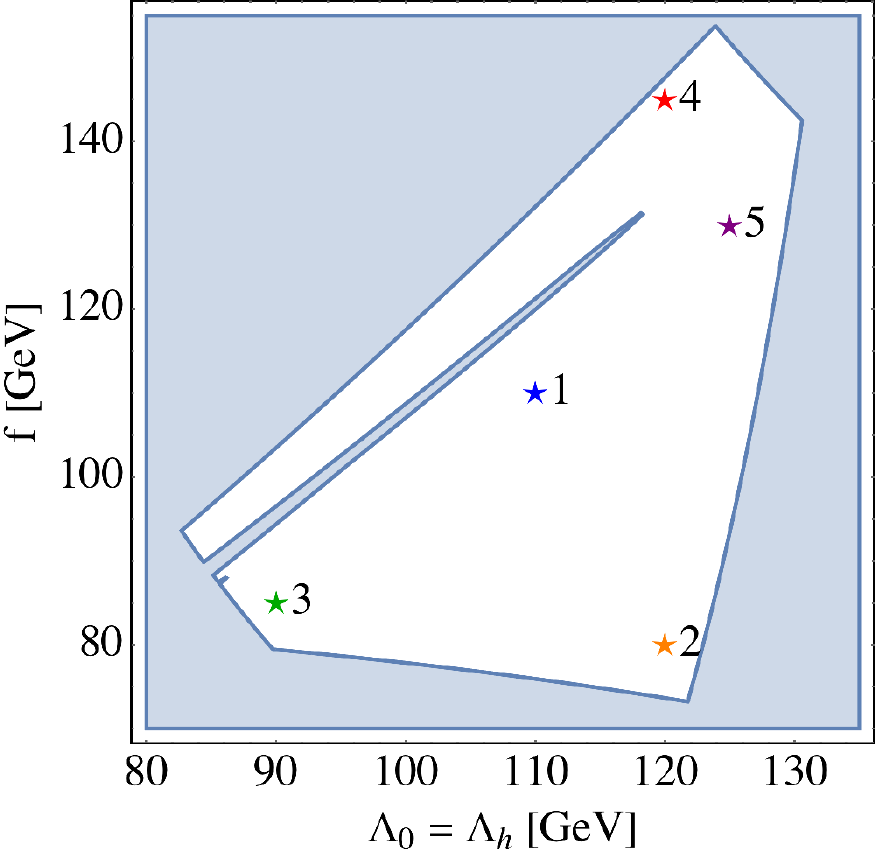}
  \end{minipage}
  \begin{minipage}{0.51\linewidth}
   \includegraphics[width=\linewidth]{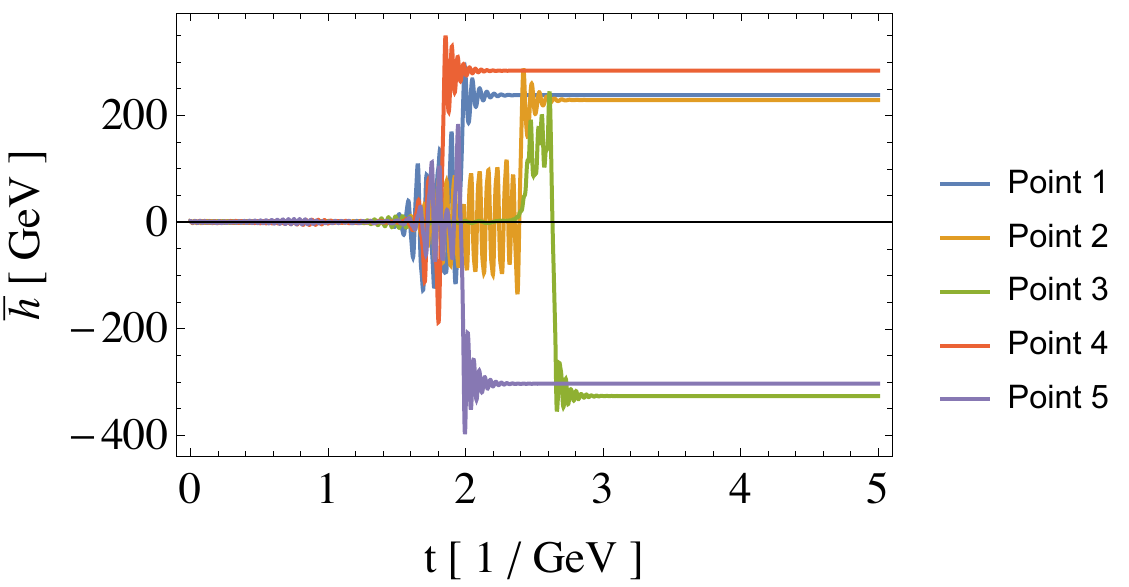}
  \end{minipage}

  \caption{The evolution of the Higgs field for each sample point. The
  locations of sample points are shown in the left panel. For each point,
  the evolution of the Higgs field is shown in the right panel.}
  \label{fig_sample}
 \end{center}
\end{figure}

We pick up some sample points within the allowed region, which are shown
with stars in the left panel of Fig.~\ref{fig_sample}. In the right
panel, we plot the evolution of the Higgs homogeneous mode for each
sample point. The Higgs oscillations are expected to start at $t=1$\,GeV$^{-1}$. As
we can see, the relaxation is successful for these sample points. We
summarize phenomenology for each example point in Table
\ref{tbl_sample}.

At Point 3, the Higgs oscillation does not start. It is simply because
the parameter set is around the boundary of
Eq.~\eqref{eq_edge_solution}. Even so, the relaxation is successful and
it corresponds to the other mechanism explained in Appendix
\ref{apx_the_other_way}.

\begin{table}[t]
 \begin{tabular}{c||c|c|c|c|c|c|c|c}
  Point& $\Lambda_0=\Lambda_h~[{\rm GeV}]$&$f~[{\rm GeV}]$&$B_{\rm thin}$&$N_\varphi^{\max}$&$N_\chi^{\max}$&$m_S~[{\rm GeV}]$&$s_\theta$&${\rm BR}(S\to 2h_{\rm 125})$\\
  \hline\hline
  1&110&110&39715&0.63&0.29&192&-0.22&0\\
  2&120&80&73916&5.18&0.12&290&-0.06&0.79\\
  3&90&85&9186&0.11&0.03&197&-0.21&0\\
  4&120&145&71693&1.33&3.49&165&-0.41&0\\
  5&125&130&101498&3.76&2.32&189&-0.22&0\\
 \end{tabular}
 \caption{Phenomenology at the example points}
 \label{tbl_sample}
\end{table}

\section{The Relaxion Potential and Cosmological History}\label{sec_cosmo}
In the preceding sections, we have discussed the dynamics of the
relaxion and the Higgs field assuming that their potential is given by
Eq.~\eqref{eq_potential}.  The most important feature of the potential
is the shift symmetry of $X$ that is broken by two distinct sources.
In this section, we give a sketch of the model in which the
relaxion is identified with a pseudo NGB.  The details of the model is
discussed in Appendix~\ref{sec:clockwork2}.  We also discuss
cosmological history that is compatible with the fast-rolling relaxion.

\subsection{Pseudo Nambu-Goldstone Relaxion}
When the relaxion appears as a pseudo NGB, we expect
that its potential has a form of
\begin{eqnarray}
\label{eq_potential2}
V(X) = \kappa_H \left(rM^2 + |\Phi|^2\right)F_H^2 
\cos\left(\frac{X}{F_H} \right)
+
\Lambda^4(|\Phi|^2)
\cos\left(\frac{X}{F_L} + \delta \right)
\ ,
\end{eqnarray}
as the NGB resides in a compact field space.
Here, we introduced two energy scales, $F_H$ and $F_L$, which are
associated with spontaneous breaking of a global $U(1)$ symmetry.
We take $F_L$ to be $f$ in Eq.\,(\ref{eq_potential}), while $F_H$ to be
very large so that it reproduces the linear potential of the relaxion.
The $X$-dependence of the potential arises from explicit breaking of the
global $U(1)$ symmetry. An arbitrary phase, $\delta$, is irrelevant in
the following discussion.

Before discussing an explicit model that generates
Eq.\,(\ref{eq_potential2}), let us first clarify the relation between
Eq.\,(\ref{eq_potential2}) and Eq.\,(\ref{eq_potential}).

To identify the first term in Eq.\,(\ref{eq_potential2}) as the linear
potential in Eq.\,(\ref{eq_potential}), we need a large enough
$F_H$. Since the relaxion needs to scan its field range of $\mathcal
O(M^2/\epsilon)$ to find the EW scale, we require
\begin{eqnarray}
\label{eq_condition1}
F_H \gtrsim \frac{M^2}{\epsilon}\ .
\end{eqnarray}
The coefficient of the linear potential in Eq.\,(\ref{eq_potential}) is
obtained by expanding $X$ around an arbitrary point, $X_0$, as
\begin{eqnarray}
X = X_0  + \delta X. 
\end{eqnarray}
Then, we have\footnote{We assume that $X_0$ is not accidentally very
close to an extremum of $\cos(X/F_H)$.}
\begin{eqnarray}
\epsilon  = \kappa_H  F_H\sin\left(\frac{X_0}{F_H}\right).
\end{eqnarray}

\begin{figure}[t]
 \begin{center}
   \includegraphics[width=0.7\linewidth]{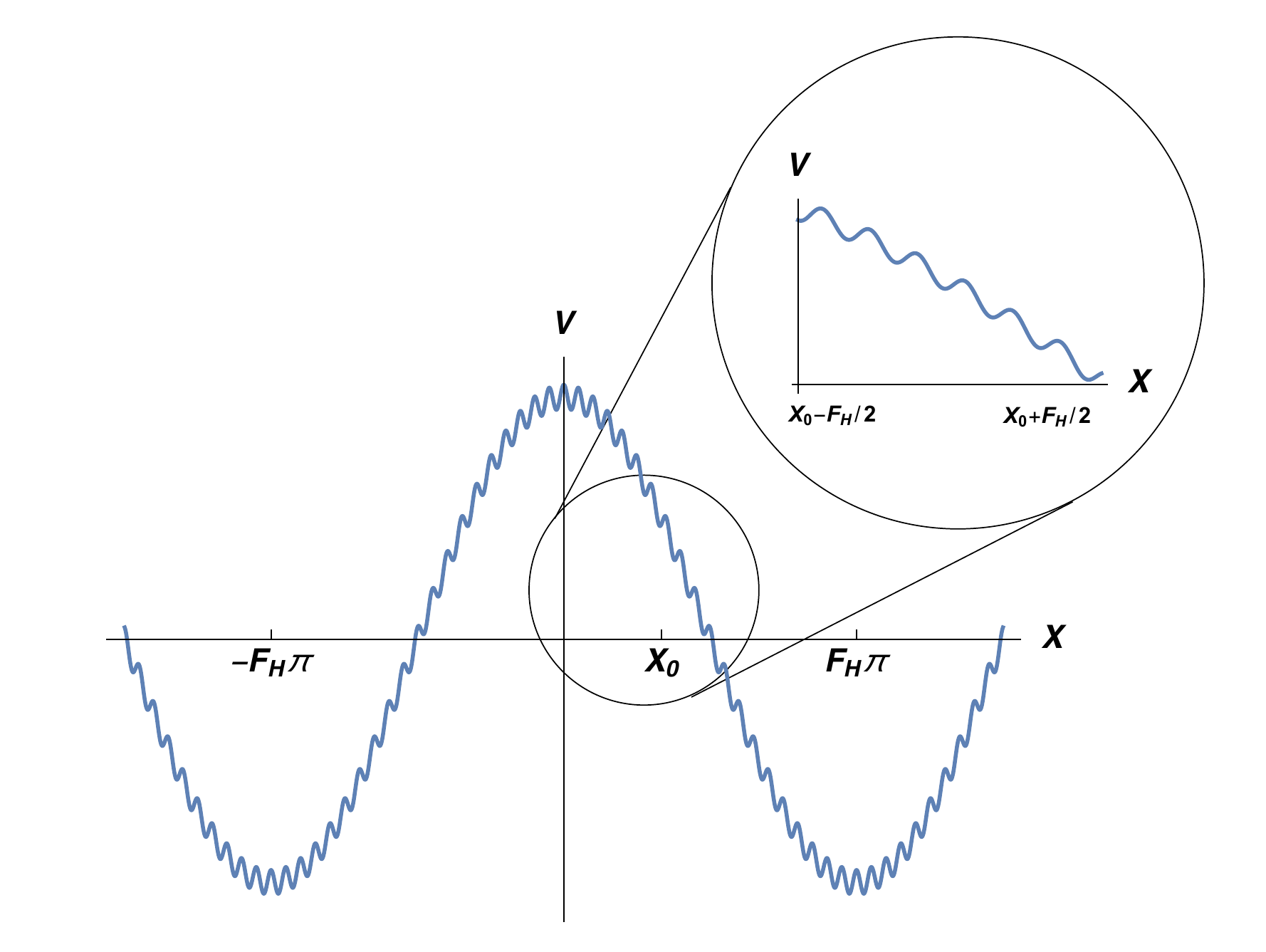} \caption{An
  illustration of the relaxion potential of Eq.\,(\ref{eq_potential2})
  with $F_H = 40F_L$.  The small wiggle comes from the second term in
  Eq.\,(\ref{eq_potential2}).  The inserted figure is a close-up of the
  potential around $X_0/F_H = \pi/3$.}  \label{fig_pNGB}
 \end{center}
\end{figure}

The first term in Eq.\,(\ref{eq_potential2}) also contributes to the Higgs mass squared by
\begin{eqnarray}
{\mit \Delta}M^2 = \epsilon F_H \cos\left(\frac{X_0}{F_H}\right)\ ,
\end{eqnarray}
at $X \simeq X_0$.  It should not exceed ${\mathcal O}(M^2)$ since we
assume the natural scale of the Higgs mass squared is $\mathcal O(M^2)$.
Together with the constraint of Eq.~\eqref{eq_condition1}, $F_H$ should
satisfy
\begin{eqnarray}
F_H \sim\frac{M^2}{\epsilon}\ .
\end{eqnarray}
For example, if we take $F_H \simeq 5\times 10^{8}$\,GeV and
$\kappa_H\simeq 10^{-9}$, we have $\epsilon \simeq 0.8\,$GeV and $M
\simeq 20\,$TeV.\footnote{A hierarchy between $F_H$ and $M$ can be
stabilized, for example, by supersymmetry with $M$ being the soft scalar
mass of the Higgs doublets.}

In Fig.\,\ref{fig_pNGB}, we show an illustration of the relaxion
potential with $F_H = 40F_L$.  The larger structure of the potential is
governed by the first term in Eq.\,(\ref{eq_potential2}), while the
small wiggle comes from the second term. We close up the potential
around $X_0 /F_H=\pi/3$ for a small interval, $\delta X \in [-F_H/2,
F_H/2]$, to illustrate that it is well approximated by a linear
potential with a small modulation.

\subsection{Relaxion Potential from Clockwork Mechanism}
\label{sec:clockwork}
Here, we discuss a model that achieves the potential in
Eq.\,(\ref{eq_potential2}).  As we have mentioned, the $X$-dependence of
the potential originates from explicit breaking of the global symmetry.
A difficulty in model building is that we need two hierarchical decay
constants, $F_{H}$ and $F_L$.

Let us first consider a global $U(1)$ symmetry that is broken by two
condensation operators; one is at $\mathcal O(F_H)$ and the other is at
$\mathcal O(F_L)$.  We assume that the $U(1)$ charges of these operators
are the same. We consider an effective scalar potential given by
\begin{eqnarray}
\label{eq_potential3}
V(X) &=& \frac{1}{2} \kappa_H \left(rM^2 + |\Phi|^2\right)F_H^2  e^{i \frac{\pi_H}{F_H}}
+
\frac{1}{2}\Lambda^4(|\Phi|^2) e^{i \frac{\pi_L}{F_L}}  - g F_H^2 F_L^2 e^{i \left(\frac{\pi_H}{F_H}- \frac{\pi_L}{F_L}\right)} + h.c.\ ,
\end{eqnarray}
where $\pi_H$ and $\pi_L$ are the phase components of the condensation
operators.  It has a similar structure to that of
Eq.\,(\ref{eq_potential2}), but we also have the third term, which is
consistent with the $U(1)$ symmetry realized by the shifts of $\pi_H$ and $\pi_L$,
\begin{eqnarray}
\frac{\pi_H}{F_H} \to \frac{\pi_H}{F_H} + \alpha\ , \quad
\frac{\pi_L}{F_L} \to \frac{\pi_L}{F_L} + \alpha\ , \quad \alpha = [0,2\pi)\ .
\end{eqnarray}
We assume that the coefficient $g$ is order one and thus the third term has the largest coefficient.  
Since one of the linear combinations of $\pi_H$ and $\pi_L$ 
becomes very heavy because of the third term, we
identify the other lighter combination as the relaxion.

The potential of Eq.\,(\ref{eq_potential3}), however, does not provide
the desirable relaxion potential of Eq.\,(\ref{eq_potential2}).  In the
limit of vanishing explicit breaking, the mass eigenstates of $\pi_H$
and $\pi_L$ are given by
\begin{eqnarray}
\left(
\begin{array}{c}
X_H  \\
X
\end{array}
\right)
=
\frac{F_H F_L}{\sqrt{F_H^2 + F_L^2}}
\left(
\begin{array}{cc}
F_H^{-1} &-F_L^{-1} \\
F_L^{-1} &F_H^{-1}   
\end{array}
\right)
\left(
\begin{array}{c}
\pi_H   \\
\pi_L
\end{array}
\right)\ .
\end{eqnarray}
Thus, we find that $\pi_H$ and $\pi_L$ include $X$ component as
\begin{eqnarray}
\frac{\pi_H}{F_H} = \frac{X }{\sqrt{F_H^2 + F_L^2} } 
\simeq  \frac{X}{F_H}
\ , \quad
\frac{\pi_L}{F_L} = \frac{X }{\sqrt{F_H^2 + F_L^2} }\simeq  \frac{X}{F_H}\ , 
\end{eqnarray}
and hence both of the first two terms in Eq.\,(\ref{eq_potential3}) give
the same decay constant of the relaxion around $F_H$.\footnote{This is
obvious from the fact that the NGB mostly resides in the phase component
of the first condensation operator.}

This difficulty can be circumvented by introducing a hierarchical $U(1)$
charge between the two condensation operators. When the operator
corresponding to $\pi_H$ has charge $Q_H \ll 1$, the third term is
modified as
\begin{eqnarray}
\label{eq_potential4}
V(X) &=& \frac{1}{2} \kappa_H \left(rM^2 + |\Phi|^2\right)F_H^2  e^{i \frac{\pi_H}{F_H}} +
\frac{1}{2}\Lambda^4(|\Phi|^2) e^{i \frac{\pi_L}{F_L}}  - g F_H^2 F_L^2 e^{i \left(\frac{\pi_H}{F_H}- \frac{Q_H\pi_L}{F_L}\right)} + h.c.\ .
\end{eqnarray}
Here, the $U(1)$ symmetry is realized by
\begin{eqnarray}
\label{eq_mixing2}
\frac{\pi_H}{F_H} \to \frac{\pi_H}{F_H} + Q_H \alpha\ , \quad
\frac{\pi_L}{F_L} \to \frac{\pi_L}{F_L} + \alpha\ , \quad \alpha = [0,2\pi)\ .
\end{eqnarray}
In this case, the relaxion component of $\pi_H$ and $\pi_L$ are given by 
\begin{eqnarray}
\frac{\pi_H}{F_H} = \frac{X }{\sqrt{F_H^2 + F_L^2/Q_H^2} } 
\ , \quad
\frac{\pi_L}{F_L} = \frac{X }{Q_H\sqrt{F_H^2 + F_L^2/Q_H^2} }\ .
\end{eqnarray}
Thus, with $F_H \sim F_L/Q_H$, we achieve
\begin{eqnarray}
\label{eq_mixing3}
\frac{\pi_H}{F_H} \sim \frac{X}{F_H}\ , \quad
\frac{\pi_L}{F_L} \sim \frac{X}{F_L} \ .
\end{eqnarray}
Substituting Eq.\,(\ref{eq_mixing3}) to the potential in
Eq.\,(\ref{eq_potential4}), we obtain the desirable potential of
Eq.\,(\ref{eq_potential2}) with $F_H\gg F_L$.

It should be noted that an exponentially small $Q_H$ can be achieved by
the clockwork
mechanism~\cite{Giudice:2016yja,Kaplan:2015fuy,Choi:2015fiu} (see
also~\cite{Harigaya:2014eta,*Harigaya:2014rga}).  For example, let us
consider $N+1$ sectors containing condensation operators. We assume the
charges of these operators decrease in geometric progression with ratio
$q^{-1}$ $(q>1)$.  Identifying the phase component of the operator in
the first sector as $\pi_L$ and that in the $(N+1)$-th sector as
$\pi_H$, we effectively get an exponentially small charge of $Q_H =
q^{-N}$.  For example, the hierarchy between $F_H \simeq 5\times
10^{8}$\,GeV and $F_L \simeq 10^{2}$\,GeV is realized with $q = 3$ and
$N=14$.\footnote{The smallness of $F_L$ does not re-introduce a
hierarchy problem if we adopt dynamical symmetry breaking in each
sector~\cite{Coy:2017yex} or supersymmetric clockwork
discussed in Appendix \ref{sec:clockwork2}.}

 There are two caveats in the construction of an explicit clockwork
model. First, the mass scales of the explicit breaking of the $U(1)$
symmetry should not exceed those of the spontaneous breaking since
otherwise the existence of the NGB is invalidated.  Second, the $F_H$
and $F_L$ should be stabilized in a way that the larger mass scale does
not interfere the lower mass scale.  In Appendix~\ref{sec:clockwork2},
we elaborate on these points and consider a clockwork mechanism with
progressively increasing VEVs to overcome these difficulties.

Finally, let us comment on the origin of the explicit breaking terms
(see Appendix D for details).  The second term in the potential in
Eq.\,(\ref{eq_potential4}) can be obtained by using the model in Section
\ref{sec_model} with $X/f\to \pi_L/F_L$.  Similarly, we can construct a
model that gives the first term in the potential in
Eq.\,(\ref{eq_potential4}). Let us consider a new strong sector that
couples to $\pi_H$, whose Lagrangian is given by
\begin{eqnarray}
\label{eq:explicitH}
{\cal L} = -\frac{1}{32\pi^2}\frac{\pi_H}{F_H}G^a_{H\mu\nu}{\tilde G}_H^{a\mu\nu}
+ M_N N_H N_H^c +M_LL_HL_H^c
 +y_H\Phi L_HN_H^c
 +\tilde y_H\Phi^\dagger L_H^cN_H\ .
\end{eqnarray}
Here, $G_H$ denotes the field strength of a new $SU(3)$ gauge group.
The new fermions, $N_H$ and $L_H$, are in the fundamental representation
of $SU(3)$ and correspond to $N$ and $L$ in Section \ref{sec_model},
respectively.  Assuming that $L_H$ is heavier than the dynamical scale,
$\Lambda_H$, and $N_H$ is lighter, we find
\begin{eqnarray}
r M^4 &\sim& \kappa_H r M^2 F_H^2 \sim  M_N \Lambda_H^3 \ ,  \\
{M^2}&\sim& \kappa_H F_H^2 \sim \frac{\tilde y_Hy_H}{M_L} \Lambda_H^3   \ .
\end{eqnarray}
With these relations, we find, for example, that $M\simeq 30$\,TeV and
$\epsilon \simeq 0.1$\,GeV can be achieved with $M_L \sim \Lambda_H
\simeq M$, $\tilde{y_H}y_H = \mathcal O(1)$, and $M_N/M_L \sim (\tilde
y_H) y_H r$.\footnote{The mass squared, $M_N$, is required to be at
least of $\mathcal O(\tilde y_Hy_H/16\pi^2 M_L )$ to be technically
natural. }
See Appendix~\ref{sec:clockwork2} for more details.

\subsection{Cosmological History}
In our mechanism, we assume that the initial velocity of the relaxion
is large enough so that it can go over the bumps created by the strong
dynamics.  In this section, we discuss cosmological history 
where the relaxion starts to roll before or during inflation.

A simple possibility is that there is a radiation dominated era 
before the relaxion mechanism takes place.  
We assume that the maximal temperature is high enough 
so that all the terms in
Eq.\,(\ref{eq_potential2}) vanish.  
The succeeding inflation era starts
typically before these terms are recreated. In the inflation era, the
temperature decreases very quickly and the first dynamics confines at
$T\simeq \Lambda_H$, which generates the first term in
Eq.\,(\ref{eq_potential2}).\footnote{Here, we assume that the global
$U(1)$ symmetry has been broken well before the inflation starts.} It
accelerates the relaxion unless the relaxion accidentally sits around an
extremum.  The relaxion reaches the terminal velocity within about one
$e$-fold after the first confinement.  Subsequently, the second dynamics
confines at $T\simeq \Lambda$, which generates the second term in
Eq.\,(\ref{eq_potential2}).  If the relaxion has been accelerated
enough, it can go over the created bumps, which explains the origin of
the fast-rolling.

\begin{figure}[t]
 \begin{center}
   \includegraphics[width=0.5\linewidth]{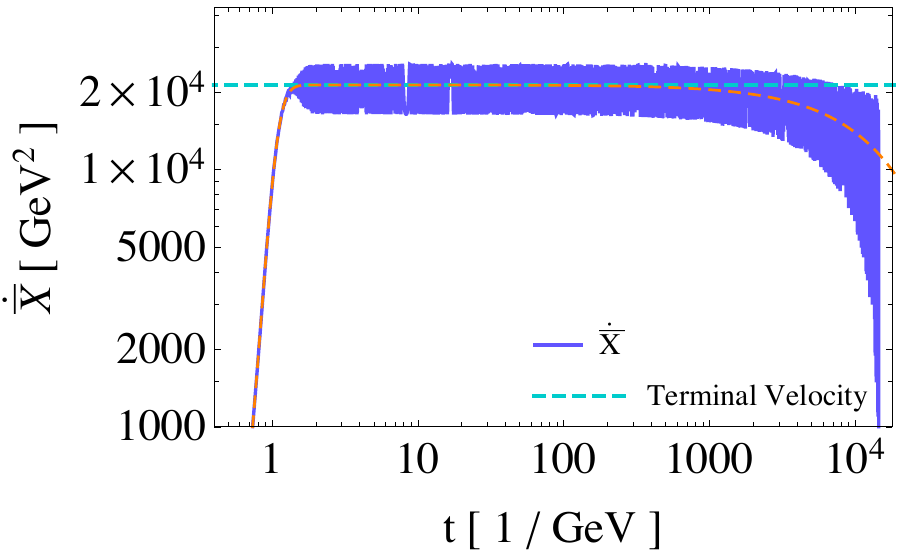} \caption{The
  evolution of $\dot X$ during inflation with $|\Phi| = 0$ (blue line).
  Low scale inflation with the Hubble constant $H = 10$\,GeV is assumed
  to take place after radiation domination.  (The time variable $t$ can
  be regarded as an $e$-folding number, ${\mathcal N}_e = Ht$.)  During
  radiation domination, the strong dynamics responsible for the relaxion
  potential in Eq.\,(\ref{eq_potential2}) does not exhibit confinement,
  and hence, the relaxion does not have any potential.  The first
  dynamics confines at $T\simeq \Lambda_H$, which generates the first
  term in Eq.\,(\ref{eq_potential2}), and the relaxion is
  accelerated. The second dynamics confines at $T\simeq \Lambda_c$,
  which modulate the relaxion potential.  The dashed cyan line shows the
  terminal velocity expected for a purely linear potential.  The dashed
  orange line shows the evolution of $\dot X$ without the second term in
  Eq.\,(\ref{eq_potential2}).  } \label{fig_Xdot}
 \end{center}
\end{figure}

In Fig.\,\ref{fig_Xdot}, we show the time evolution of $\dot {\bar X}$
during inflation (blue line), ignoring the Higgs field.  Here, we
assume that the inflation era starts at $t\sim \mathcal
O(0.1)$\,GeV$^{-1}$ with $H = 10$\,GeV.  The parameters are taken to be
$\epsilon = 0.8$\,GeV, $r = 0.002$, $M = 20$\,TeV, $F_H =
5\times10^{8}$\,GeV, $\Lambda_H = M$, and $F_L = \Lambda = 100$\,GeV.
The initial condition of $X$ is set to be $X = F_H (\pi - 1) $ and $\dot
X = 0$ at $t = 0.05$\,GeV.  In the analysis, we turn on the first and
the second terms in Eq.\,(\ref{eq_potential2}) at $T \simeq \Lambda_H$
and $T \simeq \Lambda$, respectively.  For comparison, we show $\dot
{\bar X}$ without the second term in Eq.\,(\ref{eq_potential2}) with the
dashed orange line.  We also show the terminal velocity for a purely
linear potential with the dashed cyan line.

As we can see from the figure, the relaxion reaches the terminal
velocity around $t\simeq 1$\,GeV$^{-1}$.  Soon after, the second term in
Eq.\,(\ref{eq_potential2}) is turned on and the velocity of the relaxion
starts to oscillate.  Our stopping mechanism is expected to work around
$t\sim10^3~\rm GeV^{-1}$ and the linear potential approximation is good
in this timescale. Note that the relaxion continues to roll in the
figure since we do not take into account the Higgs field.

We give another possibility for the origin of the fast-rolling, which
does not require a radiation dominated era before inflation.  Suppose
that the field value of the relaxion at the onset of the inflation is
randomly distributed in space.  Here, we assume that the potential of
Eq.\,(\ref{eq_potential2}) has already been developed.  Then, in some of
the Hubble patches, we find a small region that takes the value of the
relaxion in a different potential well.  If the vacuum energy of the
interior is higher than that of the exterior, such a region starts to
contract and eventually collapses. At the point of collapse, the
released energy is converted to the kinetic energy of the relaxion and
the relaxion starts to roll down the potential.

Let us move on to cosmological history after the relaxation. As is
discussed in \cite{Graham:2015cka}, we need to reproduce the observed
CMB fluctuations and reheat the SM particles safely. In our mechanism,
the $e$-folds during the relaxation is estimated as
\begin{equation}
 \mathcal N_e\gtrsim\frac{3H^2}{r\epsilon^2},
\end{equation}
which is about $10^5$ with the parameter sets in the previous section.
The observed CMB fluctuations can be compatible with such a number of
$e$-folds if we consider, for example, new inflation.  After the
inflation, the maximal temperature should not be raised above $\sim
100\,{\rm GeV}$, otherwise the relaxion would start to roll again. Thus,
we need another well-separated sector, into which the potential energy
of the inflaton is dumped. We do not go further in details since it
falls out of the scope of this paper.

Finally, let us comment on baryogenesis.  Although the upper bound on
the maximal temperature is similar as in \cite{Graham:2015cka}, the
situation is worse because we can not use the sphalerons to generate the
baryon asymmetry.  There are several ways to generate it with a very low
reheating temperature by using the decay of inflaton or heavy particles
\cite{Scherrer:1991yu,Pierce:2019ozl,Dimopoulos:1987rk,Babu:2006xc}, or
by using the oscillations of mesons or baryons
\cite{Nelson:2019fln,Aitken:2017wie,Elor:2018twp}.

\section{Conclusion}\label{sec_conclusion}
In the original relaxion mechanism, quantum tunneling is essential for
finding a sufficiently long-lived vacuum. However, it requires an
unacceptably large number of $e$-folds and the inflation sector seems
to be extremely fine-tuned.

We have shown that a fast-rolling relaxion can easily avoid the above
problem without extending the original relaxion model. In our mechanism,
we allow the relaxion to fly over the bumps and trigger the Higgs
oscillation around the critical point. Then, the relaxion accidentally
hits a bump and stops due to the immediate development of the Higgs
VEV. After the EW scale is relaxed, the first selected vacuum need not
decay if its lifetime is much longer than the age of the universe.

In our mechanism, the edge solution plays an important role. We have
examined it in detail and obtained the relations among the amplitude of
the Higgs oscillation, the Higgs mass, and the velocity of the
relaxion. They offer not only a physical explanation of the solution but
also a very effective way to search for viable parameters.

To keep our mechanism viable in field theory, we have discussed possible
resonant particle production. In particular, the Higgs oscillation could
produce too many particles and hence is dangerous. We discussed that it
can be avoided by limiting the number of the Higgs oscillations to be
$\mathcal O(10)$.

We have found an interesting and viable parameter region, where the
relaxion has a mass of $\mathcal O(100~{\rm GeV})$ and mixes with the
Higgs boson. The collider experiments have started to constrain the
parameter region. The cut-off scale is around $20~{\rm TeV}$ and the
fine-tuning of the Higgs mass is reduced by $10^4$.

We have also discussed a more realistic model and cosmological history,
where the relaxion resides in a compact field space and the origin of
the fast-roll is explained naturally. Importantly, the model is realized
without introducing a super-Planckian field space.

We think a higher cut-off is an interesting direction to
explore. Unfortunately, such a region typically suffers from a large
number of Higgs oscillations and the analysis becomes much more
complicated. However, our mechanism is possibly compatible with particle
production and such a region may be explored by using lattice
calculations.
\begin{acknowledgements}
  M.S. is supported in part by a JSPS Research Fellowship for Young
 Scientists Grant Number 18J12023. Y.S. is supported in part by the
 Grant-in-Aid for Innovative Areas (16H06492).  This work is also
 supported by MEXT KAKENHI Grant Nos. 15H05889 (M.I),
 No. 16H03991(M.I.), No. 17H02878(M.I.), and No. 18H05542 (M.I.), and
 World Premier International Research Center Initiative (WPI
 Initiative), MEXT, Japan. M.I. thank UC Irvine for hospitality while we
 are finalizing this work.
\end{acknowledgements}
\appendix
\section{Resonance strength}\label{apx_mathieu}
In this appendix, we give estimates of the Floquet exponents of the
Mathieu equation following \cite{Strang:aa}.

The Mathieu equation is defined as
\begin{equation}
 \frac{d^2}{dz^2}u(z)+(A-2q\cos(2z))u(z)=0\ ,\label{eq_mathieu}
\end{equation}
with real $A$ and $q$.

From the Floquet's theorem, there exist solutions that satisfy
\begin{equation}
 u(z+\pi)=e^{i\pi\nu} u(z)\ .
\end{equation}
Here, $\nu$ is a complex constant called the Floquet exponent.  Such
a function can be expanded with functions with periodicity $\pi$ as
\begin{equation}
 u(z)=e^{i\nu z}\sum_{n\in{\mathbb Z}}c_ne^{-2inz},\label{eq_ansatz}
\end{equation}
where $c_n$'s are expansion coefficients.

In the following, we evaluate
\begin{equation}
 \nu=\nu(A,q)\ ,
\end{equation}
using the Whittaker-Hill formula.

We first obtain relations among the expansion coefficients by
substituting Eq.~\eqref{eq_ansatz} to Eq.~\eqref{eq_mathieu}. We get
\begin{equation}
 c_n+\xi_n(c_{n+1}+c_{n-1})=0\ ,
\end{equation}
where
\begin{equation}
 \xi_n=\frac{q}{(\nu-2n)^2-A}\ .
\end{equation}
These relations can be expressed as
\begin{equation}
 \Xi c=0\ ,\label{eq_infeqs}
\end{equation}
where
\begin{align}
c&=(\dots,c_1,c_0,c_{-1},\dots)\ ,\\
 \Xi&=
\begin{pmatrix}
 \ddots\\
 &1&\xi_1&0\\
 &\xi_0&1&\xi_0\\
 &0&\xi_{-1}&1\\
 &&&&\ddots
\end{pmatrix}\ .
\end{align}
As shown in \cite{Strang:aa}, $(\Xi-\bf I)$ is of trace class.  It
ensures that $\det \Xi$ is finite and is independent of the choice of
basis used in the evaluation of the determinant.  In addition, it can be
shown that $\Xi$ is invertible if and only if
\begin{equation}
 \det\Xi\neq0\ .
\end{equation}

Next, we see the analytic structure of $\Delta(\nu)\equiv\det\Xi$.
As can be seen from $\Xi$, $\Delta(\nu)$ has simple poles\footnote{For
a special case, $A\in\mathbb Z$, two simple poles merges into a pole of
order 2. The following discussion is also applicable to such a case.}
at
\begin{equation}
 \nu=2n\pm\sqrt{A}\ ,
\end{equation}
and is analytic for other $\nu$.  Since $\Delta(\nu+2)=\Delta(\nu)$
and $\Delta(\nu)=\Delta(-\nu)$, it is enough to study strip
$\Re(\nu)\in[0,1]$, where we have only one pole\footnote{When $A<0$, we
slightly tilt the strip so that only one pole is in one strip.} at
\begin{equation}
 \nu=\min\left[\sqrt{A}\mod2,-\sqrt{A}\mod2\right]\ .
\end{equation}
We define a function,
\begin{equation}
 \mathcal D(\nu)=\frac{1}{\cos(\pi\nu)-\cos(\pi\sqrt{A})}\ ,
\end{equation}
which has a pole at the same position in the strip and satisfies
$\mathcal D(\nu+2)=\mathcal D(\nu)$ and $\mathcal D(\nu)=\mathcal
D(-\nu)$. Thus, for an appropriate constant, $C\neq0$,
\begin{equation}
 \Theta(\nu)=\Delta(\nu)-C\mathcal D(\nu)\ ,
\end{equation}
has no singularities for any $\nu\in\mathbb C$. Since $\Theta(\nu)$ is a
bounded entire function, it must be a constant from the Liouville's
theorem. Taking the limit of $\nu\to i\infty$, we have
\begin{align}
 \lim_{\nu\to i\infty}\mathcal D(\nu)&=0\ ,\\
 \lim_{\nu\to i\infty}\Delta(\nu)&=1\ ,
\end{align}
since $\Xi$ goes to the identity matrix. Thus, we get
\begin{equation}
 \Theta(\nu)=1\ .
\end{equation}

Using $\nu=0$ as a reference point, we have
\begin{equation}
 \Delta(\nu)=1+\frac{D(\nu)}{D(0)}(\Delta(0)-1)\ ,
\end{equation}
for $A\neq 4\ell^2$ with $\ell\in\mathbb Z$.
For Eq.~\eqref{eq_infeqs} to have a non-trivial solution, we need
$\Delta(\nu)=0$, which gives
\begin{equation}
 \sin^2\frac{\pi\nu}{2}=\Delta(0)\sin^2\frac{\pi\sqrt{A}}{2}\ ,\label{eq_nuS}
\end{equation}
which is called the Whittaker-Hill formula. Notice that we can extend it
to $A\in\mathbb R$.

Next, we evaluate $\Delta(0)$. We define finite truncated determinants,
$\Delta_p$'s, as
\begin{equation}
 \Delta_p=
 \begin{vmatrix}
  1&{\bar\xi}_p\\
  {\bar\xi}_{p-1}&1\\
  &&\ddots\\ 
  &&&1&{\bar\xi}_{p-1}\\
  &&&{\bar\xi}_p&1 
 \end{vmatrix}\ ,
\end{equation}
where ${\bar\xi}_{n}={\xi}_{n}(\nu=0)=\frac{q}{4n^2-A}$.
Then, we have
\begin{equation}
 \Delta(0)=\lim_{p\to\infty}\Delta_p\ .
\end{equation}
In the following, we obtain a recursion formula for
$\Delta_p$.  The Laplace expansions for the first row give
\begin{equation}
 \Delta_p=\Delta_p^{(-1)}-\alpha_p\Delta_p^{(-2)}\ ,
\end{equation}
where
\begin{equation}
 \alpha_p={\bar\xi}_p{\bar\xi}_{p-1}\ ,
\end{equation}
and $\Delta_p^{(-n)}$ is $\Delta_p$ without the first $n$ columns and
the first $n$ rows. In the second term, we further developed the
determinant along the first column.

Similarly, we have
\begin{align}
 \Delta_p^{(-1)}&=\Delta_{p-1}-\alpha_p\Delta_{p-1}^{(-1)}\ ,\\
 \Delta_p^{(-2)}&=\Delta_{p-1}^{(-1)}-\alpha_p\Delta_{p-2}\ .
\end{align}
After erasing $\Delta_p^{(-2)}$, we have
\begin{align}
 \Delta_p^{(-1)}+\alpha_p\Delta_{p-1}^{(-1)}&=\Delta_{p-1}\ ,\\
 \Delta_p^{(-1)}-\alpha_p\Delta_{p-1}^{(-1)}&=\Delta_p-\alpha_p^2\Delta_{p-2}\ .
\end{align}
From these, we obtain
\begin{equation}
 \Delta_p=(1-\alpha_p)\Delta_{p-1}-\alpha_p(1-\alpha_p)\Delta_{p-2}+\alpha_p\alpha_{p-1}^2\Delta_{p-3}\ .\label{eq_recS}
\end{equation}

The first three terms are given by
\begin{align}
 \Delta_{-1}=0\ ,~
 \Delta_0=1\ ,~
 \Delta_1=1-2\alpha_1\ .
\end{align}
Here, we defined $\Delta_{-1}$ so that it reproduces $\Delta_2$.

The leading terms in $q^2$ can be calculated as
\begin{align}
 \lim_{p\to\infty}\Delta_p=1-\frac{\pi\cot\left(\frac{\sqrt{A}}{2}\pi\right)}{4\sqrt{A}(A-1)}q^2+\mathcal O(q^4)\ .
\end{align}
It shows that instability of Eq.~\eqref{eq_nuS} appears around
$A=n^2,~n\in{\mathbb Z}$ for a small $q$. The most important resonance
bands are $A=1$ and $A=4$, which we call the first and the second
resonance bands, respectively.

For a more precise determination of the Floquet exponents around these
resonance bands, we solve Eq.~\eqref{eq_recS} up to a sufficiently large
$p$. We get
\begin{align}
 (\nu-1)^2&=-\frac{1}{4}q^2+
 \left(
  \frac{1}{4}(A-1)^2+\frac{3}{16}(A-1)q^2+\frac{15}{256}q^4
 \right)\nonumber\\
 &\hspace{3ex}-
 \left(
  \frac{1}{8}(A-1)^3+\frac{5}{32}(A-1)^2q^2+\frac{245}{3072}(A-1)q^4+\frac{40}{2011}q^6
 \right)\nonumber\\
 &\hspace{3ex}+\mathcal O[(A-1,q^2)^4]\ ,\\
 (\nu-2)^2&=
 \left(
  \frac{1}{16}(A-4)^2-\frac{1}{48}(A-4)q^2-\frac{5}{2304}q^4
 \right)\nonumber\\
 &\hspace{3ex}+\left(
  -\frac{1}{128}(A-4)^3+\frac{25}{2304}(A-4)^2q^2+\frac{3}{3160}(A-4)q^4+\frac{1}{94793}q^6
 \right)\nonumber\\
 &\hspace{3ex}+\mathcal O[(A-4,q^2)^4]\ .
\end{align}
Here, the rational coefficients are determined by rationalizing
numerical results with $p=1000$.

Truncating them at $\mathcal O(q^4)$, we have
\begin{align}
 \nu|_{A\simeq1}&=1\pm \frac{i}{2}\sqrt{q^2\left(1-\frac{3}{32}q^2\right)-\left(A-1+\frac{3}{8}q^2\right)^2}\ ,\\
 \nu|_{A\simeq4}&=2\pm\frac{i}{4}\sqrt{\frac{1}{16}q^4-\left(A-4-\frac{1}{6}q^2\right)^2}\ .
\end{align}
\section{Edge solution in the relaxion-Higgs system}\label{apx_edge}
In this appendix, we derive Eqs.~\eqref{eq_approx_Ah} and
\eqref{eq_approx_omega}.

We put an ansatz for the edge solution as
\begin{align}
 {\bar X}(t)&=\frac{M^2-m_{\Phi}^2}{\epsilon}+f\omega_X t+\varphi_0(t)\ ,\label{eq_ansatz_X}\\
  {\bar h}(t)&=\mathcal A_h\cos\left(\frac{\omega_X}{2}t-\alpha\right)+\chi_0(t)\ ,\label{eq_ansatz_h}
\end{align}
where $\varphi_0$ and $\chi_0$ are functions and $\alpha,~\mathcal
A_h,~m_\Phi^2$ and $\omega_X$ are constants. Here, we assume $|t|\ll |m_\Phi^2/(\epsilon f\omega_X)|$.

Substituting Eqs. \eqref{eq_ansatz_X} and \eqref{eq_ansatz_h} into
Eqs.~\eqref{eq_full_eom_X} and \eqref{eq_full_eom_h}, we get
\begin{align}
 \ddot \varphi_0+3H\dot\varphi_0&=\left[r\epsilon M^2-3fH\omega_X+\frac{\Lambda_h^2\mathcal A_h^2}{8f}\sin2\alpha\right]\nonumber\\
 &\hspace{3ex}+\left[\frac{2\Lambda_0^4+\Lambda_h^2\mathcal A_h^2}{4f}\sin\omega_Xt+\frac{\Lambda_h^2\mathcal A_h^2}{8f}\sin(2\omega_Xt-2\alpha)\right]\nonumber\\
 &\hspace{3ex}+\left[\frac{\Lambda_h^2\mathcal A_h^2}{8f^2}\cos2\alpha\right]\varphi_0\nonumber\\
 &\hspace{3ex}+\left[\frac{2\Lambda_0^4+\Lambda_h^2\mathcal A_h^2}{4f^2}\cos\omega_Xt+\frac{\Lambda_h^2\mathcal A_h^2}{8f^2}\cos(2\omega_Xt-2\alpha)\right]\varphi_0\nonumber\\
 &\hspace{3ex}+\frac{\Lambda_h^2\mathcal A_h^2}{2f}\left[\sin\left(\frac{\omega_X}{2}t+\alpha\right)+\sin\left(\frac{3\omega_X}{2}t-\alpha\right)\right]\chi_0\ ,\label{eq_eom_phi}
\end{align}
\begin{align}
 \ddot \chi_0+3H\dot\chi_0&=\left[\frac{3H\omega_X\mathcal A_h}{2}\sin\left(\frac{\omega_X}{2}t-\alpha\right)-\frac{\Lambda_h^2\mathcal A_h}{2}\cos\left(\frac{\omega_X}{2}t+\alpha\right)\right.\nonumber\\
&\hspace{6ex}+\left.\left(\frac{\omega_X^2\mathcal A_h}{4}-m_\Phi^2\mathcal A_h-\frac{3\lambda\mathcal A_h^3}{16}\right)\cos\left(\frac{\omega_X}{2}t-\alpha\right)\right]\nonumber\\
&\hspace{3ex}-\left[\frac{\Lambda_h^2\mathcal A_h}{2}\cos\left(\frac{3\omega_X}{2}t-\alpha\right)+\frac{\lambda\mathcal A_h^3}{16}\cos\left(\frac{3\omega_X}{2}t-3\alpha\right)\right]\nonumber\\
 &\hspace{3ex}-\left[m_\Phi^2+\frac{3\lambda\mathcal A_h^2}{8}\right]\chi_0\nonumber\\
 &\hspace{3ex}-\left[\Lambda_h^2\cos\omega_Xt+\frac{3\lambda\mathcal A_h^2}{8}\cos(\omega_Xt-2\alpha)\right]\chi_0\nonumber\\
 &\hspace{3ex}+\frac{\Lambda_h^2\mathcal A_h}{2f}\left[\sin\left(\frac{\omega_X}{2}t+\alpha\right)+\sin\left(\frac{3\omega_X}{2}t-\alpha\right)\right]\varphi_0\ .\label{eq_eom_chi}
\end{align}
Here, we changed the overall phase so that
$\cos[(M^2-m_\Phi^2)/(\epsilon f)]=1$.  We have ignored the other
non-linear terms and the terms with $\epsilon$ unless multiplied by
$M^2$.

In the following, we search for a special solution that is expressed as
\begin{align}
 \chi_0(t)&=\sum_{p=1}^\infty c_p^\chi\cos\left[\left(p+\frac{1}{2}\right)\omega_Xt+\alpha_p\right]\ ,\\
 \varphi_0(t)&=\sum_{p=1}^\infty c_p^\varphi\cos\left(p\omega_Xt+\beta_p\right)\ ,
\end{align}
where $c_p^\chi,c_p^\varphi,\alpha_p$ and $\beta_p$ are constants.
Notice that we can always choose $\mathcal A_h$ and $\alpha$ so that
$\chi_0$ does not contain the mode with $p=0$.

Integrating Eq.~\eqref{eq_eom_chi} over $t$ multiplying
$\sin(\omega_X t/2-\alpha)$ or $\cos(\omega_X t/2-\alpha)$, we have
\begin{align}
 \sin2\alpha&=-\frac{3H\omega_X}{\Lambda_h^2}-\frac{c^\chi_1}{\mathcal A_h}\left[\sin(\alpha+\alpha_1)+\frac{3\mathcal A_h^2}{8\Lambda_h^2}\sin(3\alpha+\alpha_1)\right]\nonumber\\
 &+\frac{c^\varphi_2}{2f}\cos(2\alpha+\beta_2)\ ,\\
 m_\Phi^2+\frac{3\lambda}{16}\mathcal A_h^2&=\frac{\omega_X^2}{4}-\frac{\Lambda_h^2}{2}\cos2\alpha\nonumber\\
&\hspace{3ex}-\frac{c^\chi_1}{\mathcal A_h}\left[\frac{\Lambda_h^2}{2}\cos(\alpha+\alpha_1)+\frac{3\lambda\mathcal A_h^2}{16}\cos(3\alpha+\alpha_1)\right]\nonumber\\
 &\hspace{3ex}-\frac{c_1^\varphi}{2f}\Lambda_h^2\sin\beta_1-\frac{c_2^\varphi}{2f}\frac{\Lambda_h^2}{2}\sin(2\alpha+\beta_2)\ .\label{eq_Ah}
\end{align}

Similarly, by integrating Eq.~\eqref{eq_eom_phi} over $t$, we have
\begin{align}
 \omega_X&=\frac{r\epsilon M^2}{3fH}+\frac{\Lambda_h^2\mathcal A_h^2}{24Hf^2}\sin2\alpha\nonumber\\
 &\hspace{3ex}+\frac{c^\varphi_1}{2f}\frac{2\Lambda_0^4+\Lambda_h^2\mathcal A_h^2}{12Hf^2}\cos\beta_1+\frac{c^\varphi_2}{2f}\frac{\Lambda_h^2\mathcal A_h^2}{24Hf^2}\cos(2\alpha+\beta_2)\nonumber\\
 &\hspace{3ex}-\frac{c^\chi_1}{\mathcal A_h}\frac{\Lambda_h^2\mathcal A_h^2}{12Hf^2}\sin(\alpha+\alpha_1)\ .\label{eq_omega_X}
\end{align}

Comparing the results with \eqref{eq_resoCondHiggs}, we
find\footnote{The other possibility corresponds to the other edge of the
resonance band.}  $\cos2\alpha<0$ and we obtain
\begin{equation}
 \alpha\simeq\pm\frac{\pi}{2}+\frac{3H\omega_X}{2\Lambda_h^2}\ ,
\end{equation}
assuming $c_1^\chi$ and $c_2^\varphi$ are small. Notice that we do not
have the edge solution when $3H\omega_X/\Lambda_h^2\gtrsim1$, where the
resonance of the Higgs field does not occur due to a large Hubble friction.

Repeating similar calculations multiplying sines and cosines with other
frequencies, we get
\begin{align}
 \alpha_1&\simeq\pm\frac{\pi}{2}-3H\left(\frac{\omega_X}{2\Lambda_h^2}\frac{3\lambda\mathcal A_h^2-8\Lambda_h^2}{\lambda\mathcal A_h^2-8\Lambda_h^2}+\frac{6\omega_X}{4m_\Phi^2-9\omega_X^2}\right)\ ,\\
 \beta_1&\simeq\pm\frac{\pi}{2}+\frac{3H}{16f^2\omega_X}(\mathcal A_h^2+16f^2)\ , \\
 \beta_2&\simeq\pm\frac{\pi}{2}-3H\left(\frac{\omega_X}{\Lambda_h^2}-\frac{1}{2\omega_X}\right)\ ,
\end{align}
and
\begin{align}
  c^\chi_1&\simeq-{\rm sign}(\alpha\alpha_1)\frac{8\Lambda_h^2-\lambda\mathcal A_h^2}{4(9\omega_X^2-4m_\Phi^2)}\mathcal A_h\left(1+\frac{18\lambda\mathcal A_h^2\omega_X^2(18\omega_X^2+3\lambda\mathcal A_h^2-8m_\Phi^2)}{\Lambda_h^2(8\Lambda_h^2-\lambda\mathcal A_h^2)^2(9\omega_X^2-4m_\Phi^2)}H^2\right)\ ,\\
 c^\varphi_1&\simeq {\rm sign}(\beta_1)\frac{2\Lambda_0^4+\Lambda_h^2\mathcal A_h^2}{4f^2\omega_X^2}f\left(1-\frac{9\mathcal A_h^2}{32f^2\Lambda_h^2}H^2\right)\ ,\\
 c^\varphi_2&\simeq- {\rm sign}(\beta_2)\frac{\Lambda_h^2\mathcal A_h^2}{32f^2\omega_X^2}f\left(1-\frac{9\mathcal A_h^2}{64f^2\Lambda_h^2}H^2\right)\ .
\end{align}
Here, we have ignored $\mathcal O(1/\omega_X^3),~\mathcal
O(H^2/\omega_X)$ and $\mathcal O(H^3)$ corrections. Plugging them into
Eqs.~\eqref{eq_Ah} and \eqref{eq_omega_X}, we obtain\footnote{The signs
of $\alpha_1,\beta_1$ and $\beta_2$ do not affect the result.}
Eqs.~\eqref{eq_approx_Ah} and \eqref{eq_approx_omega}.
\section{Another mechanism to stop the relaxion}\label{apx_the_other_way}
\begin{figure}[t]
 \begin{center}
  \begin{minipage}{0.48\linewidth}
   \includegraphics[width=\linewidth]{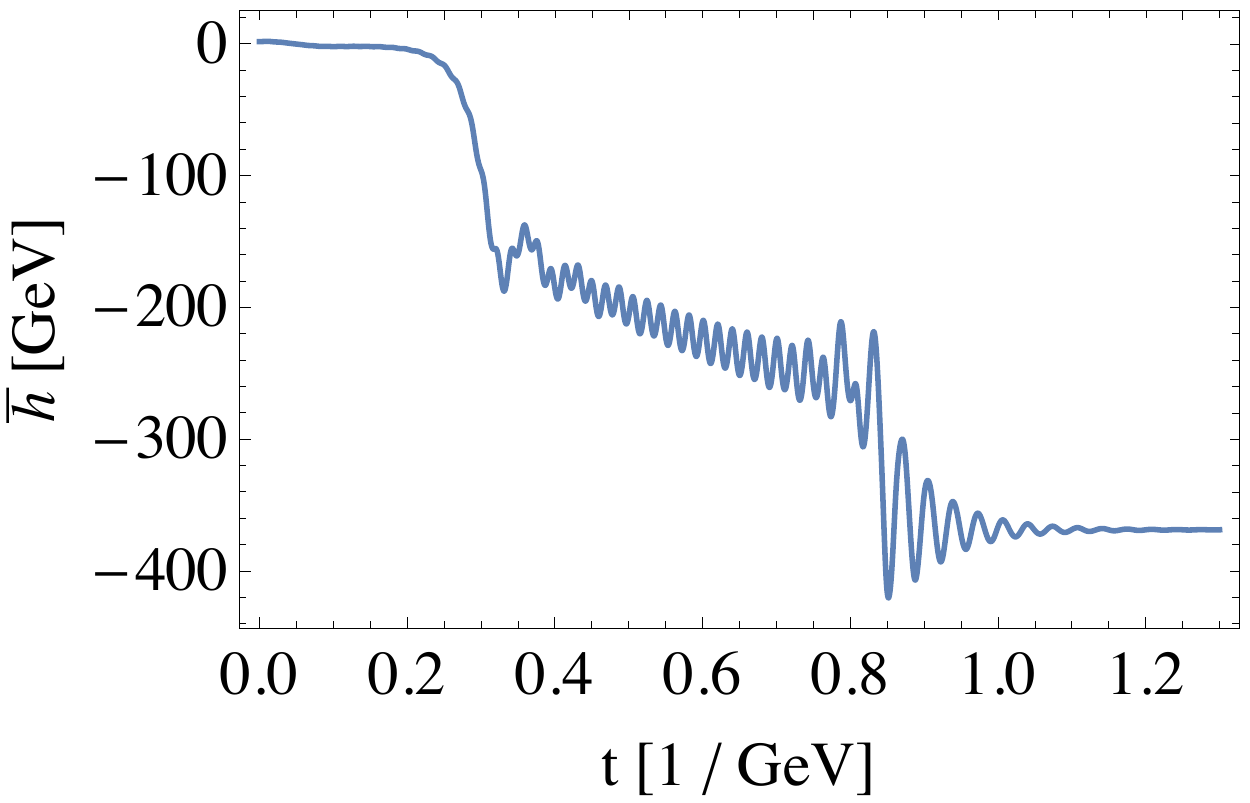}
  \end{minipage}
  \begin{minipage}{0.48\linewidth}
   \includegraphics[width=\linewidth]{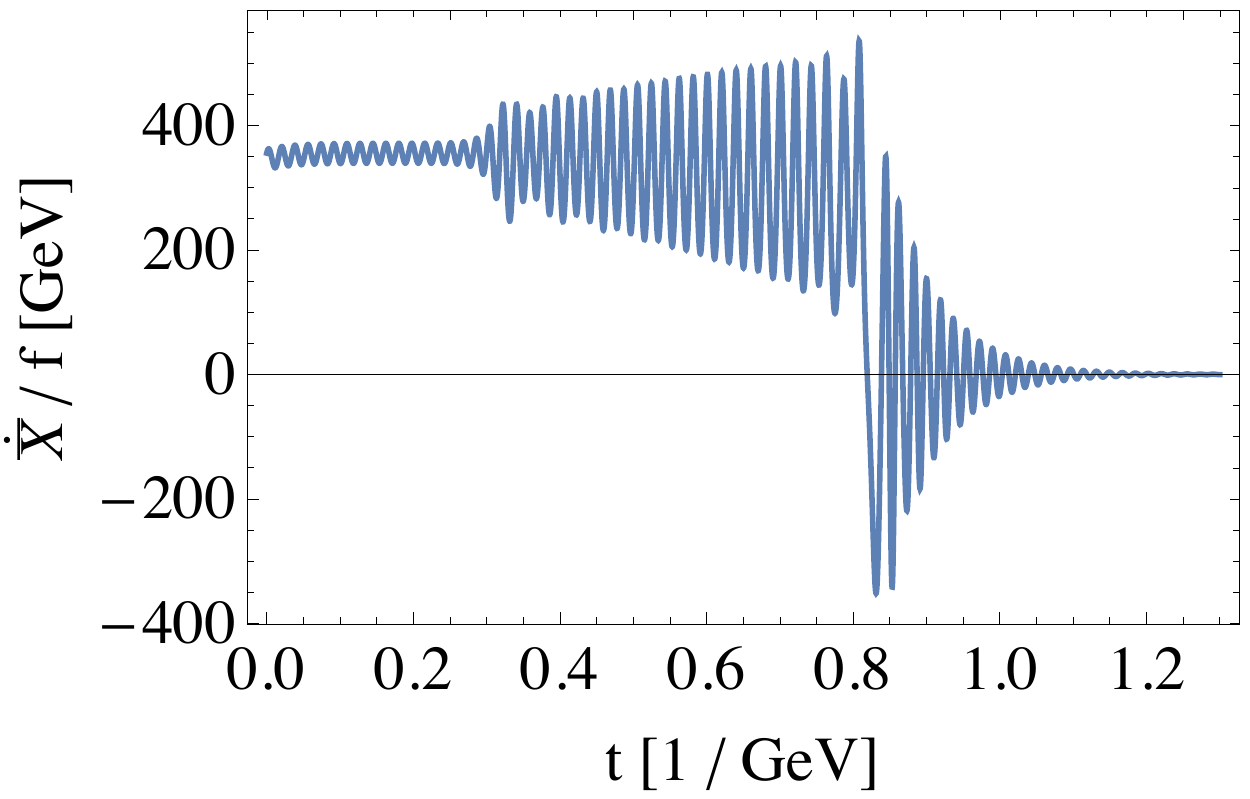}
  \end{minipage}
  \caption{Another stopping mechanism. The left and the right panels
   show the evolution of $\bar h$ and $\dot{\bar X}/f$, respectively.}
   \label{fig_ex_2}
 \end{center}
\end{figure}

Here, we present another mechanism to stop the relaxion without
slow-roll. It is summarized below.
\begin{enumerate}
 \item The initial Higgs mass is assumed to be positive and the
       relaxion rolls down the potential with its terminal velocity.
 \item[2'] The relaxion go through the critical point without suffering
       from any resonance and the Higgs field develops a VEV.
 \item[3'] The height of the bumps increases as the Higgs VEV
       increases. It enhances the difference between the maximum and the
       minimum velocities of the relaxion.
 \item[4'] The minimum velocity of the relaxion reaches zero and the
       relaxion gets trapped between bumps.
 \item[5'] The kinetic energy of the Higgs field and the relaxion dumps
       quickly by the Hubble friction.
\end{enumerate}
We show an example in Fig.~\ref{fig_ex_2}. The left and the right panels
show the evolution of $\bar h$ and $\dot{\bar X}/f$, respectively. We take
\begin{align}
 &\lambda=0.52,~H=10~{\rm GeV},~\Lambda_0=\Lambda_h=80~{\rm GeV},~\epsilon=0.8~{\rm GeV},~r=0.002\ ,\nonumber\\
 &f=60~{\rm GeV},~M=20~{\rm TeV}\ .
\end{align}
The Higgs mass becomes negative around $t\simeq0.1$. After $t\simeq0.2$,
the Higgs field develops a VEV and the range of the relaxion velocity
spreads out. The minimum of the velocity reaches zero at $t\sim0.8$ and
the relaxion gets trapped.

With this example point, one can show that the resonant particle
production is very efficient after the Higgs field obtains a VEV. Thus, it
is preferable to stop the relaxion just after the Higgs field obtains a
VEV. Such an example point is Point 3 of Fig. \ref{fig_sample}.

\newpage
\section{More on the Relaxion Potential}
\label{sec:clockwork2}

As we have mentioned in Subsection~\ref{sec:clockwork}, a naive
clockwork mechanism is not suitable for our relaxion potential.  In this
appendix, we explain the difficulties in model building and discuss how
to overcome them by using clockwork with progressively increasing VEVs.

\subsection{The clockwork potential}
To demonstrate the clockwork mechanism,
we consider a model with $N$ complex scalars, 
$\phi_k$ $(k = 0,\cdots N-1)$~\cite{Giudice:2016yja,Kaplan:2015fuy,Choi:2015fiu} 
(see also~\cite{Harigaya:2014eta,*Harigaya:2014rga}).
We assume a global $U(1)$ symmetry where the charge of 
$\phi_k$ is $3^{-k}$.
Then, the complex scalars are connected via
\begin{align}
\label{eq:connection}
    V_{\rm conn} = -\sum_{k=0}^{N-2} \kappa_k \phi_k^* \phi_{k+1}^3 +h.c.\ ,
\end{align}
where $\kappa_k$'s denote small coupling constants.  We also assume that each complex scalar
obtains a VEV via a scalar potential,
\begin{align}
    \label{eq:clock}
    V_{\rm self}=\sum_{k =0}^{N-1} \left(|\phi_k|^2 - \frac{f_k^2}{2}\right)^2\ ,
\end{align}
where we omit the coefficients for brevity.  Notice that we dropped
$|\phi_k|^2|\phi_l|^2$ with $k\neq l$ by hand and ignored the other
Planck-suppressed terms allowed by the global $U(1)$ symmetry.  For a
while, we assume that $f_k=\order{f}=\order{10^2}$\,GeV for all $k$.
Then, if $\kappa_k \ll 1$, the VEVs are given by $\langle \phi_k\rangle
\simeq f_k/\sqrt{2}$.

After all the $\phi_k$'s obtain VEVs, the connecting terms in
Eq.\,\eqref{eq:connection} lead to the potential of the phase
components of $\phi_k$ as
\begin{align}
    \label{eq:phases}
    V(\{\pi_k\})= -\frac{1}{2}\sum_{k=0}^{N-2} \kappa_kf_kf_{k+1}^3 \cos
    \left(
    \frac{\pi_k}{f_k}-3\frac{\pi_{k+1}}{f_{k+1}}
    \right)\ ,
\end{align}
where $\phi_k = f_k e^{i\pi_k/f_k}/\sqrt{2}$.  Due to this potential,
all of the phase components obtain masses of $\order{\kappa^{1/2}f}$
except for the pseudo Nambu-Goldstone boson (NGB), which we denote as
$X$.  Here, the pseudo NGB corresponds to the mode that keeps the
following relation;
\begin{align}
\label{eq:relaxionDIR}
    \frac{X}{F} \equiv \frac{\pi_0}{f_0}
    = 3 \frac{\pi_1}{f_1} = \cdots = 3^{N-1} \frac{\pi_{N-1}}{f_{N-1}}\ ,
\end{align}
where $F$ is the effective decay constant.
By substituting $\phi_k = e^{i X/(3^{k}F)}$ into 
the kinetic terms, we find that $X$ is canonically
normalized if 
\begin{align}
\label{eq:relaxionNorm}
    F^2 =  \sum_{k=0}^{N-1} \frac{1}{3^{2k}}f_k^2  = \order{f^2}\ .
\end{align}
Notice that Eq.\,\eqref{eq:relaxionDIR} shows that the relative
contribution of each phase component $\pi_k$ to the pseudo NGB is
suppressed by $3^{-(k-1)}f_{k}/F$.  This property is valid even for
$\kappa_i = \order{1}$.

An intriguing feature of the clockwork mechanism is the response to the
explicit breaking of the $U(1)$ symmetry.  For example, let us consider
a situation where the $U(1)$ symmetry is broken by two tiny parameters
with mass dimension $3$, ${\cal E}_0$ and ${\cal E}_{N-1}$, which have
the $U(1)$ charges $1$ and $3^{-(N-1)}$, respectively.  In this case,
$\phi_0$ and $\phi_{N-1}$ couple to $\cal E$'s via
\begin{align}
\label{eq:breaking}
    V_{\rm br} = {\cal E}_0^* \phi_0 + {\cal E}_{N-1}^*\phi_{N-1} 
    +h.c.\ ,
\end{align}
which leads to a non-trivial potential of $X$ as
\begin{align}
    V(X) = \sqrt{2} {\cal E}_0f_0\cos\left(\frac{X}{F}\right)
    +\sqrt{2} {\cal E}_{N-1} f_{N-1} \cos\left(\frac{X}{3^{N-1}F}\right)\ .
\end{align}
Here, we ignored the phases of ${\cal E}_0$ and ${\cal E}_{N-1}$ since
they are irrelevant for the relaxion mechanism.  For $N\gg 1$, the
periodicity of the second term, $\Delta X=3^{N-1}\times 2\pi F$, is much
longer than that of the first term, $\Delta X =2\pi F$.  In this way,
the clockwork mechanism accommodates cosine potentials of the pseudo NGB
with hierarchically different periodicity.

A caveat here is that the mass scales of the explicit breaking should be
much smaller than $f_k$'s, {\it i.e.} ${\cal E}_{0},{\cal E}_{N-1}\ll
f^3$.  Otherwise, the explicit breaking terms in
Eq.\,\eqref{eq:breaking} disturb the $U(1)$ symmetric potential of
Eq.\,\eqref{eq:clock}, which invalidates the existence of the pseudo
NGB.  In subsection~\ref{sec:clockwork}, however, we utilized the
explicit breaking term much larger than $\order{f}$.  To allow such
large explicit breaking, we need to extend the model and will discuss it
in the next subsection.

\subsection{Clockwork with progressively increasing VEVs}
To allow explicit breaking with a mass scale larger than $\order{f}$, we
consider $f_k$'s that progressively increase by a factor of $a$ $(3\ge
a>1)$, {\it i.e.} $f_k=a^{k}f_0$.\footnote{The mass scales $f_k$ do not
need to be an exact geometric series. The upper bound on $a$ comes from
the requirement that Eq.~\eqref{eq:relaxionNorm} remains of ${\cal
O}(f^2)$.}  In order to protect the mass scale of the $k$-th site
against that of the $(i+1)$-th site, we assume that the connecting terms
of Eq.~\eqref{eq:connection} are slightly suppressed as
$\kappa_k=\mathcal O(a^{-3})$. As in the previous subsection, there
appears a pseudo NGB mode, which satisfies Eqs.~\eqref{eq:relaxionDIR}
and \eqref{eq:relaxionNorm}.  The other linear combinations of the phase
components obtain masses of ${\cal O}(f_k^2)$ $(k = 0, \cdots N-2)$.

A crucial difference from the model in the previous section is that the
VEV of $\phi_{N-1}$ is not of $\order{F}$ but of ${\cal O}(a^{N-1}F)$.
Accordingly, the upper limit of the explicit breaking parameters are
much weaker;
\begin{align}
    {\cal E}_0 \ll f^3\ , \quad {\cal E}_{N-1} \ll (a^{N-1}f)^3\ .
\end{align}
Furthermore, by requiring that the explicit breaking
satisfies
\begin{align}
    {\cal E}_{N-1}/(3^{N-1}F) \ll f^2\ ,
\end{align}
we find that the pseudo NGB mode is lighter than all the other modes.
With a large enough $N$, the desired scalar potential of
Eq.\,(\ref{eq_potential2}) can be realized.

Before closing this subsection, let us comment on the relation between
the relaxion potential discussed above and the that given in
subsection~\ref{sec:clockwork}.  For that purpose, we define $\pi_L$ and
$\pi_H$ by
\begin{align}
\label{eq:piL}
    \frac{\pi_L}{F_L} \equiv \frac{\pi_0}{f_0}
    = 3 \frac{\pi_1}{f_1} = \cdots = 3^{N-2} \frac{\pi_{N-2}}{f_{N-2}}\ , \quad \quad
    \frac{\pi_H}{F_H} \equiv \frac{\pi_{N-1}}{f_{N-1}} \ ,
\end{align}
with 
\begin{align}
    F_L =  \sum_{k=0}^{N-2} \frac{1}{3^{2k}}f_k^2\ ,\quad 
    F_H =  f_{N-1}\ ,
\end{align}
respectively.
Then, the potential of $\pi_L$ and $\pi_H$ is given
\begin{align}
\label{eq:relaxionPOT2}
    V = \sqrt{2} {\cal E}_0f_0\cos\left(\frac{\pi_L}{F_L}\right)
    +\sqrt{2} {\cal E}_{H} F_{H} \cos\left(\frac{\pi_H}{F_H}\right)
    -\frac{1}{2}\kappa_{N-2}f_{N-2}f_{N-1}^3
    \cos\left(
    \frac{\pi_L}{3^{(N-2)}F_{L}}
    -3\frac{\pi_{H}}{F_{H}}
    \right)
    \ ,
\end{align}
where the third term comes from the connecting terms in Eq.\,\eqref{eq:connection}.
This potential reproduces the one in Eq.\,\eqref{eq_potential4}.

\subsection{Supersymmetric clockwork}
\label{sec:susy}

In the previous subsections, we introduced new scalars with hierarchical
VEVs, some of which are around the weak scale. They could cause the
hierarchy problem if the quadratic divergences of their mass terms are
cutoff at a high energy scale. In this subsection, we show an example of
a supersymmetric (SUSY) model to protect their VEVs against quantum
corrections. Even with SUSY, the hierarchical structure could be
destroyed by the terms such as $|\phi_{N-1}|^2 |\phi_0|^2$. For this
problem, we consider the technical naturalness of their coupling
constants.  To make our argument simple, we assume the gravity mediated
SUSY breaking in the clockwork sector where the gravitino mass,
$m_{3/2}$, is of $\order{10^2\mbox{--}10^3}$\,GeV.  We also assume that
the cutoff of the Higgs sector around $M = \order{10}$\,TeV stems from
the gauge mediated SUSY breaking to the SUSY SM (SSM) sector.

As in the previous section, let us consider a model with $N$ pairs of
chiral superfields, $(\Phi_k, \bar{\Phi}_k)$ $(k=0,\cdots, N-1)$.  The
global charges of $U(1)$ of $\Phi_k$ and $\bar{\Phi}_k$ are $2^{-k}$
and $-2^{-k}$, respectively.  Then, the connecting superpotential is
given by
\begin{align}
    \label{eq:connectionSP}
    W = \sum_{k=0}^{N-2}\kappa_k \Phi_k \bar{\Phi}_{k+1}^2\ + \bar{\kappa}_k \bar{\Phi}_k
    \Phi_{k+1}^2\ ,
\end{align}
where $\kappa_k$ and $\bar{\kappa}_k$ are coefficients of $\order{a^{-2}}$.
The VEV of each sector is obtained from the superpotential of
\begin{align}
    W = \sum_{k=0}^{N-1} Y_k \left(\Phi_k \bar{\Phi}_{k} - \frac{f_k^2}{2}  
\right)\ ,
\end{align}
where $Y_k$'s are chiral superfields and $f_k$'s are progressively
increasing constants; $f_{k+1}=af_k$ $(2\ge a > 1)$.  Due to SUSY,
$f_k$'s are stable against quantum corrections. Here, the absence of the
terms such as $Y_i f_j^2$ and $Y_i \Phi_j\bar{\Phi}_j$ ($i\neq j$) is
technically natural.\footnote{One may also assume a discrete symmetry in
each site under which the combination of $X_k$ and $\bar{\Phi}_k$ are
neutral.  Then, by assuming that each discrete symmetry is broken by
$f_k^2$, we can forbid $Y_i f_j^2$ and $Y_i \Phi_j\bar{\Phi}_j$ ($i\neq
j$).  In this case, $\bar{\kappa}_k$ is highly suppressed, though it
does not affect the clockwork mechanism.}

In this model, the NG mode corresponds to
the massless flat direction of the superpotential,
\begin{align}
        \frac{\tilde X}{F} \equiv \frac{\Pi_0}{f_0}
    = 2 \frac{\Pi_1}{f_1} = \cdots = 2^{N-1} \frac{\Pi_{N-1}}{f_{N-1}}\ ,
\end{align}
where $\Phi$'s depend on $\Pi$'s via
\begin{align}
    \Phi_k =\frac{1}{\sqrt 2}f_k e^{\Pi_k/f_k}\ ,
    \quad
    \bar{\Phi}_k = \frac{1}{\sqrt 2}\bar{f}_k e^{-\Pi_k/f_k}\ .
\end{align}
Here, the imaginary parts of the scalar components of $\Pi_k$ and
$\tilde X$ correspond to $\pi_k$ and $X$ in the previous section,
respectively.  The decay constant for the canonically normalized $X$ is
given by
\begin{align}
    F^2 =  \sum_{k=0}^{N-1} \frac{1}{2^{2k}}f_k^2  = \order{f^2}\ .
\end{align}
Similarly as in the previous section, other supermultiplets than the
relaxion direction become heavy.  Furthermore, due to the gravity
mediated SUSY breaking, the fermionic and scalar partners of the
relaxion eventually obtain the masses of $\order{m_{3/2}}$.  Thus, the
model is reduced to the clockwork mechanism in the non-supersymmetric
model below the scale of $m_{3/2}$.

\subsection{Explicit breaking in supersymmetric model}
The explicit breaking of the $U(1)$ symmetry can be implemented in the
following way.  First, let us consider the explicit breaking at the
$N$-th site by a breaking parameter ${\cal E}_{H}$, which has the $U(1)$
charge of $-2^{-(N-1)}$ and mass dimension $2$.
Then, the explicit breaking term is given by
\begin{align}
\label{eq:breakingW}
    W ={\cal E}_{H}  \Phi_{N-1} \ .
\end{align}
This superpotential leads to a runaway potential of the real part of the
scalar component of $\tilde X$, which we denote as $X_s$, as
\begin{align}
    V = \frac{1}{2}{\cal E}_{H}^2 e^{2X_s/F_H} + \frac{1}{2}m_{3/2}^2X_s^2\ . 
\end{align}
Here, the first term comes from the $F$-term potential of $\tilde X$,
while the second term comes from the gravity mediated SUSY breaking mass
squared. We take ${\cal E}_{N-1}$ to be real for simplicity and $F_H$
denotes $ 2^{N-1} F$.  In order to keep the VEVs of $\Phi_0$ and
$\bar{\Phi}_0$ around $\order{F}$, we require
\begin{align}
\label{eq:limitSUSY}
    \langle{X_s}\rangle = \frac{{\cal E}_{H}^2}{m_{3/2}^2F_H} \ll F\ .
\end{align}
The cosine potential of the relaxion is given by the SUSY breaking
$A$-term contribution as
\begin{align}
\label{eq:cosAN}
    V \simeq 4 m_{3/2} {\cal E}_{H} F_H \cos\left(\frac{X}{F_H}\right)\ .
\end{align}
The mass scale of this coefficient is $\order{10}$\,TeV for ${\cal
E}_{N-1}^{1/2}=\order{10}$\,GeV, $m_{3/2}=\order{1}$\,TeV and $F_H =
\order{10^9}$\,GeV, which satisfy the condition of
Eq.\,\eqref{eq:limitSUSY}.

Similarly, the explicit breaking at the first site is implemented by
\begin{align}
\label{eq:breakingW0}
    W ={\cal E}_{0}  \Phi_{0} \ ,
\end{align}
where the $U(1)$ charge of ${\cal E}_{0}$ is $1$ and 
its mass dimension is $2$.
To validate the notion of spontaneous $U(1)$ breaking,
${\cal E}_{0}$ should satisfy
\begin{align}
    {\cal E}_{0} \ll F^2\ .
\end{align}
Under this condition, the scalar potential of $X_s$ is dominated by the
SUSY breaking effects.

The SUSY breaking $A$-term contributions leads to the cosine potential
as
\begin{align}
\label{eq:cosA0}
    V = 4 m_{3/2} {\cal E}_{0} F \cos\left(\frac{X}{F}\right)\ .
\end{align}
The relaxion also obtains a potential from the $F$-term contributions,
\begin{align}
\label{eq:cosF0}
    V = \frac{1}{2}\left| {\cal E}_0 e^{iX/F} + {\cal E}_{N-1} e^{ iX/(2^{N-1}F)}\right|^2 \simeq {\cal E}_0 {\cal E}_{N-1}\cos\left(\frac{X}{F}\right) + {\cal E}_0^2 + {\cal E}_{N-1}^2\ .
\end{align}
They are, however, subdominant for
\begin{align}
\label{eq:nonF0}
    {\cal E}_{N-1} \ll m_{3/2} F\ ,
\end{align}
which we assume in our analysis.

Finally, let us implement the explicit breaking that depends on the
Higgs field as in Eq.\,\eqref{eq:explicitH}.  The simplest way is to
couple $\Phi_{N-1}$ and the two Higgs doublets $H$ and $\bar{H}$ in the
SSM to an $SU(N_c)$ gauge theory. We consider four flavors of the chiral
multiplets in the fundamental and anti-fundamental representation, which
we call $(Q_H,\bar{Q}_H)$, $(L_H,\bar{L}_H)$, and
$(N_H,\bar{N}_H)$. Here, $L_H$ and $(\bar{L}_H)^*$ have the same SM
gauge charges of the lepton doublets, and the others are SM gauge
singlets.  The $U(1)$ charge of the combination, $Q_H\bar{Q}_H$, is
${2^{-(N-1)}}$, while other flavors are neutral under the $U(1)$
symmetry.  Then, they couple to $\Phi_{N-1}$, $H$ and $\bar{H}$ via
\begin{align}
    W = \Phi_{N-1} Q_H\bar{Q}_H + (\bar{H}L_H\bar{N}_H)
    +  (H\bar{L}_H{N}_H) + m_L L_H\bar{L}_H + m_N N_H \bar{N}_H\ .
\end{align}
Below the scale of $\order{F_H}$, we can integrate out $Q$'s, which 
leads to
\begin{align}
    W = \frac{1}{32\pi^2}\ln\frac{\sqrt{2}\Phi_{N-1}}{F_H} W_\alpha W^\alpha + (\bar{H}L_H\bar{N}_H)
    +  (H\bar{L}_H{N}_H) + m_L L_H\bar{L}_H + m_N N_H \bar{N}_H\ ,
\end{align}
where $W_\alpha$ denotes the chiral gauge field strength superfield.

The dynamical scale of the $SU(N_c)$ gauge theory, $\Lambda_H$,
depends on $\Phi_{N-1}$ via
\begin{align}
    \tilde\Lambda_H = \Lambda_H \left(\frac{\sqrt{2}\Phi_{N-1}}{F_H}\right)^{\frac{1}{3N_c-3}}\ ,
\end{align}
where $\Lambda_H$ is the dynamical scale 
for $m_{L} = m_N= 0$ at $H=\bar{H}= 0$ and $\Phi_{N-1}=f_{N-1}/\sqrt{2}$.
For $m_{L},m_N\ll \tilde{\Lambda}_H$, the non-perturbative
superpotential~\cite{Davis:1983mz,Affleck:1983mk,Seiberg:1994bz} and the mass terms of $L_H\bar{L}_H$ and $N_H\bar{N}_H$ lead to
\begin{align}
    W \simeq \Lambda_{\rm eff}^3 \simeq \Lambda_H^3
    \left(
    \frac{m_N m_L^2 - m_L H\bar{H}}{\Lambda_H^3}
    \right)^{\frac{1}{N_c}}
    e^{\frac{X}{N_c F_H}}\ .
\end{align}
This term corresponds to the explicit breaking of the $N$-th site,
\begin{align}
    {\cal E}_{N-1} \sim  \frac{\Lambda_H^3}{F_H}
    \left(
    \frac{m_N m_L^2 - m_L H\bar{H}}{\Lambda_H^3}
    \right)^{\frac{1}{N_c}} \ .
\end{align}
By plugging into Eq.\,\eqref{eq:cosAN}, 
we obtain
\begin{align}
    V \simeq 4m_{3/2} \Lambda_H^3
    \left(
    \frac{m_N m_L^2 - m_L H\bar{H}}{\Lambda_H^3}
    \right)^{\frac{1}{N_c}}\cos\left(\frac{X}{N_c F_H}\right)\ ,
\end{align}
which leads to
\begin{align}
    \epsilon &\sim 4m_{3/2} \Lambda_H^3 \left(
    \frac{m_N m_L^2}{\Lambda_H^3}
    \right)^{\frac{1}{N_c}}
    \frac{1}{m_Nm_L}\frac{1}{N_c F_N}\ ,\\
r  &\sim \frac{m_N m_L}{M^2} \ .
\end{align}
For example, $\epsilon = 0.1$\,GeV and $r = 10^{-2}$ can be achieved for $N_c = 3$, $\Lambda_H= \order{10}$\,TeV, $m_N=m_L=\order{1}$\,GeV, 
$m_{3/2} = \order{1}$\,TeV and $F=\order{10^9}$\,GeV,
which also satisfy the conditions of Eqs.\,\eqref{eq:limitSUSY}
and \eqref{eq:nonF0}.%
\footnote{More precisely, we need to consider one of the linear combinations of the two Higgs doublets, though it does not alter 
the correspondence of the parameters significantly.}
Thus, we successfully obtain the relaxion potential based on
the clockwork mechanism with progressively increasing VEVs.

\bibliography{relaxion}

\begin{thebibliography}{74}%
\makeatletter
\providecommand \@ifxundefined [1]{%
 \@ifx{#1\undefined}
}%
\providecommand \@ifnum [1]{%
 \ifnum #1\expandafter \@firstoftwo
 \else \expandafter \@secondoftwo
 \fi
}%
\providecommand \@ifx [1]{%
 \ifx #1\expandafter \@firstoftwo
 \else \expandafter \@secondoftwo
 \fi
}%
\providecommand \natexlab [1]{#1}%
\providecommand \enquote  [1]{``#1''}%
\providecommand \bibnamefont  [1]{#1}%
\providecommand \bibfnamefont [1]{#1}%
\providecommand \citenamefont [1]{#1}%
\providecommand \href@noop [0]{\@secondoftwo}%
\providecommand \href [0]{\begingroup \@sanitize@url \@href}%
\providecommand \@href[1]{\@@startlink{#1}\@@href}%
\providecommand \@@href[1]{\endgroup#1\@@endlink}%
\providecommand \@sanitize@url [0]{\catcode `\\12\catcode `\$12\catcode
  `\&12\catcode `\#12\catcode `\^12\catcode `\_12\catcode `\%12\relax}%
\providecommand \@@startlink[1]{}%
\providecommand \@@endlink[0]{}%
\providecommand \url  [0]{\begingroup\@sanitize@url \@url }%
\providecommand \@url [1]{\endgroup\@href {#1}{\urlprefix }}%
\providecommand \urlprefix  [0]{URL }%
\providecommand \Eprint [0]{\href }%
\providecommand \doibase [0]{http://dx.doi.org/}%
\providecommand \selectlanguage [0]{\@gobble}%
\providecommand \bibinfo  [0]{\@secondoftwo}%
\providecommand \bibfield  [0]{\@secondoftwo}%
\providecommand \translation [1]{[#1]}%
\providecommand \BibitemOpen [0]{}%
\providecommand \bibitemStop [0]{}%
\providecommand \bibitemNoStop [0]{.\EOS\space}%
\providecommand \EOS [0]{\spacefactor3000\relax}%
\providecommand \BibitemShut  [1]{\csname bibitem#1\endcsname}%
\let\auto@bib@innerbib\@empty
\bibitem [{\citenamefont {Graham}\ \emph {et~al.}(2015)\citenamefont {Graham},
  \citenamefont {Kaplan},\ and\ \citenamefont {Rajendran}}]{Graham:2015cka}%
  \BibitemOpen
  \bibfield  {author} {\bibinfo {author} {\bibfnamefont {P.~W.}\ \bibnamefont
  {Graham}}, \bibinfo {author} {\bibfnamefont {D.~E.}\ \bibnamefont {Kaplan}},
  \ and\ \bibinfo {author} {\bibfnamefont {S.}~\bibnamefont {Rajendran}},\
  }\href {\doibase 10.1103/PhysRevLett.115.221801} {\bibfield  {journal}
  {\bibinfo  {journal} {Phys. Rev. Lett.}\ }\textbf {\bibinfo {volume} {115}},\
  \bibinfo {pages} {221801} (\bibinfo {year} {2015})},\ \Eprint
  {http://arxiv.org/abs/1504.07551} {arXiv:1504.07551 [hep-ph]} \BibitemShut
  {NoStop}%
\bibitem [{\citenamefont {Gupta}\ \emph {et~al.}(2019)\citenamefont {Gupta},
  \citenamefont {Reiness},\ and\ \citenamefont {Spannowsky}}]{Gupta:2019ueh}%
  \BibitemOpen
  \bibfield  {author} {\bibinfo {author} {\bibfnamefont {R.~S.}\ \bibnamefont
  {Gupta}}, \bibinfo {author} {\bibfnamefont {J.~Y.}\ \bibnamefont {Reiness}},
  \ and\ \bibinfo {author} {\bibfnamefont {M.}~\bibnamefont {Spannowsky}},\
  }\href@noop {} {\  (\bibinfo {year} {2019})},\ \Eprint
  {http://arxiv.org/abs/1902.08633} {arXiv:1902.08633 [hep-ph]} \BibitemShut
  {NoStop}%
\bibitem [{\citenamefont {Banerjee}\ \emph {et~al.}(2019)\citenamefont
  {Banerjee}, \citenamefont {Budker}, \citenamefont {Eby}, \citenamefont
  {Kim},\ and\ \citenamefont {Perez}}]{Banerjee:2019epw}%
  \BibitemOpen
  \bibfield  {author} {\bibinfo {author} {\bibfnamefont {A.}~\bibnamefont
  {Banerjee}}, \bibinfo {author} {\bibfnamefont {D.}~\bibnamefont {Budker}},
  \bibinfo {author} {\bibfnamefont {J.}~\bibnamefont {Eby}}, \bibinfo {author}
  {\bibfnamefont {H.}~\bibnamefont {Kim}}, \ and\ \bibinfo {author}
  {\bibfnamefont {G.}~\bibnamefont {Perez}},\ }\href@noop {} {\  (\bibinfo
  {year} {2019})},\ \Eprint {http://arxiv.org/abs/1902.08212} {arXiv:1902.08212
  [hep-ph]} \BibitemShut {NoStop}%
\bibitem [{\citenamefont {Wang}(2018)}]{Wang:2018ddr}%
  \BibitemOpen
  \bibfield  {author} {\bibinfo {author} {\bibfnamefont {S.-J.}\ \bibnamefont
  {Wang}},\ }\href@noop {} {\  (\bibinfo {year} {2018})},\ \Eprint
  {http://arxiv.org/abs/1811.06520} {arXiv:1811.06520 [hep-ph]} \BibitemShut
  {NoStop}%
\bibitem [{\citenamefont {Hook}\ and\ \citenamefont
  {Marques-Tavares}(2016)}]{Hook:2016mqo}%
  \BibitemOpen
  \bibfield  {author} {\bibinfo {author} {\bibfnamefont {A.}~\bibnamefont
  {Hook}}\ and\ \bibinfo {author} {\bibfnamefont {G.}~\bibnamefont
  {Marques-Tavares}},\ }\href {\doibase 10.1007/JHEP12(2016)101} {\bibfield
  {journal} {\bibinfo  {journal} {JHEP}\ }\textbf {\bibinfo {volume} {12}},\
  \bibinfo {pages} {101} (\bibinfo {year} {2016})},\ \Eprint
  {http://arxiv.org/abs/1607.01786} {arXiv:1607.01786 [hep-ph]} \BibitemShut
  {NoStop}%
\bibitem [{\citenamefont {Choi}\ and\ \citenamefont
  {Im}(2016{\natexlab{a}})}]{Choi:2016luu}%
  \BibitemOpen
  \bibfield  {author} {\bibinfo {author} {\bibfnamefont {K.}~\bibnamefont
  {Choi}}\ and\ \bibinfo {author} {\bibfnamefont {S.~H.}\ \bibnamefont {Im}},\
  }\href {\doibase 10.1007/JHEP12(2016)093} {\bibfield  {journal} {\bibinfo
  {journal} {JHEP}\ }\textbf {\bibinfo {volume} {12}},\ \bibinfo {pages} {093}
  (\bibinfo {year} {2016}{\natexlab{a}})},\ \Eprint
  {http://arxiv.org/abs/1610.00680} {arXiv:1610.00680 [hep-ph]} \BibitemShut
  {NoStop}%
\bibitem [{\citenamefont {Fonseca}\ \emph
  {et~al.}(2018{\natexlab{a}})\citenamefont {Fonseca}, \citenamefont
  {Morgante},\ and\ \citenamefont {Servant}}]{Fonseca:2018xzp}%
  \BibitemOpen
  \bibfield  {author} {\bibinfo {author} {\bibfnamefont {N.}~\bibnamefont
  {Fonseca}}, \bibinfo {author} {\bibfnamefont {E.}~\bibnamefont {Morgante}}, \
  and\ \bibinfo {author} {\bibfnamefont {G.}~\bibnamefont {Servant}},\ }\href
  {\doibase 10.1007/JHEP10(2018)020} {\bibfield  {journal} {\bibinfo  {journal}
  {JHEP}\ }\textbf {\bibinfo {volume} {10}},\ \bibinfo {pages} {020} (\bibinfo
  {year} {2018}{\natexlab{a}})},\ \Eprint {http://arxiv.org/abs/1805.04543}
  {arXiv:1805.04543 [hep-ph]} \BibitemShut {NoStop}%
\bibitem [{\citenamefont {Tangarife}\ \emph {et~al.}(2017)\citenamefont
  {Tangarife}, \citenamefont {Tobioka}, \citenamefont {Ubaldi},\ and\
  \citenamefont {Volansky}}]{Tangarife:2017vnd}%
  \BibitemOpen
  \bibfield  {author} {\bibinfo {author} {\bibfnamefont {W.}~\bibnamefont
  {Tangarife}}, \bibinfo {author} {\bibfnamefont {K.}~\bibnamefont {Tobioka}},
  \bibinfo {author} {\bibfnamefont {L.}~\bibnamefont {Ubaldi}}, \ and\ \bibinfo
  {author} {\bibfnamefont {T.}~\bibnamefont {Volansky}},\ }\href@noop {} {\
  (\bibinfo {year} {2017})},\ \Eprint {http://arxiv.org/abs/1706.00438}
  {arXiv:1706.00438 [hep-ph]} \BibitemShut {NoStop}%
\bibitem [{\citenamefont {Choi}\ \emph {et~al.}(2017)\citenamefont {Choi},
  \citenamefont {Kim},\ and\ \citenamefont {Sekiguchi}}]{Choi:2016kke}%
  \BibitemOpen
  \bibfield  {author} {\bibinfo {author} {\bibfnamefont {K.}~\bibnamefont
  {Choi}}, \bibinfo {author} {\bibfnamefont {H.}~\bibnamefont {Kim}}, \ and\
  \bibinfo {author} {\bibfnamefont {T.}~\bibnamefont {Sekiguchi}},\ }\href
  {\doibase 10.1103/PhysRevD.95.075008} {\bibfield  {journal} {\bibinfo
  {journal} {Phys. Rev.}\ }\textbf {\bibinfo {volume} {D95}},\ \bibinfo {pages}
  {075008} (\bibinfo {year} {2017})},\ \Eprint
  {http://arxiv.org/abs/1611.08569} {arXiv:1611.08569 [hep-ph]} \BibitemShut
  {NoStop}%
\bibitem [{\citenamefont {Tangarife}\ \emph {et~al.}(2018)\citenamefont
  {Tangarife}, \citenamefont {Tobioka}, \citenamefont {Ubaldi},\ and\
  \citenamefont {Volansky}}]{Tangarife:2017rgl}%
  \BibitemOpen
  \bibfield  {author} {\bibinfo {author} {\bibfnamefont {W.}~\bibnamefont
  {Tangarife}}, \bibinfo {author} {\bibfnamefont {K.}~\bibnamefont {Tobioka}},
  \bibinfo {author} {\bibfnamefont {L.}~\bibnamefont {Ubaldi}}, \ and\ \bibinfo
  {author} {\bibfnamefont {T.}~\bibnamefont {Volansky}},\ }\href {\doibase
  10.1007/JHEP02(2018)084} {\bibfield  {journal} {\bibinfo  {journal} {JHEP}\
  }\textbf {\bibinfo {volume} {02}},\ \bibinfo {pages} {084} (\bibinfo {year}
  {2018})},\ \Eprint {http://arxiv.org/abs/1706.03072} {arXiv:1706.03072
  [hep-ph]} \BibitemShut {NoStop}%
\bibitem [{\citenamefont {Matsedonskyi}\ and\ \citenamefont
  {Montull}(2018)}]{Matsedonskyi:2017rkq}%
  \BibitemOpen
  \bibfield  {author} {\bibinfo {author} {\bibfnamefont {O.}~\bibnamefont
  {Matsedonskyi}}\ and\ \bibinfo {author} {\bibfnamefont {M.}~\bibnamefont
  {Montull}},\ }\href {\doibase 10.1103/PhysRevD.98.015026} {\bibfield
  {journal} {\bibinfo  {journal} {Phys. Rev.}\ }\textbf {\bibinfo {volume}
  {D98}},\ \bibinfo {pages} {015026} (\bibinfo {year} {2018})},\ \Eprint
  {http://arxiv.org/abs/1709.09090} {arXiv:1709.09090 [hep-ph]} \BibitemShut
  {NoStop}%
\bibitem [{\citenamefont {Abel}\ \emph {et~al.}(2018)\citenamefont {Abel},
  \citenamefont {Gupta},\ and\ \citenamefont {Scholtz}}]{Abel:2018fqg}%
  \BibitemOpen
  \bibfield  {author} {\bibinfo {author} {\bibfnamefont {S.~A.}\ \bibnamefont
  {Abel}}, \bibinfo {author} {\bibfnamefont {R.~S.}\ \bibnamefont {Gupta}}, \
  and\ \bibinfo {author} {\bibfnamefont {J.}~\bibnamefont {Scholtz}},\
  }\href@noop {} {\  (\bibinfo {year} {2018})},\ \Eprint
  {http://arxiv.org/abs/1810.05153} {arXiv:1810.05153 [hep-ph]} \BibitemShut
  {NoStop}%
\bibitem [{\citenamefont {Banerjee}\ \emph {et~al.}(2018)\citenamefont
  {Banerjee}, \citenamefont {Kim},\ and\ \citenamefont
  {Perez}}]{Banerjee:2018xmn}%
  \BibitemOpen
  \bibfield  {author} {\bibinfo {author} {\bibfnamefont {A.}~\bibnamefont
  {Banerjee}}, \bibinfo {author} {\bibfnamefont {H.}~\bibnamefont {Kim}}, \
  and\ \bibinfo {author} {\bibfnamefont {G.}~\bibnamefont {Perez}},\
  }\href@noop {} {\  (\bibinfo {year} {2018})},\ \Eprint
  {http://arxiv.org/abs/1810.01889} {arXiv:1810.01889 [hep-ph]} \BibitemShut
  {NoStop}%
\bibitem [{\citenamefont {Geller}\ \emph {et~al.}(2018)\citenamefont {Geller},
  \citenamefont {Hochberg},\ and\ \citenamefont {Kuflik}}]{Geller:2018xvz}%
  \BibitemOpen
  \bibfield  {author} {\bibinfo {author} {\bibfnamefont {M.}~\bibnamefont
  {Geller}}, \bibinfo {author} {\bibfnamefont {Y.}~\bibnamefont {Hochberg}}, \
  and\ \bibinfo {author} {\bibfnamefont {E.}~\bibnamefont {Kuflik}},\
  }\href@noop {} {\  (\bibinfo {year} {2018})},\ \Eprint
  {http://arxiv.org/abs/1809.07338} {arXiv:1809.07338 [hep-ph]} \BibitemShut
  {NoStop}%
\bibitem [{\citenamefont {Fonseca}\ and\ \citenamefont
  {Morgante}(2018)}]{Fonseca:2018kqf}%
  \BibitemOpen
  \bibfield  {author} {\bibinfo {author} {\bibfnamefont {N.}~\bibnamefont
  {Fonseca}}\ and\ \bibinfo {author} {\bibfnamefont {E.}~\bibnamefont
  {Morgante}},\ }\href@noop {} {\  (\bibinfo {year} {2018})},\ \Eprint
  {http://arxiv.org/abs/1809.04534} {arXiv:1809.04534 [hep-ph]} \BibitemShut
  {NoStop}%
\bibitem [{\citenamefont {Frugiuele}\ \emph {et~al.}(2018)\citenamefont
  {Frugiuele}, \citenamefont {Fuchs}, \citenamefont {Perez},\ and\
  \citenamefont {Schlaffer}}]{Frugiuele:2018coc}%
  \BibitemOpen
  \bibfield  {author} {\bibinfo {author} {\bibfnamefont {C.}~\bibnamefont
  {Frugiuele}}, \bibinfo {author} {\bibfnamefont {E.}~\bibnamefont {Fuchs}},
  \bibinfo {author} {\bibfnamefont {G.}~\bibnamefont {Perez}}, \ and\ \bibinfo
  {author} {\bibfnamefont {M.}~\bibnamefont {Schlaffer}},\ }\href {\doibase
  10.1007/JHEP10(2018)151} {\bibfield  {journal} {\bibinfo  {journal} {JHEP}\
  }\textbf {\bibinfo {volume} {10}},\ \bibinfo {pages} {151} (\bibinfo {year}
  {2018})},\ \Eprint {http://arxiv.org/abs/1807.10842} {arXiv:1807.10842
  [hep-ph]} \BibitemShut {NoStop}%
\bibitem [{\citenamefont {Davidi}\ \emph {et~al.}(2018)\citenamefont {Davidi},
  \citenamefont {Gupta}, \citenamefont {Perez}, \citenamefont {Redigolo},\ and\
  \citenamefont {Shalit}}]{Davidi:2018sii}%
  \BibitemOpen
  \bibfield  {author} {\bibinfo {author} {\bibfnamefont {O.}~\bibnamefont
  {Davidi}}, \bibinfo {author} {\bibfnamefont {R.~S.}\ \bibnamefont {Gupta}},
  \bibinfo {author} {\bibfnamefont {G.}~\bibnamefont {Perez}}, \bibinfo
  {author} {\bibfnamefont {D.}~\bibnamefont {Redigolo}}, \ and\ \bibinfo
  {author} {\bibfnamefont {A.}~\bibnamefont {Shalit}},\ }\href {\doibase
  10.1007/JHEP08(2018)153} {\bibfield  {journal} {\bibinfo  {journal} {JHEP}\
  }\textbf {\bibinfo {volume} {08}},\ \bibinfo {pages} {153} (\bibinfo {year}
  {2018})},\ \Eprint {http://arxiv.org/abs/1806.08791} {arXiv:1806.08791
  [hep-ph]} \BibitemShut {NoStop}%
\bibitem [{\citenamefont {Gupta}(2018)}]{Gupta:2018wif}%
  \BibitemOpen
  \bibfield  {author} {\bibinfo {author} {\bibfnamefont {R.~S.}\ \bibnamefont
  {Gupta}},\ }\href {\doibase 10.1103/PhysRevD.98.055023} {\bibfield  {journal}
  {\bibinfo  {journal} {Phys. Rev.}\ }\textbf {\bibinfo {volume} {D98}},\
  \bibinfo {pages} {055023} (\bibinfo {year} {2018})},\ \Eprint
  {http://arxiv.org/abs/1805.09316} {arXiv:1805.09316 [hep-ph]} \BibitemShut
  {NoStop}%
\bibitem [{\citenamefont {Fonseca}\ \emph
  {et~al.}(2018{\natexlab{b}})\citenamefont {Fonseca}, \citenamefont
  {Von~Harling}, \citenamefont {De~Lima},\ and\ \citenamefont
  {Machado}}]{Fonseca:2017crh}%
  \BibitemOpen
  \bibfield  {author} {\bibinfo {author} {\bibfnamefont {N.}~\bibnamefont
  {Fonseca}}, \bibinfo {author} {\bibfnamefont {B.}~\bibnamefont
  {Von~Harling}}, \bibinfo {author} {\bibfnamefont {L.}~\bibnamefont
  {De~Lima}}, \ and\ \bibinfo {author} {\bibfnamefont {C.~S.}\ \bibnamefont
  {Machado}},\ }\href {\doibase 10.1007/JHEP07(2018)033} {\bibfield  {journal}
  {\bibinfo  {journal} {JHEP}\ }\textbf {\bibinfo {volume} {07}},\ \bibinfo
  {pages} {033} (\bibinfo {year} {2018}{\natexlab{b}})},\ \Eprint
  {http://arxiv.org/abs/1712.07635} {arXiv:1712.07635 [hep-ph]} \BibitemShut
  {NoStop}%
\bibitem [{\citenamefont {Jeong}\ and\ \citenamefont
  {Shin}(2018)}]{Jeong:2017gdy}%
  \BibitemOpen
  \bibfield  {author} {\bibinfo {author} {\bibfnamefont {K.~S.}\ \bibnamefont
  {Jeong}}\ and\ \bibinfo {author} {\bibfnamefont {C.~S.}\ \bibnamefont
  {Shin}},\ }\href {\doibase 10.1007/JHEP01(2018)121} {\bibfield  {journal}
  {\bibinfo  {journal} {JHEP}\ }\textbf {\bibinfo {volume} {01}},\ \bibinfo
  {pages} {121} (\bibinfo {year} {2018})},\ \Eprint
  {http://arxiv.org/abs/1709.10025} {arXiv:1709.10025 [hep-ph]} \BibitemShut
  {NoStop}%
\bibitem [{\citenamefont {Nelson}\ and\ \citenamefont
  {Prescod-Weinstein}(2017)}]{Nelson:2017cfv}%
  \BibitemOpen
  \bibfield  {author} {\bibinfo {author} {\bibfnamefont {A.}~\bibnamefont
  {Nelson}}\ and\ \bibinfo {author} {\bibfnamefont {C.}~\bibnamefont
  {Prescod-Weinstein}},\ }\href {\doibase 10.1103/PhysRevD.96.113007}
  {\bibfield  {journal} {\bibinfo  {journal} {Phys. Rev.}\ }\textbf {\bibinfo
  {volume} {D96}},\ \bibinfo {pages} {113007} (\bibinfo {year} {2017})},\
  \Eprint {http://arxiv.org/abs/1708.00010} {arXiv:1708.00010 [hep-ph]}
  \BibitemShut {NoStop}%
\bibitem [{\citenamefont {Batell}\ \emph {et~al.}(2017)\citenamefont {Batell},
  \citenamefont {Fedderke},\ and\ \citenamefont {Wang}}]{Batell:2017kho}%
  \BibitemOpen
  \bibfield  {author} {\bibinfo {author} {\bibfnamefont {B.}~\bibnamefont
  {Batell}}, \bibinfo {author} {\bibfnamefont {M.~A.}\ \bibnamefont
  {Fedderke}}, \ and\ \bibinfo {author} {\bibfnamefont {L.-T.}\ \bibnamefont
  {Wang}},\ }\href {\doibase 10.1007/JHEP12(2017)139} {\bibfield  {journal}
  {\bibinfo  {journal} {JHEP}\ }\textbf {\bibinfo {volume} {12}},\ \bibinfo
  {pages} {139} (\bibinfo {year} {2017})},\ \Eprint
  {http://arxiv.org/abs/1705.09666} {arXiv:1705.09666 [hep-ph]} \BibitemShut
  {NoStop}%
\bibitem [{\citenamefont {You}(2017)}]{You:2017kah}%
  \BibitemOpen
  \bibfield  {author} {\bibinfo {author} {\bibfnamefont {T.}~\bibnamefont
  {You}},\ }\href {\doibase 10.1088/1475-7516/2017/09/019} {\bibfield
  {journal} {\bibinfo  {journal} {JCAP}\ }\textbf {\bibinfo {volume} {1709}},\
  \bibinfo {pages} {019} (\bibinfo {year} {2017})},\ \Eprint
  {http://arxiv.org/abs/1701.09167} {arXiv:1701.09167 [hep-ph]} \BibitemShut
  {NoStop}%
\bibitem [{\citenamefont {Lalak}\ and\ \citenamefont
  {Markiewicz}(2018)}]{Lalak:2016mbv}%
  \BibitemOpen
  \bibfield  {author} {\bibinfo {author} {\bibfnamefont {Z.}~\bibnamefont
  {Lalak}}\ and\ \bibinfo {author} {\bibfnamefont {A.}~\bibnamefont
  {Markiewicz}},\ }\href {\doibase 10.1088/1361-6471/aaa4c7} {\bibfield
  {journal} {\bibinfo  {journal} {J. Phys.}\ }\textbf {\bibinfo {volume}
  {G45}},\ \bibinfo {pages} {035002} (\bibinfo {year} {2018})},\ \Eprint
  {http://arxiv.org/abs/1612.09128} {arXiv:1612.09128 [hep-ph]} \BibitemShut
  {NoStop}%
\bibitem [{\citenamefont {McAllister}\ \emph {et~al.}(2018)\citenamefont
  {McAllister}, \citenamefont {Schwaller}, \citenamefont {Servant},
  \citenamefont {Stout},\ and\ \citenamefont {Westphal}}]{McAllister:2016vzi}%
  \BibitemOpen
  \bibfield  {author} {\bibinfo {author} {\bibfnamefont {L.}~\bibnamefont
  {McAllister}}, \bibinfo {author} {\bibfnamefont {P.}~\bibnamefont
  {Schwaller}}, \bibinfo {author} {\bibfnamefont {G.}~\bibnamefont {Servant}},
  \bibinfo {author} {\bibfnamefont {J.}~\bibnamefont {Stout}}, \ and\ \bibinfo
  {author} {\bibfnamefont {A.}~\bibnamefont {Westphal}},\ }\href {\doibase
  10.1007/JHEP02(2018)124} {\bibfield  {journal} {\bibinfo  {journal} {JHEP}\
  }\textbf {\bibinfo {volume} {02}},\ \bibinfo {pages} {124} (\bibinfo {year}
  {2018})},\ \Eprint {http://arxiv.org/abs/1610.05320} {arXiv:1610.05320
  [hep-th]} \BibitemShut {NoStop}%
\bibitem [{\citenamefont {Flacke}\ \emph {et~al.}(2017)\citenamefont {Flacke},
  \citenamefont {Frugiuele}, \citenamefont {Fuchs}, \citenamefont {Gupta},\
  and\ \citenamefont {Perez}}]{Flacke:2016szy}%
  \BibitemOpen
  \bibfield  {author} {\bibinfo {author} {\bibfnamefont {T.}~\bibnamefont
  {Flacke}}, \bibinfo {author} {\bibfnamefont {C.}~\bibnamefont {Frugiuele}},
  \bibinfo {author} {\bibfnamefont {E.}~\bibnamefont {Fuchs}}, \bibinfo
  {author} {\bibfnamefont {R.~S.}\ \bibnamefont {Gupta}}, \ and\ \bibinfo
  {author} {\bibfnamefont {G.}~\bibnamefont {Perez}},\ }\href {\doibase
  10.1007/JHEP06(2017)050} {\bibfield  {journal} {\bibinfo  {journal} {JHEP}\
  }\textbf {\bibinfo {volume} {06}},\ \bibinfo {pages} {050} (\bibinfo {year}
  {2017})},\ \Eprint {http://arxiv.org/abs/1610.02025} {arXiv:1610.02025
  [hep-ph]} \BibitemShut {NoStop}%
\bibitem [{\citenamefont {Kobayashi}\ \emph {et~al.}(2017)\citenamefont
  {Kobayashi}, \citenamefont {Seto}, \citenamefont {Shimomura},\ and\
  \citenamefont {Urakawa}}]{Kobayashi:2016bue}%
  \BibitemOpen
  \bibfield  {author} {\bibinfo {author} {\bibfnamefont {T.}~\bibnamefont
  {Kobayashi}}, \bibinfo {author} {\bibfnamefont {O.}~\bibnamefont {Seto}},
  \bibinfo {author} {\bibfnamefont {T.}~\bibnamefont {Shimomura}}, \ and\
  \bibinfo {author} {\bibfnamefont {Y.}~\bibnamefont {Urakawa}},\ }\href
  {\doibase 10.1142/S0217732317501425} {\bibfield  {journal} {\bibinfo
  {journal} {Mod. Phys. Lett.}\ }\textbf {\bibinfo {volume} {A32}},\ \bibinfo
  {pages} {1750142} (\bibinfo {year} {2017})},\ \Eprint
  {http://arxiv.org/abs/1605.06908} {arXiv:1605.06908 [astro-ph.CO]}
  \BibitemShut {NoStop}%
\bibitem [{\citenamefont {Evans}\ \emph {et~al.}(2016)\citenamefont {Evans},
  \citenamefont {Gherghetta}, \citenamefont {Nagata},\ and\ \citenamefont
  {Thomas}}]{Evans:2016htp}%
  \BibitemOpen
  \bibfield  {author} {\bibinfo {author} {\bibfnamefont {J.~L.}\ \bibnamefont
  {Evans}}, \bibinfo {author} {\bibfnamefont {T.}~\bibnamefont {Gherghetta}},
  \bibinfo {author} {\bibfnamefont {N.}~\bibnamefont {Nagata}}, \ and\ \bibinfo
  {author} {\bibfnamefont {Z.}~\bibnamefont {Thomas}},\ }\href {\doibase
  10.1007/JHEP09(2016)150} {\bibfield  {journal} {\bibinfo  {journal} {JHEP}\
  }\textbf {\bibinfo {volume} {09}},\ \bibinfo {pages} {150} (\bibinfo {year}
  {2016})},\ \Eprint {http://arxiv.org/abs/1602.04812} {arXiv:1602.04812
  [hep-ph]} \BibitemShut {NoStop}%
\bibitem [{\citenamefont {Fowlie}\ \emph {et~al.}(2016)\citenamefont {Fowlie},
  \citenamefont {Balazs}, \citenamefont {White}, \citenamefont {Marzola},\ and\
  \citenamefont {Raidal}}]{Fowlie:2016jlx}%
  \BibitemOpen
  \bibfield  {author} {\bibinfo {author} {\bibfnamefont {A.}~\bibnamefont
  {Fowlie}}, \bibinfo {author} {\bibfnamefont {C.}~\bibnamefont {Balazs}},
  \bibinfo {author} {\bibfnamefont {G.}~\bibnamefont {White}}, \bibinfo
  {author} {\bibfnamefont {L.}~\bibnamefont {Marzola}}, \ and\ \bibinfo
  {author} {\bibfnamefont {M.}~\bibnamefont {Raidal}},\ }\href {\doibase
  10.1007/JHEP08(2016)100} {\bibfield  {journal} {\bibinfo  {journal} {JHEP}\
  }\textbf {\bibinfo {volume} {08}},\ \bibinfo {pages} {100} (\bibinfo {year}
  {2016})},\ \Eprint {http://arxiv.org/abs/1602.03889} {arXiv:1602.03889
  [hep-ph]} \BibitemShut {NoStop}%
\bibitem [{\citenamefont {Ibanez}\ \emph {et~al.}(2016)\citenamefont {Ibanez},
  \citenamefont {Montero}, \citenamefont {Uranga},\ and\ \citenamefont
  {Valenzuela}}]{Ibanez:2015fcv}%
  \BibitemOpen
  \bibfield  {author} {\bibinfo {author} {\bibfnamefont {L.~E.}\ \bibnamefont
  {Ibanez}}, \bibinfo {author} {\bibfnamefont {M.}~\bibnamefont {Montero}},
  \bibinfo {author} {\bibfnamefont {A.}~\bibnamefont {Uranga}}, \ and\ \bibinfo
  {author} {\bibfnamefont {I.}~\bibnamefont {Valenzuela}},\ }\href {\doibase
  10.1007/JHEP04(2016)020} {\bibfield  {journal} {\bibinfo  {journal} {JHEP}\
  }\textbf {\bibinfo {volume} {04}},\ \bibinfo {pages} {020} (\bibinfo {year}
  {2016})},\ \Eprint {http://arxiv.org/abs/1512.00025} {arXiv:1512.00025
  [hep-th]} \BibitemShut {NoStop}%
\bibitem [{\citenamefont {Di~Chiara}\ \emph {et~al.}(2016)\citenamefont
  {Di~Chiara}, \citenamefont {Kannike}, \citenamefont {Marzola}, \citenamefont
  {Racioppi}, \citenamefont {Raidal},\ and\ \citenamefont
  {Spethmann}}]{DiChiara:2015euo}%
  \BibitemOpen
  \bibfield  {author} {\bibinfo {author} {\bibfnamefont {S.}~\bibnamefont
  {Di~Chiara}}, \bibinfo {author} {\bibfnamefont {K.}~\bibnamefont {Kannike}},
  \bibinfo {author} {\bibfnamefont {L.}~\bibnamefont {Marzola}}, \bibinfo
  {author} {\bibfnamefont {A.}~\bibnamefont {Racioppi}}, \bibinfo {author}
  {\bibfnamefont {M.}~\bibnamefont {Raidal}}, \ and\ \bibinfo {author}
  {\bibfnamefont {C.}~\bibnamefont {Spethmann}},\ }\href {\doibase
  10.1103/PhysRevD.93.103527} {\bibfield  {journal} {\bibinfo  {journal} {Phys.
  Rev.}\ }\textbf {\bibinfo {volume} {D93}},\ \bibinfo {pages} {103527}
  (\bibinfo {year} {2016})},\ \Eprint {http://arxiv.org/abs/1511.02858}
  {arXiv:1511.02858 [hep-ph]} \BibitemShut {NoStop}%
\bibitem [{\citenamefont {Marzola}\ and\ \citenamefont
  {Raidal}(2016)}]{Marzola:2015dia}%
  \BibitemOpen
  \bibfield  {author} {\bibinfo {author} {\bibfnamefont {L.}~\bibnamefont
  {Marzola}}\ and\ \bibinfo {author} {\bibfnamefont {M.}~\bibnamefont
  {Raidal}},\ }\href {\doibase 10.1142/S0217732316502151} {\bibfield  {journal}
  {\bibinfo  {journal} {Mod. Phys. Lett.}\ }\textbf {\bibinfo {volume} {A31}},\
  \bibinfo {pages} {1650215} (\bibinfo {year} {2016})},\ \Eprint
  {http://arxiv.org/abs/1510.00710} {arXiv:1510.00710 [hep-ph]} \BibitemShut
  {NoStop}%
\bibitem [{\citenamefont {Gupta}\ \emph {et~al.}(2016)\citenamefont {Gupta},
  \citenamefont {Komargodski}, \citenamefont {Perez},\ and\ \citenamefont
  {Ubaldi}}]{Gupta:2015uea}%
  \BibitemOpen
  \bibfield  {author} {\bibinfo {author} {\bibfnamefont {R.~S.}\ \bibnamefont
  {Gupta}}, \bibinfo {author} {\bibfnamefont {Z.}~\bibnamefont {Komargodski}},
  \bibinfo {author} {\bibfnamefont {G.}~\bibnamefont {Perez}}, \ and\ \bibinfo
  {author} {\bibfnamefont {L.}~\bibnamefont {Ubaldi}},\ }\href {\doibase
  10.1007/JHEP02(2016)166} {\bibfield  {journal} {\bibinfo  {journal} {JHEP}\
  }\textbf {\bibinfo {volume} {02}},\ \bibinfo {pages} {166} (\bibinfo {year}
  {2016})},\ \Eprint {http://arxiv.org/abs/1509.00047} {arXiv:1509.00047
  [hep-ph]} \BibitemShut {NoStop}%
\bibitem [{\citenamefont {Jaeckel}\ \emph {et~al.}(2016)\citenamefont
  {Jaeckel}, \citenamefont {Mehta},\ and\ \citenamefont
  {Witkowski}}]{Jaeckel:2015txa}%
  \BibitemOpen
  \bibfield  {author} {\bibinfo {author} {\bibfnamefont {J.}~\bibnamefont
  {Jaeckel}}, \bibinfo {author} {\bibfnamefont {V.~M.}\ \bibnamefont {Mehta}},
  \ and\ \bibinfo {author} {\bibfnamefont {L.~T.}\ \bibnamefont {Witkowski}},\
  }\href {\doibase 10.1103/PhysRevD.93.063522} {\bibfield  {journal} {\bibinfo
  {journal} {Phys. Rev.}\ }\textbf {\bibinfo {volume} {D93}},\ \bibinfo {pages}
  {063522} (\bibinfo {year} {2016})},\ \Eprint
  {http://arxiv.org/abs/1508.03321} {arXiv:1508.03321 [hep-ph]} \BibitemShut
  {NoStop}%
\bibitem [{\citenamefont {Antipin}\ and\ \citenamefont
  {Redi}(2015)}]{Antipin:2015jia}%
  \BibitemOpen
  \bibfield  {author} {\bibinfo {author} {\bibfnamefont {O.}~\bibnamefont
  {Antipin}}\ and\ \bibinfo {author} {\bibfnamefont {M.}~\bibnamefont {Redi}},\
  }\href {\doibase 10.1007/JHEP12(2015)031} {\bibfield  {journal} {\bibinfo
  {journal} {JHEP}\ }\textbf {\bibinfo {volume} {12}},\ \bibinfo {pages} {031}
  (\bibinfo {year} {2015})},\ \Eprint {http://arxiv.org/abs/1508.01112}
  {arXiv:1508.01112 [hep-ph]} \BibitemShut {NoStop}%
\bibitem [{\citenamefont {Patil}\ and\ \citenamefont
  {Schwaller}(2016)}]{Patil:2015oxa}%
  \BibitemOpen
  \bibfield  {author} {\bibinfo {author} {\bibfnamefont {S.~P.}\ \bibnamefont
  {Patil}}\ and\ \bibinfo {author} {\bibfnamefont {P.}~\bibnamefont
  {Schwaller}},\ }\href {\doibase 10.1007/JHEP02(2016)077} {\bibfield
  {journal} {\bibinfo  {journal} {JHEP}\ }\textbf {\bibinfo {volume} {02}},\
  \bibinfo {pages} {077} (\bibinfo {year} {2016})},\ \Eprint
  {http://arxiv.org/abs/1507.08649} {arXiv:1507.08649 [hep-ph]} \BibitemShut
  {NoStop}%
\bibitem [{\citenamefont {Hardy}(2015)}]{Hardy:2015laa}%
  \BibitemOpen
  \bibfield  {author} {\bibinfo {author} {\bibfnamefont {E.}~\bibnamefont
  {Hardy}},\ }\href {\doibase 10.1007/JHEP11(2015)077} {\bibfield  {journal}
  {\bibinfo  {journal} {JHEP}\ }\textbf {\bibinfo {volume} {11}},\ \bibinfo
  {pages} {077} (\bibinfo {year} {2015})},\ \Eprint
  {http://arxiv.org/abs/1507.07525} {arXiv:1507.07525 [hep-ph]} \BibitemShut
  {NoStop}%
\bibitem [{\citenamefont {Kobakhidze}(2015)}]{Kobakhidze:2015jya}%
  \BibitemOpen
  \bibfield  {author} {\bibinfo {author} {\bibfnamefont {A.}~\bibnamefont
  {Kobakhidze}},\ }\href {\doibase 10.1140/epjc/s10052-015-3621-4} {\bibfield
  {journal} {\bibinfo  {journal} {Eur. Phys. J.}\ }\textbf {\bibinfo {volume}
  {C75}},\ \bibinfo {pages} {384} (\bibinfo {year} {2015})},\ \Eprint
  {http://arxiv.org/abs/1506.04840} {arXiv:1506.04840 [hep-ph]} \BibitemShut
  {NoStop}%
\bibitem [{\citenamefont {Iso}\ \emph {et~al.}(2016)\citenamefont {Iso},
  \citenamefont {Kohri},\ and\ \citenamefont {Shimada}}]{Iso:2015wsf}%
  \BibitemOpen
  \bibfield  {author} {\bibinfo {author} {\bibfnamefont {S.}~\bibnamefont
  {Iso}}, \bibinfo {author} {\bibfnamefont {K.}~\bibnamefont {Kohri}}, \ and\
  \bibinfo {author} {\bibfnamefont {K.}~\bibnamefont {Shimada}},\ }\href
  {\doibase 10.1103/PhysRevD.93.084009} {\bibfield  {journal} {\bibinfo
  {journal} {Phys. Rev.}\ }\textbf {\bibinfo {volume} {D93}},\ \bibinfo {pages}
  {084009} (\bibinfo {year} {2016})},\ \Eprint
  {http://arxiv.org/abs/1511.05923} {arXiv:1511.05923 [hep-ph]} \BibitemShut
  {NoStop}%
\bibitem [{\citenamefont {Dine}\ and\ \citenamefont
  {Pack}(2012)}]{Dine:2011ws}%
  \BibitemOpen
  \bibfield  {author} {\bibinfo {author} {\bibfnamefont {M.}~\bibnamefont
  {Dine}}\ and\ \bibinfo {author} {\bibfnamefont {L.}~\bibnamefont {Pack}},\
  }\href {\doibase 10.1088/1475-7516/2012/06/033} {\bibfield  {journal}
  {\bibinfo  {journal} {JCAP}\ }\textbf {\bibinfo {volume} {1206}},\ \bibinfo
  {pages} {033} (\bibinfo {year} {2012})},\ \Eprint
  {http://arxiv.org/abs/1109.2079} {arXiv:1109.2079 [hep-ph]} \BibitemShut
  {NoStop}%
\bibitem [{\citenamefont {German}\ \emph {et~al.}(2001)\citenamefont {German},
  \citenamefont {Ross},\ and\ \citenamefont {Sarkar}}]{German:2001tz}%
  \BibitemOpen
  \bibfield  {author} {\bibinfo {author} {\bibfnamefont {G.}~\bibnamefont
  {German}}, \bibinfo {author} {\bibfnamefont {G.~G.}\ \bibnamefont {Ross}}, \
  and\ \bibinfo {author} {\bibfnamefont {S.}~\bibnamefont {Sarkar}},\ }\href
  {\doibase 10.1016/S0550-3213(01)00258-9} {\bibfield  {journal} {\bibinfo
  {journal} {Nucl. Phys.}\ }\textbf {\bibinfo {volume} {B608}},\ \bibinfo
  {pages} {423} (\bibinfo {year} {2001})},\ \Eprint
  {http://arxiv.org/abs/hep-ph/0103243} {arXiv:hep-ph/0103243 [hep-ph]}
  \BibitemShut {NoStop}%
\bibitem [{\citenamefont {Beauchesne}\ \emph {et~al.}(2017)\citenamefont
  {Beauchesne}, \citenamefont {Bertuzzo},\ and\ \citenamefont {Grilli~di
  Cortona}}]{Beauchesne:2017ukw}%
  \BibitemOpen
  \bibfield  {author} {\bibinfo {author} {\bibfnamefont {H.}~\bibnamefont
  {Beauchesne}}, \bibinfo {author} {\bibfnamefont {E.}~\bibnamefont
  {Bertuzzo}}, \ and\ \bibinfo {author} {\bibfnamefont {G.}~\bibnamefont
  {Grilli~di Cortona}},\ }\href {\doibase 10.1007/JHEP08(2017)093} {\bibfield
  {journal} {\bibinfo  {journal} {JHEP}\ }\textbf {\bibinfo {volume} {08}},\
  \bibinfo {pages} {093} (\bibinfo {year} {2017})},\ \Eprint
  {http://arxiv.org/abs/1705.06325} {arXiv:1705.06325 [hep-ph]} \BibitemShut
  {NoStop}%
\bibitem [{\citenamefont {Matsui}\ and\ \citenamefont
  {Takahashi}(2019)}]{Matsui:2018bsy}%
  \BibitemOpen
  \bibfield  {author} {\bibinfo {author} {\bibfnamefont {H.}~\bibnamefont
  {Matsui}}\ and\ \bibinfo {author} {\bibfnamefont {F.}~\bibnamefont
  {Takahashi}},\ }\href {\doibase 10.1103/PhysRevD.99.023533} {\bibfield
  {journal} {\bibinfo  {journal} {Phys. Rev.}\ }\textbf {\bibinfo {volume}
  {D99}},\ \bibinfo {pages} {023533} (\bibinfo {year} {2019})},\ \Eprint
  {http://arxiv.org/abs/1807.11938} {arXiv:1807.11938 [hep-th]} \BibitemShut
  {NoStop}%
\bibitem [{\citenamefont {Dimopoulos}(2018)}]{Dimopoulos:2018upl}%
  \BibitemOpen
  \bibfield  {author} {\bibinfo {author} {\bibfnamefont {K.}~\bibnamefont
  {Dimopoulos}},\ }\href {\doibase 10.1103/PhysRevD.98.123516} {\bibfield
  {journal} {\bibinfo  {journal} {Phys. Rev.}\ }\textbf {\bibinfo {volume}
  {D98}},\ \bibinfo {pages} {123516} (\bibinfo {year} {2018})},\ \Eprint
  {http://arxiv.org/abs/1810.03438} {arXiv:1810.03438 [gr-qc]} \BibitemShut
  {NoStop}%
\bibitem [{\citenamefont {Kinney}(2019)}]{Kinney:2018kew}%
  \BibitemOpen
  \bibfield  {author} {\bibinfo {author} {\bibfnamefont {W.~H.}\ \bibnamefont
  {Kinney}},\ }\href {\doibase 10.1103/PhysRevLett.122.081302} {\bibfield
  {journal} {\bibinfo  {journal} {Phys. Rev. Lett.}\ }\textbf {\bibinfo
  {volume} {122}},\ \bibinfo {pages} {081302} (\bibinfo {year} {2019})},\
  \Eprint {http://arxiv.org/abs/1811.11698} {arXiv:1811.11698 [astro-ph.CO]}
  \BibitemShut {NoStop}%
\bibitem [{\citenamefont {Allahverdi}\ \emph {et~al.}(2010)\citenamefont
  {Allahverdi}, \citenamefont {Brandenberger}, \citenamefont {Cyr-Racine},\
  and\ \citenamefont {Mazumdar}}]{Allahverdi:2010xz}%
  \BibitemOpen
  \bibfield  {author} {\bibinfo {author} {\bibfnamefont {R.}~\bibnamefont
  {Allahverdi}}, \bibinfo {author} {\bibfnamefont {R.}~\bibnamefont
  {Brandenberger}}, \bibinfo {author} {\bibfnamefont {F.-Y.}\ \bibnamefont
  {Cyr-Racine}}, \ and\ \bibinfo {author} {\bibfnamefont {A.}~\bibnamefont
  {Mazumdar}},\ }\href {\doibase 10.1146/annurev.nucl.012809.104511} {\bibfield
   {journal} {\bibinfo  {journal} {Ann. Rev. Nucl. Part. Sci.}\ }\textbf
  {\bibinfo {volume} {60}},\ \bibinfo {pages} {27} (\bibinfo {year} {2010})},\
  \Eprint {http://arxiv.org/abs/1001.2600} {arXiv:1001.2600 [hep-th]}
  \BibitemShut {NoStop}%
\bibitem [{\citenamefont {Coleman}(1977)}]{Coleman:1977py}%
  \BibitemOpen
  \bibfield  {author} {\bibinfo {author} {\bibfnamefont {S.~R.}\ \bibnamefont
  {Coleman}},\ }\href {\doibase 10.1103/PhysRevD.15.2929,
  10.1103/PhysRevD.16.1248} {\bibfield  {journal} {\bibinfo  {journal} {Phys.
  Rev.}\ }\textbf {\bibinfo {volume} {D15}},\ \bibinfo {pages} {2929} (\bibinfo
  {year} {1977})},\ \bibinfo {note} {[Erratum: Phys.
  Rev.D16,1248(1977)]}\BibitemShut {NoStop}%
\bibitem [{\citenamefont {Callan}\ and\ \citenamefont
  {Coleman}(1977)}]{Callan:1977pt}%
  \BibitemOpen
  \bibfield  {author} {\bibinfo {author} {\bibfnamefont {C.~G.}\ \bibnamefont
  {Callan}, \bibfnamefont {Jr.}}\ and\ \bibinfo {author} {\bibfnamefont
  {S.~R.}\ \bibnamefont {Coleman}},\ }\href {\doibase 10.1103/PhysRevD.16.1762}
  {\bibfield  {journal} {\bibinfo  {journal} {Phys. Rev.}\ }\textbf {\bibinfo
  {volume} {D16}},\ \bibinfo {pages} {1762} (\bibinfo {year}
  {1977})}\BibitemShut {NoStop}%
\bibitem [{\citenamefont {Coy}\ \emph {et~al.}(2017)\citenamefont {Coy},
  \citenamefont {Frigerio},\ and\ \citenamefont {Ibe}}]{Coy:2017yex}%
  \BibitemOpen
  \bibfield  {author} {\bibinfo {author} {\bibfnamefont {R.}~\bibnamefont
  {Coy}}, \bibinfo {author} {\bibfnamefont {M.}~\bibnamefont {Frigerio}}, \
  and\ \bibinfo {author} {\bibfnamefont {M.}~\bibnamefont {Ibe}},\ }\href
  {\doibase 10.1007/JHEP10(2017)002} {\bibfield  {journal} {\bibinfo  {journal}
  {JHEP}\ }\textbf {\bibinfo {volume} {10}},\ \bibinfo {pages} {002} (\bibinfo
  {year} {2017})},\ \Eprint {http://arxiv.org/abs/1706.04529} {arXiv:1706.04529
  [hep-ph]} \BibitemShut {NoStop}%
\bibitem [{\citenamefont {Bechtle}\ \emph
  {et~al.}(2014{\natexlab{a}})\citenamefont {Bechtle}, \citenamefont {Brein},
  \citenamefont {Heinemeyer}, \citenamefont {Stl}, \citenamefont {Stefaniak},
  \citenamefont {Weiglein},\ and\ \citenamefont {Williams}}]{Bechtle:2013wla}%
  \BibitemOpen
  \bibfield  {author} {\bibinfo {author} {\bibfnamefont {P.}~\bibnamefont
  {Bechtle}}, \bibinfo {author} {\bibfnamefont {O.}~\bibnamefont {Brein}},
  \bibinfo {author} {\bibfnamefont {S.}~\bibnamefont {Heinemeyer}}, \bibinfo
  {author} {\bibfnamefont {O.}~\bibnamefont {Stl}}, \bibinfo {author}
  {\bibfnamefont {T.}~\bibnamefont {Stefaniak}}, \bibinfo {author}
  {\bibfnamefont {G.}~\bibnamefont {Weiglein}}, \ and\ \bibinfo {author}
  {\bibfnamefont {K.~E.}\ \bibnamefont {Williams}},\ }\href {\doibase
  10.1140/epjc/s10052-013-2693-2} {\bibfield  {journal} {\bibinfo  {journal}
  {Eur. Phys. J.}\ }\textbf {\bibinfo {volume} {C74}},\ \bibinfo {pages} {2693}
  (\bibinfo {year} {2014}{\natexlab{a}})},\ \Eprint
  {http://arxiv.org/abs/1311.0055} {arXiv:1311.0055 [hep-ph]} \BibitemShut
  {NoStop}%
\bibitem [{\citenamefont {Bechtle}\ \emph {et~al.}(2012)\citenamefont
  {Bechtle}, \citenamefont {Brein}, \citenamefont {Heinemeyer}, \citenamefont
  {Stal}, \citenamefont {Stefaniak}, \citenamefont {Weiglein},\ and\
  \citenamefont {Williams}}]{Bechtle:2013gu}%
  \BibitemOpen
  \bibfield  {author} {\bibinfo {author} {\bibfnamefont {P.}~\bibnamefont
  {Bechtle}}, \bibinfo {author} {\bibfnamefont {O.}~\bibnamefont {Brein}},
  \bibinfo {author} {\bibfnamefont {S.}~\bibnamefont {Heinemeyer}}, \bibinfo
  {author} {\bibfnamefont {O.}~\bibnamefont {Stal}}, \bibinfo {author}
  {\bibfnamefont {T.}~\bibnamefont {Stefaniak}}, \bibinfo {author}
  {\bibfnamefont {G.}~\bibnamefont {Weiglein}}, \ and\ \bibinfo {author}
  {\bibfnamefont {K.}~\bibnamefont {Williams}},\ }\bibfield  {booktitle} {\emph
  {\bibinfo {booktitle} {{Proceedings, 4th International Workshop on Prospects
  for Charged Higgs Discovery at Colliders (CHARGED 2012): Uppsala, Sweden,
  October 8-11, 2012}}},\ }\href {\doibase 10.22323/1.156.0024} {\bibfield
  {journal} {\bibinfo  {journal} {PoS}\ }\textbf {\bibinfo {volume}
  {CHARGED2012}},\ \bibinfo {pages} {024} (\bibinfo {year} {2012})},\ \Eprint
  {http://arxiv.org/abs/1301.2345} {arXiv:1301.2345 [hep-ph]} \BibitemShut
  {NoStop}%
\bibitem [{\citenamefont {Bechtle}\ \emph {et~al.}(2011)\citenamefont
  {Bechtle}, \citenamefont {Brein}, \citenamefont {Heinemeyer}, \citenamefont
  {Weiglein},\ and\ \citenamefont {Williams}}]{Bechtle:2011sb}%
  \BibitemOpen
  \bibfield  {author} {\bibinfo {author} {\bibfnamefont {P.}~\bibnamefont
  {Bechtle}}, \bibinfo {author} {\bibfnamefont {O.}~\bibnamefont {Brein}},
  \bibinfo {author} {\bibfnamefont {S.}~\bibnamefont {Heinemeyer}}, \bibinfo
  {author} {\bibfnamefont {G.}~\bibnamefont {Weiglein}}, \ and\ \bibinfo
  {author} {\bibfnamefont {K.~E.}\ \bibnamefont {Williams}},\ }\href {\doibase
  10.1016/j.cpc.2011.07.015} {\bibfield  {journal} {\bibinfo  {journal}
  {Comput. Phys. Commun.}\ }\textbf {\bibinfo {volume} {182}},\ \bibinfo
  {pages} {2605} (\bibinfo {year} {2011})},\ \Eprint
  {http://arxiv.org/abs/1102.1898} {arXiv:1102.1898 [hep-ph]} \BibitemShut
  {NoStop}%
\bibitem [{\citenamefont {Bechtle}\ \emph {et~al.}(2010)\citenamefont
  {Bechtle}, \citenamefont {Brein}, \citenamefont {Heinemeyer}, \citenamefont
  {Weiglein},\ and\ \citenamefont {Williams}}]{Bechtle:2008jh}%
  \BibitemOpen
  \bibfield  {author} {\bibinfo {author} {\bibfnamefont {P.}~\bibnamefont
  {Bechtle}}, \bibinfo {author} {\bibfnamefont {O.}~\bibnamefont {Brein}},
  \bibinfo {author} {\bibfnamefont {S.}~\bibnamefont {Heinemeyer}}, \bibinfo
  {author} {\bibfnamefont {G.}~\bibnamefont {Weiglein}}, \ and\ \bibinfo
  {author} {\bibfnamefont {K.~E.}\ \bibnamefont {Williams}},\ }\href {\doibase
  10.1016/j.cpc.2009.09.003} {\bibfield  {journal} {\bibinfo  {journal}
  {Comput. Phys. Commun.}\ }\textbf {\bibinfo {volume} {181}},\ \bibinfo
  {pages} {138} (\bibinfo {year} {2010})},\ \Eprint
  {http://arxiv.org/abs/0811.4169} {arXiv:0811.4169 [hep-ph]} \BibitemShut
  {NoStop}%
\bibitem [{\citenamefont {Bechtle}\ \emph {et~al.}(2015)\citenamefont
  {Bechtle}, \citenamefont {Heinemeyer}, \citenamefont {Stal}, \citenamefont
  {Stefaniak},\ and\ \citenamefont {Weiglein}}]{Bechtle:2015pma}%
  \BibitemOpen
  \bibfield  {author} {\bibinfo {author} {\bibfnamefont {P.}~\bibnamefont
  {Bechtle}}, \bibinfo {author} {\bibfnamefont {S.}~\bibnamefont {Heinemeyer}},
  \bibinfo {author} {\bibfnamefont {O.}~\bibnamefont {Stal}}, \bibinfo {author}
  {\bibfnamefont {T.}~\bibnamefont {Stefaniak}}, \ and\ \bibinfo {author}
  {\bibfnamefont {G.}~\bibnamefont {Weiglein}},\ }\href {\doibase
  10.1140/epjc/s10052-015-3650-z} {\bibfield  {journal} {\bibinfo  {journal}
  {Eur. Phys. J.}\ }\textbf {\bibinfo {volume} {C75}},\ \bibinfo {pages} {421}
  (\bibinfo {year} {2015})},\ \Eprint {http://arxiv.org/abs/1507.06706}
  {arXiv:1507.06706 [hep-ph]} \BibitemShut {NoStop}%
\bibitem [{\citenamefont {Bechtle}\ \emph
  {et~al.}(2014{\natexlab{b}})\citenamefont {Bechtle}, \citenamefont
  {Heinemeyer}, \citenamefont {Stl}, \citenamefont {Stefaniak},\ and\
  \citenamefont {Weiglein}}]{Bechtle:2014ewa}%
  \BibitemOpen
  \bibfield  {author} {\bibinfo {author} {\bibfnamefont {P.}~\bibnamefont
  {Bechtle}}, \bibinfo {author} {\bibfnamefont {S.}~\bibnamefont {Heinemeyer}},
  \bibinfo {author} {\bibfnamefont {O.}~\bibnamefont {Stl}}, \bibinfo {author}
  {\bibfnamefont {T.}~\bibnamefont {Stefaniak}}, \ and\ \bibinfo {author}
  {\bibfnamefont {G.}~\bibnamefont {Weiglein}},\ }\href {\doibase
  10.1007/JHEP11(2014)039} {\bibfield  {journal} {\bibinfo  {journal} {JHEP}\
  }\textbf {\bibinfo {volume} {11}},\ \bibinfo {pages} {039} (\bibinfo {year}
  {2014}{\natexlab{b}})},\ \Eprint {http://arxiv.org/abs/1403.1582}
  {arXiv:1403.1582 [hep-ph]} \BibitemShut {NoStop}%
\bibitem [{\citenamefont {Bechtle}\ \emph
  {et~al.}(2014{\natexlab{c}})\citenamefont {Bechtle}, \citenamefont
  {Heinemeyer}, \citenamefont {Stl}, \citenamefont {Stefaniak},\ and\
  \citenamefont {Weiglein}}]{Bechtle:2013xfa}%
  \BibitemOpen
  \bibfield  {author} {\bibinfo {author} {\bibfnamefont {P.}~\bibnamefont
  {Bechtle}}, \bibinfo {author} {\bibfnamefont {S.}~\bibnamefont {Heinemeyer}},
  \bibinfo {author} {\bibfnamefont {O.}~\bibnamefont {Stl}}, \bibinfo {author}
  {\bibfnamefont {T.}~\bibnamefont {Stefaniak}}, \ and\ \bibinfo {author}
  {\bibfnamefont {G.}~\bibnamefont {Weiglein}},\ }\href {\doibase
  10.1140/epjc/s10052-013-2711-4} {\bibfield  {journal} {\bibinfo  {journal}
  {Eur. Phys. J.}\ }\textbf {\bibinfo {volume} {C74}},\ \bibinfo {pages} {2711}
  (\bibinfo {year} {2014}{\natexlab{c}})},\ \Eprint
  {http://arxiv.org/abs/1305.1933} {arXiv:1305.1933 [hep-ph]} \BibitemShut
  {NoStop}%
\bibitem [{\citenamefont {collaboration}(2018)}]{ATLAS:2018doi}%
  \BibitemOpen
  \bibfield  {author} {\bibinfo {author} {\bibfnamefont {T.~A.}\ \bibnamefont
  {collaboration}} (\bibinfo {collaboration} {ATLAS}),\ }\href@noop {}
  {\bibfield  {journal} {\bibinfo  {journal} {ATLAS-CONF-2018-031}\ } (\bibinfo
  {year} {2018})}\BibitemShut {NoStop}%
\bibitem [{\citenamefont {Collaboration}(2018)}]{CMS:2018lkl}%
  \BibitemOpen
  \bibfield  {author} {\bibinfo {author} {\bibfnamefont {C.}~\bibnamefont
  {Collaboration}} (\bibinfo {collaboration} {CMS}),\ }\href@noop {} {\bibfield
   {journal} {\bibinfo  {journal} {CMS-PAS-HIG-17-031}\ } (\bibinfo {year}
  {2018})}\BibitemShut {NoStop}%
\bibitem [{\citenamefont {Giudice}\ and\ \citenamefont
  {McCullough}(2017)}]{Giudice:2016yja}%
  \BibitemOpen
  \bibfield  {author} {\bibinfo {author} {\bibfnamefont {G.~F.}\ \bibnamefont
  {Giudice}}\ and\ \bibinfo {author} {\bibfnamefont {M.}~\bibnamefont
  {McCullough}},\ }\href {\doibase 10.1007/JHEP02(2017)036} {\bibfield
  {journal} {\bibinfo  {journal} {JHEP}\ }\textbf {\bibinfo {volume} {02}},\
  \bibinfo {pages} {036} (\bibinfo {year} {2017})},\ \Eprint
  {http://arxiv.org/abs/1610.07962} {arXiv:1610.07962 [hep-ph]} \BibitemShut
  {NoStop}%
\bibitem [{\citenamefont {Kaplan}\ and\ \citenamefont
  {Rattazzi}(2016)}]{Kaplan:2015fuy}%
  \BibitemOpen
  \bibfield  {author} {\bibinfo {author} {\bibfnamefont {D.~E.}\ \bibnamefont
  {Kaplan}}\ and\ \bibinfo {author} {\bibfnamefont {R.}~\bibnamefont
  {Rattazzi}},\ }\href {\doibase 10.1103/PhysRevD.93.085007} {\bibfield
  {journal} {\bibinfo  {journal} {Phys. Rev.}\ }\textbf {\bibinfo {volume}
  {D93}},\ \bibinfo {pages} {085007} (\bibinfo {year} {2016})},\ \Eprint
  {http://arxiv.org/abs/1511.01827} {arXiv:1511.01827 [hep-ph]} \BibitemShut
  {NoStop}%
\bibitem [{\citenamefont {Choi}\ and\ \citenamefont
  {Im}(2016{\natexlab{b}})}]{Choi:2015fiu}%
  \BibitemOpen
  \bibfield  {author} {\bibinfo {author} {\bibfnamefont {K.}~\bibnamefont
  {Choi}}\ and\ \bibinfo {author} {\bibfnamefont {S.~H.}\ \bibnamefont {Im}},\
  }\href {\doibase 10.1007/JHEP01(2016)149} {\bibfield  {journal} {\bibinfo
  {journal} {JHEP}\ }\textbf {\bibinfo {volume} {01}},\ \bibinfo {pages} {149}
  (\bibinfo {year} {2016}{\natexlab{b}})},\ \Eprint
  {http://arxiv.org/abs/1511.00132} {arXiv:1511.00132 [hep-ph]} \BibitemShut
  {NoStop}%
\bibitem [{\citenamefont {Harigaya}\ and\ \citenamefont
  {Ibe}(2014{\natexlab{a}})}]{Harigaya:2014eta}%
  \BibitemOpen
  \bibfield  {author} {\bibinfo {author} {\bibfnamefont {K.}~\bibnamefont
  {Harigaya}}\ and\ \bibinfo {author} {\bibfnamefont {M.}~\bibnamefont {Ibe}},\
  }\href {\doibase 10.1016/j.physletb.2014.09.061} {\bibfield  {journal}
  {\bibinfo  {journal} {Phys. Lett.}\ }\textbf {\bibinfo {volume} {B738}},\
  \bibinfo {pages} {301} (\bibinfo {year} {2014}{\natexlab{a}})},\ \Eprint
  {http://arxiv.org/abs/1404.3511} {arXiv:1404.3511 [hep-ph]} \BibitemShut
  {NoStop}%
\bibitem [{\citenamefont {Harigaya}\ and\ \citenamefont
  {Ibe}(2014{\natexlab{b}})}]{Harigaya:2014rga}%
  \BibitemOpen
  \bibfield  {author} {\bibinfo {author} {\bibfnamefont {K.}~\bibnamefont
  {Harigaya}}\ and\ \bibinfo {author} {\bibfnamefont {M.}~\bibnamefont {Ibe}},\
  }\href {\doibase 10.1007/JHEP11(2014)147} {\bibfield  {journal} {\bibinfo
  {journal} {JHEP}\ }\textbf {\bibinfo {volume} {11}},\ \bibinfo {pages} {147}
  (\bibinfo {year} {2014}{\natexlab{b}})},\ \Eprint
  {http://arxiv.org/abs/1407.4893} {arXiv:1407.4893 [hep-ph]} \BibitemShut
  {NoStop}%
\bibitem [{\citenamefont {Scherrer}\ \emph {et~al.}(1991)\citenamefont
  {Scherrer}, \citenamefont {Cline}, \citenamefont {Raby},\ and\ \citenamefont
  {Seckel}}]{Scherrer:1991yu}%
  \BibitemOpen
  \bibfield  {author} {\bibinfo {author} {\bibfnamefont {R.~J.}\ \bibnamefont
  {Scherrer}}, \bibinfo {author} {\bibfnamefont {J.~M.}\ \bibnamefont {Cline}},
  \bibinfo {author} {\bibfnamefont {S.}~\bibnamefont {Raby}}, \ and\ \bibinfo
  {author} {\bibfnamefont {D.}~\bibnamefont {Seckel}},\ }\href {\doibase
  10.1103/PhysRevD.44.3760} {\bibfield  {journal} {\bibinfo  {journal} {Phys.
  Rev.}\ }\textbf {\bibinfo {volume} {D44}},\ \bibinfo {pages} {3760} (\bibinfo
  {year} {1991})}\BibitemShut {NoStop}%
\bibitem [{\citenamefont {Pierce}\ and\ \citenamefont
  {Shakya}(2019)}]{Pierce:2019ozl}%
  \BibitemOpen
  \bibfield  {author} {\bibinfo {author} {\bibfnamefont {A.}~\bibnamefont
  {Pierce}}\ and\ \bibinfo {author} {\bibfnamefont {B.}~\bibnamefont
  {Shakya}},\ }\href@noop {} {\  (\bibinfo {year} {2019})},\ \Eprint
  {http://arxiv.org/abs/1901.05493} {arXiv:1901.05493 [hep-ph]} \BibitemShut
  {NoStop}%
\bibitem [{\citenamefont {Dimopoulos}\ and\ \citenamefont
  {Hall}(1987)}]{Dimopoulos:1987rk}%
  \BibitemOpen
  \bibfield  {author} {\bibinfo {author} {\bibfnamefont {S.}~\bibnamefont
  {Dimopoulos}}\ and\ \bibinfo {author} {\bibfnamefont {L.~J.}\ \bibnamefont
  {Hall}},\ }\href {\doibase 10.1016/0370-2693(87)90593-4} {\bibfield
  {journal} {\bibinfo  {journal} {Phys. Lett.}\ }\textbf {\bibinfo {volume}
  {B196}},\ \bibinfo {pages} {135} (\bibinfo {year} {1987})}\BibitemShut
  {NoStop}%
\bibitem [{\citenamefont {Babu}\ \emph {et~al.}(2006)\citenamefont {Babu},
  \citenamefont {Mohapatra},\ and\ \citenamefont {Nasri}}]{Babu:2006xc}%
  \BibitemOpen
  \bibfield  {author} {\bibinfo {author} {\bibfnamefont {K.~S.}\ \bibnamefont
  {Babu}}, \bibinfo {author} {\bibfnamefont {R.~N.}\ \bibnamefont {Mohapatra}},
  \ and\ \bibinfo {author} {\bibfnamefont {S.}~\bibnamefont {Nasri}},\ }\href
  {\doibase 10.1103/PhysRevLett.97.131301} {\bibfield  {journal} {\bibinfo
  {journal} {Phys. Rev. Lett.}\ }\textbf {\bibinfo {volume} {97}},\ \bibinfo
  {pages} {131301} (\bibinfo {year} {2006})},\ \Eprint
  {http://arxiv.org/abs/hep-ph/0606144} {arXiv:hep-ph/0606144 [hep-ph]}
  \BibitemShut {NoStop}%
\bibitem [{\citenamefont {Nelson}\ and\ \citenamefont
  {Xiao}(2019)}]{Nelson:2019fln}%
  \BibitemOpen
  \bibfield  {author} {\bibinfo {author} {\bibfnamefont {A.~E.}\ \bibnamefont
  {Nelson}}\ and\ \bibinfo {author} {\bibfnamefont {H.}~\bibnamefont {Xiao}},\
  }\href@noop {} {\  (\bibinfo {year} {2019})},\ \Eprint
  {http://arxiv.org/abs/1901.08141} {arXiv:1901.08141 [hep-ph]} \BibitemShut
  {NoStop}%
\bibitem [{\citenamefont {Aitken}\ \emph {et~al.}(2017)\citenamefont {Aitken},
  \citenamefont {McKeen}, \citenamefont {Neder},\ and\ \citenamefont
  {Nelson}}]{Aitken:2017wie}%
  \BibitemOpen
  \bibfield  {author} {\bibinfo {author} {\bibfnamefont {K.}~\bibnamefont
  {Aitken}}, \bibinfo {author} {\bibfnamefont {D.}~\bibnamefont {McKeen}},
  \bibinfo {author} {\bibfnamefont {T.}~\bibnamefont {Neder}}, \ and\ \bibinfo
  {author} {\bibfnamefont {A.~E.}\ \bibnamefont {Nelson}},\ }\href {\doibase
  10.1103/PhysRevD.96.075009} {\bibfield  {journal} {\bibinfo  {journal} {Phys.
  Rev.}\ }\textbf {\bibinfo {volume} {D96}},\ \bibinfo {pages} {075009}
  (\bibinfo {year} {2017})},\ \Eprint {http://arxiv.org/abs/1708.01259}
  {arXiv:1708.01259 [hep-ph]} \BibitemShut {NoStop}%
\bibitem [{\citenamefont {Elor}\ \emph {et~al.}(2019)\citenamefont {Elor},
  \citenamefont {Escudero},\ and\ \citenamefont {Nelson}}]{Elor:2018twp}%
  \BibitemOpen
  \bibfield  {author} {\bibinfo {author} {\bibfnamefont {G.}~\bibnamefont
  {Elor}}, \bibinfo {author} {\bibfnamefont {M.}~\bibnamefont {Escudero}}, \
  and\ \bibinfo {author} {\bibfnamefont {A.}~\bibnamefont {Nelson}},\ }\href
  {\doibase 10.1103/PhysRevD.99.035031} {\bibfield  {journal} {\bibinfo
  {journal} {Phys. Rev.}\ }\textbf {\bibinfo {volume} {D99}},\ \bibinfo {pages}
  {035031} (\bibinfo {year} {2019})},\ \Eprint
  {http://arxiv.org/abs/1810.00880} {arXiv:1810.00880 [hep-ph]} \BibitemShut
  {NoStop}%
\bibitem [{\citenamefont {Str{\"a}ng}(2005)}]{Strang:aa}%
  \BibitemOpen
  \bibfield  {author} {\bibinfo {author} {\bibfnamefont {J.~E.}\ \bibnamefont
  {Str{\"a}ng}},\ }\href@noop {} {\bibfield  {journal} {\bibinfo  {journal}
  {Acad. Roy. Belg. Bull. Cl. Sci. (6) 16 (2005), no. 7-12, 269--287}\ }
  (\bibinfo {year} {2005})}\BibitemShut {NoStop}%
\bibitem [{\citenamefont {Davis}\ \emph {et~al.}(1983)\citenamefont {Davis},
  \citenamefont {Dine},\ and\ \citenamefont {Seiberg}}]{Davis:1983mz}%
  \BibitemOpen
  \bibfield  {author} {\bibinfo {author} {\bibfnamefont {A.~C.}\ \bibnamefont
  {Davis}}, \bibinfo {author} {\bibfnamefont {M.}~\bibnamefont {Dine}}, \ and\
  \bibinfo {author} {\bibfnamefont {N.}~\bibnamefont {Seiberg}},\ }\href
  {\doibase 10.1016/0370-2693(83)91332-1} {\bibfield  {journal} {\bibinfo
  {journal} {Phys. Lett.}\ }\textbf {\bibinfo {volume} {125B}},\ \bibinfo
  {pages} {487} (\bibinfo {year} {1983})}\BibitemShut {NoStop}%
\bibitem [{\citenamefont {Affleck}\ \emph {et~al.}(1984)\citenamefont
  {Affleck}, \citenamefont {Dine},\ and\ \citenamefont
  {Seiberg}}]{Affleck:1983mk}%
  \BibitemOpen
  \bibfield  {author} {\bibinfo {author} {\bibfnamefont {I.}~\bibnamefont
  {Affleck}}, \bibinfo {author} {\bibfnamefont {M.}~\bibnamefont {Dine}}, \
  and\ \bibinfo {author} {\bibfnamefont {N.}~\bibnamefont {Seiberg}},\ }\href
  {\doibase 10.1016/0550-3213(84)90058-0} {\bibfield  {journal} {\bibinfo
  {journal} {Nucl. Phys.}\ }\textbf {\bibinfo {volume} {B241}},\ \bibinfo
  {pages} {493} (\bibinfo {year} {1984})}\BibitemShut {NoStop}%
\bibitem [{\citenamefont {Seiberg}(1994)}]{Seiberg:1994bz}%
  \BibitemOpen
  \bibfield  {author} {\bibinfo {author} {\bibfnamefont {N.}~\bibnamefont
  {Seiberg}},\ }\href {\doibase 10.1103/PhysRevD.49.6857} {\bibfield  {journal}
  {\bibinfo  {journal} {Phys. Rev.}\ }\textbf {\bibinfo {volume} {D49}},\
  \bibinfo {pages} {6857} (\bibinfo {year} {1994})},\ \Eprint
  {http://arxiv.org/abs/hep-th/9402044} {arXiv:hep-th/9402044 [hep-th]}
  \BibitemShut {NoStop}%
\end{thebibliography}%
\end{document}